\documentclass[10pt]{article}

\usepackage[T1]{fontenc}
\usepackage{amsmath,amssymb}

\DeclareMathOperator\ric{Ric}
\usepackage{empheq}
\usepackage{graphicx}
\usepackage{soul}
\usepackage{float}
\usepackage{slashed}
\usepackage{amsthm}
\usepackage{amsfonts}
\usepackage{amsthm}
\usepackage{amsmath}
\usepackage{amscd}
\usepackage{bbm}
\newtheorem{theorem}{Theorem}[section]
\newtheorem*{theorem*}{Theorem}
\graphicspath{ {Writings/} }
\usepackage{geometry}
\usepackage{enumerate}
\usepackage{pgf,tikz}
\usepackage{csquotes}
\usepackage[bottom]{footmisc}

\usepackage{caption}
\usepackage{subcaption}

\usepackage{color}
\usepackage{xcolor}
\newcommand{\bsy}{\ensuremath{\boldsymbol}}

\DeclareMathOperator{\chih}{\widehat{\chi}}
\DeclareMathOperator{\omegah}{\hat{\omega}}

\DeclareMathOperator{\nablasl}{\slashed{\nabla}}
\DeclareMathOperator{\divsl}{\slashed{\divv}}
\DeclareMathOperator{\curlsl}{\slashed{\curl}}

\DeclareMathOperator{\chibh}{\widehat{\underline{\chi}}}
\DeclareMathOperator{\omegabh}{\hat{\underline{\omega}}}
\DeclareMathOperator{\etab}{\underline{\eta}}
\DeclareMathOperator{\omegab}{\underline{\omega}}
\DeclareMathOperator{\yb}{\underline{\xi}}

\DeclareMathOperator{\alphab}{\underline{\alpha}}
\DeclareMathOperator{\betab}{\underline{\beta}}
\DeclareMathOperator{\chib}{\underline{\chi}}

\DeclareMathOperator{\fo}{\overset{\text{\scalebox{.6}{$(1)$}}}{\mathfrak{f}}}
\DeclareMathOperator{\Gammao}{\overset{\text{\scalebox{.6}{$(1)$}}}{\Gamma}}
\DeclareMathOperator{\psio}{\overset{\text{\scalebox{.6}{$(1)$}}}{\psi}}
\DeclareMathOperator{\etao}{\overset{\text{\scalebox{.6}{$(1)$}}}{\eta}}
\DeclareMathOperator{\zetao}{\overset{\text{\scalebox{.6}{$(1)$}}}{\zeta}}

\DeclareMathOperator{\omegabo}{\overset{\text{\scalebox{.6}{$(1)$}}}{\hat{\omegab}}}
\DeclareMathOperator{\trchio}{\overset{\text{\scalebox{.6}{$(1)$}}}{\text{tr}\chi}}
\DeclareMathOperator{\trchibo}{\overset{\text{\scalebox{.6}{$(1)$}}}{\text{tr}\chib}}
\DeclareMathOperator{\chiho}{\overset{\text{\scalebox{.6}{$(1)$}}}{\chih}}
\DeclareMathOperator{\chibho}{\overset{\text{\scalebox{.6}{$(1)$}}}{\chibh}}

\DeclareMathOperator{\alphao}{\overset{\text{\scalebox{.6}{$(1)$}}}{\alpha}}
\DeclareMathOperator{\alphabo}{\overset{\text{\scalebox{.6}{$(1)$}}}{\alphab}}
\DeclareMathOperator{\betao}{\overset{\text{\scalebox{.6}{$(1)$}}}{\beta}}
\DeclareMathOperator{\betabo}{\overset{\text{\scalebox{.6}{$(1)$}}}{\betab}}
\DeclareMathOperator{\rhoo}{\overset{\text{\scalebox{.6}{$(1)$}}}{\rho}}
\DeclareMathOperator{\sigmao}{\overset{\text{\scalebox{.6}{$(1)$}}}{\sigma}}
\DeclareMathOperator{\ybo}{\overset{\text{\scalebox{.6}{$(1)$}}}{\yb}}

\DeclareMathOperator{\divv}{\text{div}}
\DeclareMathOperator{\curl}{\text{curl}}

\newtheorem{prop}[theorem]{Proposition}
\newtheorem{remark}[theorem]{Remark}

\newtheorem{lemma}[theorem]{Lemma}

\newtheorem*{maintheorem}{Main Theorem}

\newtheorem*{remark*}{Remark}

\usepackage[affil-it]{authblk}
\usepackage[linktoc=all]{hyperref}
\hypersetup{
    colorlinks = true,
    linkcolor = {black},
    citecolor = {black}
}

\usepackage{verbatim}
\usepackage{cleveref}
\usepackage{mathtools}

\begin{document}

\title{Elliptic curvature estimates for linearised gravitational perturbations of Kerr in the full sub-extremal range $|a|<M$}

\author[1]{Gabriele Benomio}
\author[2]{Rita Teixeira da Costa}

\affil[1]{\small Gran Sasso Science Institute, Viale Francesco Crispi 7, L'Aquila (AQ), 67100, Italy}
\affil[2]{\small University of Cambridge, Department of Pure Mathematics and Mathematical Statistics, 

Wilberforce Road, Cambridge CB3 0WA, United Kingdom}

\maketitle

\begin{abstract}

We consider the linearised vacuum Einstein equations around a Kerr exterior solution and present a scheme to prove elliptic $L^2(\mathbb{S}^2)$-estimates for the linearised curvature quantities in the equations.~The scheme employs the linearised system of equations derived in \cite{Benomio_thesis_doi_2} and applies to the full sub-extremal range of Kerr parameters $0\leq |a|<M$.

\end{abstract}

\tableofcontents

\section{Introduction}

In the analysis of the vacuum Einstein equations
\begin{equation*}
\ric (g)=0 \, ,
\end{equation*}
the study of the \emph{linearised} system of equations has served as a fundamental building block.~In this paper, we focus on the linearised vacuum Einstein equations around Kerr black hole spacetimes, as derived in \cite{Benomio_thesis_doi_2}.~We introduce a scheme to prove elliptic estimates for the linearised curvature quantities on the entire exterior region (including the event horizon) of the Kerr solution \emph{in the full sub-extremal range $0\leq |a|<M$}.~As we shall outline in this introduction (see Sections \ref{sec_intro_elliptic_schwarzschild} and \ref{sec_intro_elliptic_kerr}), to prove elliptic estimates on the Kerr exterior region, one has to overcome certain difficulties which are, in most part, already manifest at (and close to) the event horizon, and which are absent for analogous estimates on a Schwarzschild ($|a|=0$) exterior solution. 

\medskip

The motivation for the present paper is to provide an important ingredient in the linear stability problem for the system of \cite{Benomio_thesis_doi_2} in the full sub-extremal range $0\leq |a|<M$.~As we shall explain (see Section \ref{sec_intro_applications}), our main result will likely play a crucial role in proving new linear stability statements which we hope will be directly applied to the nonlinear stability problem for the Kerr solution in the full sub-extremal range $0\leq |a|<M$.

\medskip

The literature on the Kerr stability problem is vast and includes several recent developments. In Section \ref{sec_intro_applications}, we provide a survey of those recent works which are more closely related to the subject of the present paper. For further background on the problem and a more thorough discussion of previous works, the reader may refer to the introduction of \cite{benomio_kerr_system_2}.

\subsection{Elliptic estimates for linearised curvature on Schwarzschild} \label{sec_intro_elliptic_schwarzschild}

The elliptic estimates of the form presented in this paper can be more easily derived on a Schwarzschild ($|a|=0$) exterior solution (see \cite{DHR, benomio_schwarzschild_stability} and Section \ref{sec_intro_applications}). For a direct connection to our scheme for Kerr, we briefly review here the Schwarzschild elliptic estimates for the linearised curvature quantities 
\begin{equation} \label{intro_list_curvature}
    \psio =\left\lbrace \betao,\betabo,(\rhoo,\sigmao) \right\rbrace
\end{equation}
as proven for the linearised system of equations of \cite{Benomio_thesis_doi_2} (with $|a|=0$, see also \cite{benomio_schwarzschild_stability} and the system reported here in Section \ref{sec_lin_system}).

\medskip

To establish an elliptic estimate for $\beta$, one starts by considering the linearised Bianchi equation (lower order terms are henceforth denoted by ``l.o.t.'')
\begin{equation} \label{intro_aux_00}
\nablasl_3\alphao=-2\, \slashed{\mathcal{D}}{}_2^{\star}  \betao +\,\text{l.o.t.} \, ,
\end{equation}
where the objects appearing in the equation are tensors and differential operators defined over the foliation (round) spheres $\mathbb{S}^2$ (see Sections \ref{sec_kerr_exterior_manifold} and \ref{sec_lin_eqns_Bianchi} with $|a|=0$). One can commute equation \eqref{intro_aux_00} with the angular operator $\slashed{\mathcal{D}}{}_2^{\star}\slashed{\text{div}}$ and obtain the equation
\begin{equation*}
\nablasl_3\slashed{\mathcal{D}}{}_2^{\star}\slashed{\text{div}}\alphao=-2\, \slashed{\mathcal{D}}{}_2^{\star}\slashed{\text{div}}\slashed{\mathcal{D}}{}_2^{\star} \betao +\,\text{l.o.t.} \, ,
\end{equation*}
which immediately allows for an $L^2(\mathbb{S}^2)$-estimate of the form
\begin{equation} \label{intro_aux_000}
\|\slashed{\mathcal{D}}{}_2^{\star}\slashed{\text{div}}\slashed{\mathcal{D}}{}_2^{\star} \betao\|_{L^2(\mathbb{S}^2)}\lesssim \|\nablasl_3\slashed{\mathcal{D}}{}_2^{\star}\slashed{\text{div}}\alphao\|_{L^2(\mathbb{S}^2)} + \|\text{l.o.t.}\|_{L^2(\mathbb{S}^2)} \, .
\end{equation}
By the standard ellipticity properties on (round) spheres of the angular operator on the left hand side of \eqref{intro_aux_000}, one then derives the estimate (see the notation \eqref{schematic_notation_th})
\begin{equation*} 
\|\nablasl{}^3\betao\|_{L^2(\mathbb{S}^2)}\lesssim \sum_{0\leq i_1+i_2+i_3\leq 3}\|\nablasl{}^{i_1}_4\nablasl{}_3^{i_2}\nablasl{}^{i_3}(\alphao,\alphabo)\|_{L^2(\mathbb{S}^2)} + \|\text{l.o.t.}\|_{L^2(\mathbb{S}^2)} \, .
\end{equation*}
By employing the other linearised Bianchi equations in a similar fashion, one can show that analogous estimates hold for all the linearised curvature quantities \eqref{intro_list_curvature}, i.e.
\begin{equation}
\|\nablasl{}^3\psio\|_{L^2(\mathbb{S}^2)} \lesssim \sum_{0\leq i_1+i_2+i_3\leq 3}\|\nablasl{}^{i_1}_4\nablasl{}_3^{i_2}\nablasl{}^{i_3}(\alphao,\alphabo)\|_{L^2(\mathbb{S}^2)}+\|\text{l.o.t.}\|_{L^2(\mathbb{S}^2)}   \, . \label{intro_aux_0000}
\end{equation}
The estimates \eqref{intro_aux_0000} for angular derivatives then lead, by commuting the linearised Bianchi equations, to control over all third order derivatives of the linearised curvature quantities \eqref{intro_list_curvature}. Moreover, a higher order commutation procedure yields control over all higher (than third) order derivatives.

\medskip

We remark that the quantities $\alpha$ and $\alphab$ are also linearised, sometimes dubbed ``extremal'', curvature quantities which, together with the quantities $\psi$ listed in \eqref{intro_list_curvature}, exhaust all the linearised curvature quantities in the system of equations. The meaning of the appearance of the quantities $\alpha$ and $\alphab$ on the right hand side of both the estimates \eqref{intro_aux_0000} and our Kerr elliptic estimates (see below) is discussed in Section \ref{sec_intro_applications}.

\subsection{Elliptic estimates for linearised curvature on Kerr} \label{sec_intro_elliptic_kerr}

The core of the present paper is the derivation of estimates of the form of \eqref{intro_aux_0000} for all the non-extremal linearised curvature quantities, i.e.
\begin{equation} \label{intro_list_curvature_c}
    \psio =\left\lbrace \betao,\betabo,(\rhoo,\sigmao) \right\rbrace \, ,
\end{equation}
in the system of \cite{Benomio_thesis_doi_2} (reported here in Section \ref{sec_lin_system}) on a Kerr exterior solution. Such estimates are achieved on the entire exterior region (including the event horizon) for the full sub-extremal range of parameters $|a|<M$.

\medskip

As in the Schwarzschild $|a|=0$ case, to establish an elliptic estimate for $\beta$, one may start by considering the linearised Bianchi equation
\begin{equation} \label{intro_aux_0}
\nablasl_3\alphao=-2\, \slashed{\mathcal{D}}{}_2^{\star} \betao + \,\text{l.o.t.} \, ,
\end{equation}
where the objects appearing in the equation are tensors and differential operators defined over the (regular) \emph{non-integrable} distribution 
\begin{equation*}
\mathfrak{D}_{\mathcal{N}_{\text{as}}}=\left\langle e_4^{\text{as}},e_3^{\text{as}} \right\rangle^{\perp} 
\end{equation*}
induced by the algebraically special frame $\mathcal{N}_{\text{as}}$ of the Kerr metric $g$ (see Section \ref{sec_as_frame}). While, by commuting equation \eqref{intro_aux_0} with the differential operator $\slashed{\mathcal{D}}{}_2^{\star}\slashed{\text{div}}$, one can derive the $L^2(\mathbb{S}^2)$-estimate\footnote{Although one deals with tensors over a non-integrable distribution, one can define a natural pointwise norm of tensors and still derive $L^2(\mathbb{S}^2)$-estimates by integrating the pointwise norm over the foliation spheres.}
\begin{equation} \label{intro_aux_1}
\|\slashed{\mathcal{D}}{}_2^{\star}\slashed{\text{div}}\slashed{\mathcal{D}}{}_2^{\star} \betao\|_{L^2(\mathbb{S}^2)}\lesssim \|\nablasl_3\slashed{\mathcal{D}}{}_2^{\star}\slashed{\text{div}}\alphao\|_{L^2(\mathbb{S}^2)} + \|\text{l.o.t.}\|_{L^2(\mathbb{S}^2)}
\end{equation}
on the foliation spheres $\mathbb{S}^2$, it remains unclear how control over the differential operator on the left hand side of \eqref{intro_aux_1} can possibly lead to control over derivatives of the linearised curvature quantities. In other words, contrary to the Schwarzschild $|a|=0$ case, there is no direct elliptic structure over non-integrable distributions that one can exploit to obtain the desired estimates.

\medskip

To overcome this difficulty, we start by performing a $\mathbb{S}^2$-projection of the system of linearised Bianchi equations over the foliation (Boyer--Lindquist, in our case) spheres (see Section \ref{sec_proj_formulae}). Once projected, equation \eqref{intro_aux_0} becomes an equation of the form (see Section \ref{sec_proj_lin_system})
\begin{equation} \label{intro_aux_2}
\check{\nablasl}_3\widetilde{\alphao}=-2\, \check{\slashed{\mathcal{D}}}{}_2^{\star}  \widetilde{\betao} +\mathfrak{k}\,\widehat{\otimes} \check{\nablasl}_4 \widetilde{\betao} +\mathfrak{h}\,\widehat{\otimes} \check{\nablasl}_3 \widetilde{\betao}+\text{l.o.t.} \, .
\end{equation}
The objects appearing in equation \eqref{intro_aux_2} are now (tilded) tensors and (checked) differential operators defined over the foliation spheres $\mathbb{S}^2$, rather than $\mathfrak{D}_{\mathcal N_{\rm as}}$. Note, however, that equation \eqref{intro_aux_2}, as well as the other $\mathbb{S}^2$-projected linearised Bianchi equations, now involve \emph{top-order error terms}, where the $\mathbb{S}^2$ one-forms $\mathfrak{k}$ and $\mathfrak{h}$ multiplying all the top-order error terms are \emph{explicit} background one-forms defined as
\begin{align*}
    \mathfrak{k}(X)&=\frac{1}{2}\, g(e_3^{\text{as}},X) \, , & \mathfrak{h}(X)&=\frac{1}{2}\, g(e_4^{\text{as}},X) 
\end{align*}
for any vector field $X$ tangent to the foliation spheres, and thus such that
\begin{align} \label{intro_aux_2b}
\mathfrak{h}|_{\mathcal{H}^+}&=0 \, , &  |\mathfrak{k}| \, , \, &|\mathfrak{h}|  \xrightarrow[]{r\rightarrow \infty} 0  
\end{align}
and identically vanishing for $|a|=0$. The former of the identities \eqref{intro_aux_2b} is tied to the properties of the frame $\mathcal{N}_{\text{as}}$ on the event horizon, where $e_4^{\text{as}}$ is the Killing generator of the event horizon and can be (locally) completed to an \emph{integrable} distribution $\left\langle e_4^{\text{as}},e_1^{\text{as}},e_2^{\text{as}}\right\rangle =T\mathcal{H}^+$ (see Section \ref{sec_as_frame}).

\medskip

The $\mathbb{S}^2$ one-forms $\mathfrak{k}$ and $\mathfrak{h}$ satisfy the identity
\begin{equation*}
    \mathfrak{h}=\frac{\Delta}{\Sigma}\, \mathfrak{k} \, ,
\end{equation*}
with $\Delta$ and $\Sigma$ the standard Kerr functions of the Boyer--Lindquist coordinates $(r,\theta)$ (see Section \ref{sec_kerr_exterior_manifold}), which, together with the monotonicity of the radial function
\begin{equation} \label{aux_k_intro}
    \sup_{\mathbb{S}^2}|\mathfrak{k}|=\frac{ar}{2(r^2+a^2)} \, ,
\end{equation}
play a crucial role in the problem. Indeed, suppose that one is provided with a scheme which, by using the linearised Bianchi equations of the form of \eqref{intro_aux_2} and exploiting the size of $\mathfrak{k}$ and $\mathfrak{h}$ to absorb the top-order error terms, proves the desired elliptic estimates over foliation spheres $\mathbb{S}^2$ \emph{on and close to the event horizon}, say for radius $r_+\leq r \leq r_+ +\epsilon$ with sufficiently small $\epsilon>0$. Then, in the region with $r\geq r_+ +\epsilon$, one can rewrite the error terms of equation \eqref{intro_aux_2}, as well as the error terms in the other linearised Bianchi equations, in the form
\begin{equation} \label{intro_aux_2_bis}
\check{\nablasl}_3\widetilde{\alphao}=-2\, \check{\slashed{\mathcal{D}}}{}_2^{\star}  \widetilde{\betao} +\mathfrak{k}\,\widehat{\otimes} \check{\nablasl}_4 \widetilde{\betao} + c_{\epsilon}\mathfrak{k}\,\widehat{\otimes} \,\frac{\Delta}{c_{\epsilon}\Sigma}\,\check{\nablasl}_3 \widetilde{\betao}+\text{l.o.t.} \, ,
\end{equation}
with nowhere vanishing angular function $c_{\epsilon}(\theta)=\Delta(r_++\epsilon)/\Sigma(r_++\epsilon,\theta)$. The one-form $c_{\epsilon}\mathfrak{k}$ equals the one-form $\mathfrak{h}$ at $r=r_++\epsilon$ and is strictly \emph{decreasing} for all $r\geq r_+$ (in the sense that \eqref{aux_k_intro} is a strictly decreasing radial function), thus suggesting that the \emph{same} scheme applied to prove elliptic estimates on and close to the event horizon (i.e. for $r_+\leq r \leq r_+ +\epsilon$) starting from the Bianchi equations in the form of \eqref{intro_aux_2} would allow one to establish the desired estimates on the remaining part of the exterior region (i.e. for $r\geq r_+ + \epsilon$) starting from the Bianchi equations in the form of \eqref{intro_aux_2_bis},\footnote{Away from the event horizon, one can estimate the non-degenerate quantity $(\Delta/c_{\epsilon}\Sigma)\check{\nablasl}_3\,\beta$ in \eqref{intro_aux_2_bis} and analogous rescaled quantities appearing in the other linearised Bianchi equations.} in that the factors multiplying the top-order error terms in the latter equations are strictly smaller.

\medskip

While one may entertain the idea of proving (and then combining) elliptic estimates in two distinct regions as described above, it turns out that the desired estimates can be achieved on the entire exterior region \emph{at once}. This requires one to exploit the linearised Bianchi equations with top-order error terms in the form of \eqref{intro_aux_2} and to compute explicitly all the factors multiplying the top-order error terms in the estimates, as carried out by the scheme presented in the paper. In view of the discussed monotonicity of the multiplying factors, one may nonetheless view the applicability of our scheme on and close to the event horizon as the central ingredient. Moreover, though computationally simpler, the estimates on and close to the event horizon already exhibit the essential difficulties of the problem and features of our scheme. In fact, the closeness to the event horizon introduces a smallness parameter (originating from the former of the identities \eqref{intro_aux_2b}) in the estimates which, while allowing to dispense with computing explicitly the multiplying factors, does not \emph{alone} guarantee the closure of our scheme. Indeed, certain potentially problematic error terms already arise in the estimates \emph{at} the event horizon, and thus come \emph{without} smallness parameter. To deal with such error terms, our scheme crucially relies on some additional structure in the estimates which only becomes apparent once all linearised curvature quantities are estimated at the same time. We shall now give an overview of the proof of these estimates.

\medskip

To establish elliptic estimates for the $\mathbb{S}^2$-projected linearised curvature quantities
\begin{equation} \label{intro_list_curvature_b}
    \widetilde{\psio} =\left\lbrace \widetilde{\betao},\widetilde{\betabo},(\rhoo,\sigmao) \right\rbrace \, ,
\end{equation}
we start by commuting equation \eqref{intro_aux_2} with $\check{\slashed{\mathcal{D}}}{}_2^{\star}\,\check{\slashed{\text{div}}}$ and obtain
\begin{equation*}
\check{\nablasl}_3\check{\slashed{\mathcal{D}}}{}_2^{\star}\check{\slashed{\text{div}}}\widetilde{\alphao}=-2\, \check{\slashed{\mathcal{D}}}{}_2^{\star}\check{\slashed{\text{div}}}\check{\slashed{\mathcal{D}}}{}_2^{\star}  \widetilde{\betao} +\mathfrak{k}\,\widehat{\otimes} \check{\nablasl}_4 \check{\slashed{\mathcal{D}}}{}_2^{\star}\check{\slashed{\text{div}}}\widetilde{\betao} +\mathfrak{h}\,\widehat{\otimes} \check{\nablasl}_3 \check{\slashed{\mathcal{D}}}{}_2^{\star}\check{\slashed{\text{div}}}\widetilde{\betao}+\text{l.o.t.} \, ,
\end{equation*} 
from which we derive the estimate (cf. \eqref{est_ang_op_beta_inter} in the proof of Proposition \ref{prop_estimates_covariant_angular_derivatives})
\begin{align}
\|\check{\slashed{\mathcal{D}}}{}_2^{\star}\check{\slashed{\textup{div}}}\check{\slashed{\mathcal{D}}}{}_2^{\star}\widetilde{\betao}\|_{L^2(\mathbb{S}^2)}\lesssim& \, \boxed{\|\check{\nablasl}_4\check{\nablasl}{}^2\widetilde{\betao}\|_{L^2(\mathbb{S}^2)}}+h\|\check{\nablasl}_3\check{\nablasl}{}^2\widetilde{\betao}\|_{L^2(\mathbb{S}^2)}  \label{intro_aux_3}\\
&+\sum_{0\leq i_1+i_2+i_3\leq 3}\|\check{\nablasl}{}^{i_1}_4\check{\nablasl}{}_3^{i_2}\check{\nablasl}{}^{i_3}(\widetilde{\alphao},\widetilde{\alphabo})\|_{L^2(\mathbb{S}^2)}+\|\text{l.o.t.}\|_{L^2(\mathbb{S}^2)} \nonumber
\end{align}
for some positive radial function $h$ which identically vanishes on the event horizon. The presence of the $h$-factor originates from the former of the properties \eqref{intro_aux_2b}. It turns out (cf. Propositions \ref{prop_commutation_1} and \ref{prop_commutation_2} below) that, by using the $\mathbb{S}^2$-projected linearised Bianchi equations, one can estimate the top-order error terms by top-order angular derivatives of curvature quantities and/or top-order mixed derivatives of the extremal curvature quantities (plus additional lower order terms). Crucially, the only error term on the right hand side of \eqref{intro_aux_3} which is \emph{not} multiplied by a $h$-factor (i.e. the boxed error term) can be controlled, at top order, by the extremal curvature quantities, i.e.
\begin{align}  \label{intro_aux_4}
\|\check{\nablasl}_4 \check{\nablasl}{}^2\widetilde{\betao}\|_{L^2(\mathbb{S}^2)}\lesssim& \,  \sum_{0\leq i_1+i_2+i_3\leq 3}\|\check{\nablasl}{}^{i_1}_4\check{\nablasl}{}_3^{i_2}\check{\nablasl}{}^{i_3}(\widetilde{\alphao},\widetilde{\alphabo})\|_{L^2(\mathbb{S}^2)}+\|\text{l.o.t.}\|_{L^2(\mathbb{S}^2)} \, .
\end{align}
By establishing ellipticity properties on (now not anymore round) spheres of the angular operator on the left hand side of \eqref{intro_aux_3} (see Lemma \ref{lemma_angular_operators} below), one can obtain the estimate (cf. \eqref{eq:est_before_assorbing_beta} in Proposition \ref{prop_angular_before_assorbing})
\begin{align}
\|\check{\nablasl}{}^3\widetilde{\betao}\|_{L^2(\mathbb{S}^2)}\lesssim & \,  h\|\check{\nablasl}{}^3(\rhoo,\sigmao)\|_{L^2(\mathbb{S}^2)}+h\|\check{\nablasl}{}^3\widetilde{\betabo}\|_{L^2(\mathbb{S}^2)} \label{intro_aux_5}\\
&+\sum_{0\leq i_1+i_2+i_3\leq 3}\|\check{\nablasl}{}^{i_1}_4\check{\nablasl}{}_3^{i_2}\check{\nablasl}{}^{i_3}(\widetilde{\alphao},\widetilde{\alphabo})\|_{L^2(\mathbb{S}^2)}+\|\text{l.o.t.}\|_{L^2(\mathbb{S}^2)} \nonumber
\end{align}
over foliation spheres which are on or sufficiently close to the event horizon (i.e. for $h$ sufficiently small). The property that \emph{all} top-order non-extremal curvature terms on the right hand side of \eqref{intro_aux_5} come with a $h$-factor is tied to the fact that the top-order error terms in \eqref{intro_aux_3} are either multiplied by a $h$-factor or controlled, at top order, by the extremal curvature quantities. An estimate with analogous structure can also be obtained for $\betab$ (cf. \eqref{eq:est_before_assorbing_betab} in Proposition \ref{prop_angular_before_assorbing}).

\medskip

The $\mathbb{S}^2$-projected linearised Bianchi equations also allow us to derive $L^2(\mathbb{S}^2)$-estimates for $\rho$ and $\sigma$. One obtains the estimate (cf. \eqref{est_ang_op_rho_sigma_inter} in the proof of Proposition \ref{prop_estimates_covariant_angular_derivatives})
\begin{align}
\|\check{\slashed{\textup{div}}}\check{\slashed{\mathcal{D}}}{}_2^{\star}\check{\slashed{\mathcal{D}}}{}_1^{\star}(\rhoo,\sigmao)\|_{L^2(\mathbb{S}^2)} \lesssim& \, \boxed{\|\check{\nablasl}_4\check{\nablasl}{}^2(\rhoo,\sigmao)\|_{L^2(\mathbb{S}^2)}+\|\check{\nablasl}{}_4^2\check{\nablasl}\widetilde{\betabo}\|_{L^2(\mathbb{S}^2)}}  \label{intro_aux_60}\\
&+h\|\check{\nablasl}_3\check{\nablasl}{}^2(\rhoo,\sigmao)\|_{L^2(\mathbb{S}^2)}+h\|\check{\nablasl}_4\check{\nablasl}_3\check{\nablasl}\widetilde{\betabo}\|_{L^2(\mathbb{S}^2)} \nonumber\\
&+\sum_{0\leq i_1+i_2+i_3\leq 3}\|\check{\nablasl}{}^{i_1}_4\check{\nablasl}{}_3^{i_2}\check{\nablasl}{}^{i_3}(\widetilde{\alphao},\widetilde{\alphabo})\|_{L^2(\mathbb{S}^2)}+\|\text{l.o.t.}\|_{L^2(\mathbb{S}^2)} \nonumber\, .
\end{align}
The boxed top-order error terms in the estimate come \emph{without} $h$-factor. It is again the case (cf. Propositions \ref{prop_commutation_1} and \ref{prop_commutation_2} below) that, by using the $\mathbb{S}^2$-projected linearised Bianchi equations, one can estimate the top-order error terms by top-order angular derivatives of curvature quantities and/or top-order mixed derivatives of the extremal curvature quantities (plus additional lower order terms). In particular, the boxed error terms can be controlled as follows
\begin{align}
\|\check{\nablasl}_4 \check{\nablasl}{}^2 (\rhoo,\sigmao)\|_{L^2(\mathbb{S}^2)} +\|\check{\nablasl}{}_4^2\check{\nablasl}\widetilde{\betabo}\|_{L^2(\mathbb{S}^2)}\lesssim& \, h\|\check{\nablasl}{}^3(\rhoo,\sigmao)\|_{L^2(\mathbb{S}^2)}+\boxed{\|\check{\nablasl}{}^3\widetilde{\betao}\|_{L^2(\mathbb{S}^2)}}+h\|\check{\nablasl}{}^3\widetilde{\betabo}\|_{L^2(\mathbb{S}^2)} \label{intro_aux_6}\\
&+\sum_{0\leq i_1+i_2+i_3\leq 3}\|\check{\nablasl}{}^{i_1}_4\check{\nablasl}{}_3^{i_2}\check{\nablasl}{}^{i_3}(\widetilde{\alphao},\widetilde{\alphabo})\|_{L^2(\mathbb{S}^2)}+\|\text{l.o.t.}\|_{L^2(\mathbb{S}^2)} \, , \nonumber
\end{align}
where the boxed term in \eqref{intro_aux_6} crucially comes \emph{without} $h$-factor. As a result, by appealing to the ellipticity properties of the angular operator on the left hand side of \eqref{intro_aux_60} (see again Lemma \ref{lemma_angular_operators} below), one derives an estimate of the form (cf. \eqref{eq:est_before_assorbing_rho_sigma} in Proposition \ref{prop_angular_before_assorbing})
\begin{align}
\|\check{\nablasl}{}^3(\rhoo,\sigmao)\|_{L^2(\mathbb{S}^2)} \lesssim& \,   \boxed{\|\check{\nablasl}{}^3\widetilde{\betao}\|_{L^2(\mathbb{S}^2)}}+h\|\check{\nablasl}{}^3\widetilde{\betabo}\|_{L^2(\mathbb{S}^2)} \label{intro_aux_7}\\
&+\sum_{0\leq i_1+i_2+i_3\leq 3}\|\check{\nablasl}{}^{i_1}_4\check{\nablasl}{}_3^{i_2}\check{\nablasl}{}^{i_3}(\widetilde{\alphao},\widetilde{\alphabo})\|_{L^2(\mathbb{S}^2)}+\|\text{l.o.t.}\|_{L^2(\mathbb{S}^2)}  \nonumber
\end{align}
over foliation spheres which are on or sufficiently close to the event horizon (i.e.\ for $h$ sufficiently small). Contrary to the estimate \eqref{intro_aux_5}, the right hand side of the estimate \eqref{intro_aux_7} possesses a top-order non-extremal curvature term which comes \emph{without} $h$-factor (boxed term).

\medskip

Upon combining the estimate \eqref{intro_aux_5}, the analogue of \eqref{intro_aux_5} for $\betab$ and the estimate \eqref{intro_aux_7}, and by taking $h$ sufficiently small, the ``good'' structure of the estimate \eqref{intro_aux_5} allows us to absorb the ``bad'' (boxed) term in estimate \eqref{intro_aux_7} and achieve the desired estimates for all the linearised curvature quantities \eqref{intro_list_curvature_b}, i.e.\ (cf.\ Proposition \ref{main_th_angular_version})
\begin{equation}
\|\check{\nablasl}{}^3\widetilde{\psio}\|_{L^2(\mathbb{S}^2)} \lesssim \sum_{0\leq i_1+i_2+i_3\leq 3}\|\check{\nablasl}{}^{i_1}_4\check{\nablasl}{}_3^{i_2}\check{\nablasl}{}^{i_3}(\widetilde{\alphao},\widetilde{\alphabo})\|_{L^2(\mathbb{S}^2)}+\|\text{l.o.t.}\|_{L^2(\mathbb{S}^2)}    \label{intro_aux_8}
\end{equation}
for any foliation sphere on or sufficiently close to the event horizon in the full sub-extremal range of parameters $|a|<M$. With the estimates \eqref{intro_aux_8} at hand, one can obtain analogous estimates for all third order derivatives of the linearised curvature quantities \eqref{intro_list_curvature_b}. One can also iterate the scheme presented in the proof to derive higher (than third) order elliptic estimates. Moreover, by the nature of our $\mathbb{S}^2$-projection procedure, control over the $\mathbb{S}^2$-projected curvature quantities \eqref{intro_list_curvature_b} directly implies control over the original curvature quantities \eqref{intro_list_curvature_c} (see Section \ref{sec_proj_th} and the considerations therein). 

\medskip

By computing explicitly all the factors multiplying the top-order error terms in the estimates, one can apply the \emph{same} scheme as the one illustrated above to prove the desired estimates \emph{on the entire exterior region} for the full sub-extremal range of parameters $|a|<M$. In other words, the smallness $h$-factor is, in fact, not needed to absorb the top-order error terms in the estimates. As a result, one obtains the main theorem of the paper (see Theorem \ref{main_th} for a more precise statement).

\medskip

\begin{maintheorem}
Let $0\leq |a|<M$, $k\in\mathbb{N}$ with $k\geq 3$, and finite constant $R>r_+$.~Then, there exists a universal constant $C_{k,R}>0$ such that, for any solution to the linearised system of equations, the elliptic $L^2(\mathbb{S}^2)$-estimates 
\begin{equation*}
\sum_{0\leq i_1+i_2+i_3\leq k}\|\nablasl_4^{i_3}\nablasl^{i_2}_3\nablasl^{i_3}\psio\|_{L^2(\mathbb{S}^2)} \leq C_{k,R} \left[ \sum_{0\leq i_1+i_2+i_3\leq k}\|\nablasl_4^{i_3}\nablasl^{i_2}_3\nablasl^{i_3}(\alphao,\alphabo)\|_{L^2(\mathbb{S}^2)}  +\|\textup{l.o.t.}\|_{L^2(\mathbb{S}^2)} \right]  \, ,
\end{equation*}
with $\psio=\left\lbrace \betao,\betabo,(\rhoo,\sigmao)\right\rbrace$, hold for all Boyer--Lindquist $\mathbb{S}^2$-spheres with $r_+\leq r \leq R$.
\end{maintheorem}

\medskip

We remark that the universal constant in the Main Theorem is, in fact, uniform in $0\leq |a|\leq M$ (i.e. up to, and including, $|a|=M$). We also point out that, in the very slowly rotating regime of parameters $|a| \ll M$, the Main Theorem can be proven without computing explicitly the factors multiplying the top-order error terms. Indeed, the vanishing of the one-forms $\mathfrak{k}$ and $\mathfrak{h}$ for $|a|=0$ implies that, in the very slowly rotating regime $|a| \ll M$, there are smallness parameters multiplying (and allowing to immediately absorb) \emph{all} the top-order error terms in the estimates.

\medskip

Going from the Main Theorem to proving elliptic estimates for $k+1$ derivatives of all the linearised connection coefficients would require one to exploit the transport and elliptic equations for the linearised connection coefficients present in the system. These estimates typically rely on an additional structural property of the system, namely the fact that the top-order coupling terms appearing in the elliptic equations solve ``good'' transport equations (i.e.\ without linearised curvature quantities on the right hand side). Elliptic estimates for the linearised connection coefficients are not pursued here.

\medskip

\begin{remark*}
The top-order structure of the linearised Bianchi equations which is exploited to prove the Main Theorem is common to any linearised (around the Kerr exterior manifold) system of vacuum Einstein equations for frame quantities obtained, like the system of \cite{Benomio_thesis_doi_2} employed here, by linearising the equations relative to the algebraically special frame $\mathcal{N}_{\textup{as}}$ of the Kerr metric. The other features of the gauge in which the system of \cite{Benomio_thesis_doi_2} is derived only affect the linearised Bianchi equations in the lower order terms, and are therefore not directly relevant for the proof of the Main Theorem. In this respect, our scheme to prove elliptic estimates for the linearised curvature quantities may be applied to linearised systems formulated in any geometric gauge which embeds the frame $\mathcal{N}_{\textup{as}}$ as the background reference frame. On the other hand, more specific features of the gauge determine the top-order structure of the remaining linearised equations of \cite{Benomio_thesis_doi_2} and, in particular, would be directly relevant to prove elliptic estimates for the linearised connection coefficients. 
\end{remark*}

\subsection{Previous works and future applications} \label{sec_intro_applications}

The motivation for the present paper is the linear stability problem for the Kerr solution in the full sub-extremal range $|a|<M$ employing the system of \cite{Benomio_thesis_doi_2} (see also \cite{benomio_kerr_system_2}).

\medskip

In the Schwarzschild ($|a|=0$) case, elliptic estimates for linearised curvature quantities are derived in various recent works. In a similar spirit to the present work, we mention the references \cite{DHR}, which established the linear stability of the Schwarzschild solution, and \cite{benomio_schwarzschild_stability}, which revisited the latter proof. In both cases, estimates of the form of \eqref{intro_aux_0000} played a crucial role in obtaining \emph{orbital stability} statements, meaning uniform boundedness statements \emph{without loss of regularity}, for all the linearised curvature quantities in the system.  Building upon the orbital stability result of \cite{DHR}, the work \cite{DHRT} proves the full finite co-dimension nonlinear (orbital and asymptotic) stability of the Schwarzschild $|a|=0$ solution.

\medskip

Though the two works \cite{benomio_schwarzschild_stability,DHR} differ in their choice of geometric gauge---while \cite{DHR} derives the linearised vacuum Einstein equations in a double null gauge, \cite{benomio_schwarzschild_stability} makes use of the system in the present paper---the proofs proceed in broadly similar steps:
\begin{enumerate}
    \item One observes that the extremal linearised curvature quantities are gauge invariant and satisfy second order equations, the so-called Teukolsky equations, which decouple from the rest of the systems. Uniform boundedness  and decay statements are obtained for these quantities in a self-contained fashion which exploits the Teukolsky equations' hyperbolic structure.
    \item The remaining quantities in the system are gauge dependent. A novel analysis, introduced in \cite{DHR} and revisited in \cite{benomio_schwarzschild_stability}, combines the Bianchi equations for the curvature quantities with both transport and elliptic equations for the connection coefficients to yield control, with loss of regularity, over all the gauge dependent linearised quantities.
    \item Elliptic estimates of the form of \eqref{intro_aux_0000}, which in \cite{DHR} are derived by exploiting the Bianchi equations similarly to the way described in Section~\ref{sec_intro_elliptic_schwarzschild} above for \cite{benomio_schwarzschild_stability}, eventually allow one to improve on regularity and prove top-order orbital stability statements for all the gauge dependent linearised curvature quantities.\footnote{In \cite{DHR, benomio_schwarzschild_stability}, elliptic estimates of the form of \eqref{intro_aux_0000} are, in the notation of the Main Theorem, derived for $k\geq 5$. This also corresponds to the order of differentiability at which the orbital stability statements are proven.}
\end{enumerate}

A remarkable property of the linearised system of equations employed to prove the Main Theorem is that, in the rotating ($|a|\neq 0$) Kerr case, the extremal linearised curvature quantities remain gauge invariant, still exactly decouple into Teukolsky equations and can therefore be analysed independently from the rest of the system.~In other words, step 1 can still be carried out as described:~uniform boundedness---in the form of true orbital stability---and decay statements have been recently established for general solutions to the Teukolsky equations, and thus for the extremal curvature quantities in our system, in the full sub-extremal\footnote{The extremal $|a|=M$ case is still open and
understood to be far more challenging.} range $|a|<M$ \cite{Shl_Teix_teuk_1, Shl_Teix_teuk_2}. In the notation of the Main Theorem, the order of differentiability at which the orbital stability statements are proven corresponds to $k\geq 3$, thus matching the top order of the extremal curvature terms on the right hand side of \eqref{intro_aux_8} and implying control over those terms \emph{without loss of regularity}. 

\medskip

The analysis of step 2 for our linearised system of equations in the Kerr ($|a|\neq 0$) case is left for future work. We note however that, in the very slowly rotating regime $|a|\ll M$, step 2 can already be approached with the scheme of \cite{benomio_schwarzschild_stability}. Provided that step 2 is carried out in the full sub-extremal range $|a|<M$ so as to achieve, as one expects, suitable control over the lower order terms on the right hand side of \eqref{intro_aux_8}, the Main Theorem applies to resolve step 3, i.e.\ it provides an \emph{orbital stability} statement for all the gauge dependent linearised curvature quantities in the full sub-extremal range $|a|<M$.

\medskip

Uniform decay results for linearised gravity on the Kerr solution have been shown in \cite{Andersson_Blue_Lin_Kerr} and \cite{Hintz_Vasy_Kerr} in the very slowly rotating $|a|\ll M$ case. The former work formulates the system of equations in a geometric gauge employing non-integrable null tetrads (within the Geroch--Held--Penrose formalism) and it extends to the full sub-extremal range $|a|<M$ conditional on the later decay estimates of \cite{Shl_Teix_teuk_1, Shl_Teix_teuk_2}. The latter work performs a microlocal analysis of the system of equations, therein formulated for metric perturbations in a harmonic gauge, and it has been recently extended to $|a|<M$ in \cite{Andersson_Hafner_Whiting_mode_linearised_kerr,Hafner2025}. We also highlight the work \cite{Millet2023} (see also \cite{Ma_Zhang_Teukolsky_Kerr_sharp}), which obtains uniform sharp decay results for the Teukolsky equations in the full sub-extremal range $|a|<M$. The results mentioned in this paragraph do not include orbital stability statements in the above sense of the phrase.

\medskip

The series of works \cite{Klain_form_kerr,Giorgi_Klainerman_Szeftel_wave_estimates,Klain_GCM_kerr_1,Klain_GCM_kerr_2,Klainerman_Szeftel_Kerr_small_a_1,Shen_gcm_spheres_kerr} proves the nonlinear stability of the Kerr solution in the very slowly rotating regime $|a|\ll M$ (we note that the subset of works \cite{Klain_GCM_kerr_1,Klain_GCM_kerr_2,Klainerman_Szeftel_Kerr_small_a_1,Shen_gcm_spheres_kerr} include results which apply to the full sub-extremal range $|a|<M$). In these works, the equations are formulated in a geometric gauge which makes use of various non-integrable frames. Elliptic estimates play a relevant role in the analysis.

\medskip

The nonlinear stability of the Kerr solution in the full sub-extremal range $|a|<M$ remains a major open problem in the subject. Orbital stability for the linearised theory in the full sub-extremal range $|a|<M$, which remains even more compellingly open and to which the present paper hopes to contribute in the way described, will likely serve as an important ingredient in proving nonlinear (orbital and asymptotic) stability of the Kerr solution in the full sub-extremal range $|a|<M$.

\subsection{Outline of the paper} \label{sec_intro_outline}

In Section \ref{sec_kerr_exterior_manifold}, we recall some essential facts about the Kerr exterior manifolds, the algebraically special frame of Kerr and introduce the relevant differentiable structures. In Section \ref{sec_lin_system}, we report on the system of linearised vacuum Einstein equations around the Kerr exterior manifold, as derived in \cite{Benomio_thesis_doi_2}. In Section \ref{sec_main_theorem}, we state the main theorem of the paper. In Section \ref{sec_proj_formulae}, by first recalling and then employing the projection procedure constructed in \cite{Benomio_thesis_doi_2} (see also \cite{benomio_kerr_system_2}), we derive the $\mathbb{S}^2$-projected linearised system of equations and state the $\mathbb{S}^2$-projected version of the main theorem. The $\mathbb{S}^2$-projected version of the main theorem, which we shall explain is equivalent to its original version, is proven in Section \ref{sec_proof_main_th}. In Appendix \ref{sec_appendix_lemma}, we prove a general lemma on the ellipticity of the angular operators considered in the proof of the main theorem.

\medskip

\emph{Notation}:~The symbols $\lesssim$ and $\gtrsim$ denote inequalities up to a positive multiplicative constant. When relevant, the parameters on which the implicit constant depends will appear as subscripts (e.g.~$\lesssim_{k}$).

\subsection*{Acknowledgments}

The authors thank Mihalis Dafermos and Gustav Holzegel for useful discussions and comments. GB is particularly indebted to Gustav Holzegel for numerous conversations over the years. RTdC gratefully acknowledges the hospitality of the Gran Sasso Science Institute, where part of this research was conducted.

\section{The Kerr exterior manifold} \label{sec_kerr_exterior_manifold}

We define the manifold-with-boundary 
\begin{equation*}  
\mathcal{M}:= (-\infty,\infty) \times [0,\infty) \times \mathbb{S}^2 
\end{equation*}
with coordinates $\bar{t}\in (-\infty,\infty)$, $y\in [0,\infty)$ and standard (local) spherical coordinates $(\bar{\theta},\bar{\phi})\in\mathbb{S}^2$.~We define the vector fields
\begin{align*}
T& :=\partial_{\bar{t}} \, ,  &  \Phi& :=\partial_{\bar{\phi}}
\end{align*}
on $\mathcal{M}$, where the latter vector field can be smoothly extended (so as to vanish at the poles of $\mathbb{S}^2$) to a global vector field on $\mathbb{S}^2$.

\medskip

We define the \emph{(future) event horizon} as the boundary
\begin{equation*}
\mathcal{H}^+:=\partial\mathcal{M}  
\end{equation*} 
of $\mathcal{M}$ and \emph{(future) null infinity} as the formal hypersurface
\begin{equation*}
\mathcal{I}^+ :=\left\lbrace \bar{t}=\infty \right\rbrace \, .
\end{equation*}
We note that $\mathcal{H}^+=\left\lbrace y=0 \right\rbrace$. 

\medskip

Given real parameters $a$ and $M$, with $|a|<M$, we define the positive constants
\begin{equation*}
r_{\pm}:=M\pm\sqrt{M^2-a^2}
\end{equation*}
and a new coordinate $\bar{r}_{a,M}=\bar{r}_{a,M}(y)$ such that
\begin{equation*}
\bar{r}_{a,M}:[0,\infty)\rightarrow [r_+,\infty) \, ,
\end{equation*}
with $\bar{r}_{a,M}(3M)=y(3M)$.~We will simply denote $\bar{r}_{a,M}$ by $\bar{r}$.~We define the two-spheres
\begin{equation*}
\mathbb{S}^2_{\bar{t},\bar{r}} :=\left\lbrace \bar{t},\bar{r} \right\rbrace \times \mathbb{S}^2 \, .
\end{equation*} 
The coordinates 
\begin{equation} \label{kerr_coords}
(\bar{t},\bar{r},\bar{\theta},\bar{\phi})
\end{equation}
are such that $\mathcal{H}^+=\left\lbrace \bar{r}=r_+ \right\rbrace$. 

\medskip

For fixed $|a|<M$, we define the smooth scalar functions
\begin{align*}
\Delta(\bar{r}) &:= (\bar{r}-r_+)(\bar{r}-r_-) \, , &
\Sigma(\bar{r},\bar{\theta}) &:= \bar{r}^2+a^2\cos^2\bar{\theta} 
\end{align*}
on $\mathcal{M}$.~We define the \emph{Kerr family of metrics} as the two-parameter family of Lorentzian metrics $g_{a,M}$ on $\mathcal{M}$ such that 
\begin{align}
g_{a,M}= &-\left(1-\frac{2M\bar{r}}{\Sigma}\right){d\bar{t}}{}^2+2\,d\bar{t}\, d\bar{r}+\Sigma\, {d\bar{\theta}}{}^2 +\frac{(\bar{r}^2+a^2)^2-a^2\Delta \sin^2\bar{\theta}}{\Sigma}\,\sin^2\bar{\theta} \, {d\bar{\phi}}{}^{2}   \label{def_kerr_metric} \\
&-2\,a\sin^2\bar{\theta} \, d\bar{r} \,d\bar{\phi}-\frac{4aM\bar{r}}{\Sigma}\,\sin^2\bar{\theta} \, d\bar{t}\, d\bar{\phi} \, .   \nonumber
\end{align}
The metric \eqref{def_kerr_metric} is manifestly smooth on $\mathcal{M}$, including on $\mathcal{H}^+$.~The event horizon $\mathcal{H}^+$ can be checked to be a \emph{null} hypersurface relative to $g_{a,M}$.~The vector fields $T$ and $\Phi$ are Killing vector fields of $g_{a,M}$.~The smooth Lorentzian manifold $(\mathcal{M},g_{a,M})$ will be referred to as the \emph{Kerr exterior manifold}.~One can check that $(\mathcal{M},g_{a,M})$ solves the vacuum Einstein equations.

\medskip

The coordinates \eqref{kerr_coords} can be related to \emph{Boyer--Lindquist coordinates} $$(t,r,\theta,\phi)$$ on $\mathcal{M}\setminus\mathcal{H}^+$ by the coordinate transformation
\begin{align*}
\bar{t}(t,r)&=t+\int_{r_0}^{r}\frac{{r^{\prime}}^2+a^2}{\Delta(r^{\prime})}\,dr^{\prime}  \, ,  & \bar{r}(r)&=r \, ,  \\
\bar{\theta}(\theta)&=\theta \, , & \bar{\phi}(\phi,r)&=\phi +\int_{r_0}^{r}\frac{a}{\Delta(r^{\prime})}\,dr^{\prime} \quad \textup{mod} \, 2\pi \, .
\end{align*}
We define the two-spheres
\begin{equation*}
    \mathbb{S}^2_{t,r}:=\left\lbrace t,r \right\rbrace \times \mathbb{S}^2
\end{equation*}
and we note that, for any $(t,r)$, we have $\mathbb{S}^2_{t,r}=\mathbb{S}^2_{\bar{t}(t,r),\bar{r}(r)}$. With a slight abuse of notation, we will not distinguish between the two foliations by spheres and denote the $\mathbb{S}^2_{\bar{t},\bar{r}}$-spheres by $\mathbb{S}^2_{t,r}$ (although, strictly speaking, the latter spheres do not foliate the event horizon).

\medskip

In Boyer--Lindquist coordinates, the Kerr metric $g_{a,M}$ reads
\begin{equation}
g_{a,M}=-\frac{\Delta}{\Sigma}\,(dt-a\,\sin^2\theta\,d\phi)^2+\frac{\Sigma}{\Delta}\,dr^2+\Sigma \,d\theta^2+\frac{\sin^2\theta}{\Sigma}\,(a\, dt-(r^2+a^2)\,d\phi)^2 \, . \label{kerr_metric_bl}
\end{equation}
The Kerr metric will henceforth be denoted by $g$.

\subsection{The algebraically special frame} \label{sec_as_frame}

We define the \emph{algebraically special} null vector fields (in Boyer--Lindquist coordinates)
\begin{align}  
e_4^{\textup{as}}&=\frac{r^2+a^2}{\Sigma}\,\partial_{t}+\frac{\Delta}{\Sigma}\,\partial_{r}+\frac{a}{\Sigma}\,\partial_{\phi} \, ,   &  e_3^{\textup{as}}&= \frac{r^2+a^2}{\Delta}\,\partial_{t}-\partial_{r}+\frac{a}{\Delta}\,\partial_{\phi} \, . \label{as_null_frame_vectors}
\end{align}
The vector fields $e_4^{\text{as}}$ and $e_3^{\text{as}}$ are global, \emph{regular} and non-degenerate vector fields on the whole manifold $\mathcal{M}$, including on $\mathcal{H}^+$.~One can check that the null vector field $e^{\text{as}}_3$ is geodesic, i.e.~the identity
\begin{equation*}
\nabla_{e^{\text{as}}_3} e^{\text{as}}_3=0 
\end{equation*} 
holds on $\mathcal{M}$, with $\nabla$ the Levi-Civita connection of $g$.

\medskip

We define the induced global, regular horizontal distribution
\begin{equation*}
\mathfrak{D}_{\mathcal{N}_{\text{as}}}:=\left\langle e^{\text{as}}_3,e^{\text{as}}_4 \right\rangle^{\perp} 
\end{equation*}
on $\mathcal{M}$.~One can complete the null vector fields \eqref{as_null_frame_vectors} with a local orthonormal basis
\begin{equation*}
(e^{\text{as}}_1,e^{\text{as}}_2)
\end{equation*}
of $\mathfrak{D}_{\mathcal{N}_{\text{as}}}$ to form the local null frame $$\mathcal{N}_{\text{as}}=(e^{\text{as}}_1,e^{\text{as}}_2,e^{\text{as}}_3,e^{\text{as}}_4) \, ,$$ called the \emph{algebraically special frame} of the Kerr exterior manifold $(\mathcal{M},g)$.~Crucially, the frame $\mathcal{N}_{\text{as}}$ is \emph{non-integrable} for $|a|>0$, meaning that $\mathfrak{D}_{\mathcal{N}_{\text{as}}}$ is a \emph{non-integrable} distribution.

\medskip

We note three properties of the frame $\mathcal{N}_{\text{as}}$ which hold if one restricts to the event horizon $\mathcal{H}^+$:
\begin{itemize}
\item We have
\begin{equation} \label{intro_e4_horizon_killing}
e_4^{\text{as}}|_{\mathcal{H}^+} = 2\,\frac{r_+^2+a^2}{\Sigma(r_+)}\,\partial_{t} +2\,\frac{a}{\Sigma(r_+)}\,\partial_{\phi} \, , 
\end{equation}
i.e.~the vector field $e_4^{\text{as}}$ is tangent to $\mathcal{H}^+$ and in the span of the Killing vector fields $T$ and $\Phi$. 
\item We have
\begin{equation} \label{e4_distribution_horizon}
\mathfrak{D}_{\mathcal{N}_{\text{as}}}|_{\mathcal{H}^+}\subset T\mathcal{H}^+ \, ,
\end{equation}
i.e.~any local frame $(e^{\text{as}}_1,e^{\text{as}}_2)$ of $\mathfrak{D}_{\mathcal{N}_{\text{as}}}$ is tangent to $\mathcal{H}^+$.~Thus, the vector fields $$(e^{\text{as}}_4,e^{\text{as}}_1,e^{\text{as}}_2)$$ form a local basis of $T\mathcal{H}^+$ (so, in particular, generate an \emph{integrable} distribution).
\item We have
\begin{align} \label{eA_kerr_fermi_horizon}
g(\nabla_{e_4^{\text{as}}}e_A^{\text{as}},e_B^{\text{as}})|_{\mathcal{H}^+}&=0 \, , &  A&=\left\lbrace 1,2 \right\rbrace
\end{align}
for any local frame $(e^{\text{as}}_1,e^{\text{as}}_2)$ of $\mathfrak{D}_{\mathcal{N}_{\text{as}}}$.
\end{itemize}

\medskip

The metric $g$ induces a metric $\slashed{g}$ on $\mathfrak{D}_{\mathcal{N}_{\text{as}}}$, with associated inverse metric $\slashed{g}{}^{-1}$ and standard volume form $\slashed{\varepsilon}_{\slashed{g}}$.~The induced metric $\slashed{g}$, as well as the connection coefficients and curvature components of $g$ relative to $\mathcal{N}_{\text{as}}$, are defined as \emph{$\mathfrak{D}_{\mathcal{N}_{\text{as}}}$ tensors} and are given below.\footnote{See Section 4.1 of \cite{benomio_kerr_system_2} for the definition of \emph{$\mathfrak{D}_{\mathcal{N}_{\text{as}}}$ tensors} (i.e.~Definition 4.9 with the identification $\bsy{\mathfrak{D}}_{\bsy{\mathcal{N}}}= \mathfrak{D}_{\mathcal{N}_{\text{as}}}$) and of the connection coefficients and curvature components which follow.}~Since $\mathfrak{D}_{\mathcal{N}_{\text{as}}}$ is a global, regular distribution on $\mathcal{M}$, the $\mathfrak{D}_{\mathcal{N}_{\text{as}}}$ tensors are defined globally on $\mathcal{M}$ and are regular quantities on the whole manifold $\mathcal{M}$, including on $\mathcal{H}^+$.

\medskip

In Boyer--Lindquist coordinates, the induced metric reads
\begin{equation} \label{kerr_induced_metric}
\slashed{g}=\frac{a^2}{\Sigma}\,\sin^2\theta\,dt^2-2\,\frac{a\,(r^2+a^2)}{\Sigma}\,\sin^2\theta\,dt\,d\phi+\Sigma\,d\theta^2+\frac{(r^2+a^2)^2}{\Sigma}\,\sin^2\theta\,d\phi^2 
\end{equation}
and the connection coefficients $\eta$, $\etab$ and $\zeta$ of $g$ relative to $\mathcal{N}_{\text{as}}$ read
\begin{align}
\eta&=-\frac{a^2r}{\Sigma^2}\,\sin^2\theta\,dt-\frac{a^2\sin(2\theta)}{2\,\Sigma}\,d\theta+\frac{a\,r(r^2+a^2)}{\Sigma^2}\,\sin^2\theta\,d\phi \, , \label{kerr_eta}\\
\etab&= \frac{a^2r}{\Sigma^2}\,\sin^2\theta\,dt-\frac{a^2\sin(2\theta)}{2\,\Sigma}\,d\theta - \frac{a\,r(r^2+a^2)}{\Sigma^2}\,\sin^2\theta\,d\phi \, , \label{kerr_etab}\\
\zeta&=\eta \, . \label{kerr_zeta}
\end{align}
The coordinate expressions \eqref{kerr_induced_metric}--\eqref{kerr_etab} are given for the natural extension of the induced metric $\slashed{g}$ and connection coefficients $\eta$ and $\etab$ to spacetime tensors which identically vanish when evaluated on either $e_4^{\text{as}}$ or $e_3^{\text{as}}$.~In fact, the non-integrability of the distribution $\mathfrak{D}_{\mathcal{N}_{\textup{as}}}$ does not allow to express $\mathfrak{D}_{\mathcal{N}_{\textup{as}}}$ tensors in coordinate form, in that $\mathfrak{D}_{\mathcal{N}_{\textup{as}}}$ does not admit coordinate (co-)bases.~The remaining connection coefficients of $g$ relative to $\mathcal{N}_{\text{as}}$ read
\begin{align*}
{\chih} &=0 \, , & {\chibh} &= 0 \, ,\\
(\slashed{\varepsilon}\cdot\chi) &= \frac{2 a \Delta\cos \theta }{\Sigma^2} \, ,   & (\slashed{\varepsilon}\cdot\chib)&= \frac{ 2a \cos \theta}{\Sigma} \, , \\
(\text{tr}\chi) &= \frac{2r \Delta}{\Sigma^2} \, ,  &  (\text{tr}\chib) &= -\frac{2 r }{\Sigma} 
\end{align*}
and
\begin{align*}
\hat{\omega} &= -\frac{ 2\left(a^2   (M-r)\cos^2  \theta +a^2 r- M r^2\right)}{\Sigma^2} \, , &
\omegabh &= 0 \, , \\
\xi &=0 \, , &  \yb &=0 \, .
\end{align*}
The curvature components of $g$ relative to $\mathcal{N}_{\text{as}}$ read
\begin{align*}
\alpha&=0\, , & \alphab&=0 \, ,\\
\beta &= 0 \, , & \betab &= 0 \, ,\\
\rho&=\frac{ 2 M r \left(3a^2 \cos^2 \theta - r^2\right)}{\Sigma^3}  \, , &  \sigma &= \frac{ 2a  M \cos \theta  \left(3  r^2-a^2 \cos^2 \theta \right)}{\Sigma^3} \, .
\end{align*}

\medskip

The connection $\nabla$ induces a connection $\nablasl$ on $\mathfrak{D}_{\mathcal{N}_{\text{as}}}$. Since $\mathfrak{D}_{\mathcal{N}_{\text{as}}}$ is non-integrable, the induced connection $\nablasl$ over the bundle of $\mathfrak{D}_{\mathcal{N}_{\text{as}}}$ tensors is \emph{not} the Levi-Civita connection of the induced metric $\slashed{g}$.\footnote{See Remark 4.34 of \cite{benomio_kerr_system_2} with the identification $\bsy{\mathfrak{D}}_{\bsy{\mathcal{N}}}= \mathfrak{D}_{\mathcal{N}_{\text{as}}}$.}~In particular, the connection $\nablasl$ is compatible with $\slashed{g}$ but fails to be torsion-free.

\medskip

On the event horizon $\mathcal{H}^+$, the identities \eqref{intro_e4_horizon_killing} and \eqref{eA_kerr_fermi_horizon} imply
\begin{align*}
\nablasl_4\Gamma|_{\mathcal{H}^+}&=0 \, , & \nablasl_4 \psi|_{\mathcal{H}^+}&=0  
\end{align*}
for any connection coefficient $\Gamma$ and curvature component $\psi$ of $g$ relative to $\mathcal{N}_{\text{as}}$.

\subsection{Products and horizontal differential operators} \label{sec_def_operators}

For any $\mathfrak{D}_{\mathcal{N}_{\text{as}}}$ one-forms $\varsigma,\tilde{\varsigma}$ and $\mathfrak{D}_{\mathcal{N}_{\text{as}}}$ two-tensors $\theta,\tilde{\theta}$, we define
\begin{align*}
\varsigma\cdot\tilde{\varsigma}&:=\slashed{g}{}^{AB}\varsigma_A\,\tilde{\varsigma}_B \, , & \varsigma \wedge \tilde{\varsigma}&:=\slashed{\varepsilon}{}^{AB}\varsigma_A\,\tilde{\varsigma}_B 
\end{align*}
and
\begin{align*}
(\theta,\tilde{\theta})&:=\slashed{g}{}^{AD}\slashed{g}{}^{CB}\theta_{AC}\,\tilde{\theta}_{BD} \, , &
\theta \wedge \tilde{\theta}&:=\slashed{\varepsilon}{}^{AD}\slashed{g}{}^{CB}\theta_{AC}\,\tilde{\theta}_{BD} \, , &
(\theta\times \tilde{\theta})_{AB}&:=\theta^{\sharp_2}{}_{A}^C\,\tilde{\theta}_{CB} \, .
\end{align*} 
We also define
\begin{align*}
\varsigma \, \widehat{\otimes} \, \tilde{\varsigma} &:=(\varsigma \otimes \tilde{\varsigma})+(\tilde{\varsigma} \otimes \varsigma)-(\varsigma,\tilde{\varsigma})\,\slashed{g} \, .
\end{align*} 
The $\mathfrak{D}_{\mathcal{N}_{\text{as}}}$ two-tensor $\varsigma \,\widehat{\otimes} \,\tilde{\varsigma}$ is symmetric and traceless relative to $\slashed{g}$.

\medskip

For any $\mathfrak{D}_{\mathcal{N}_{\text{as}}}$ one-form $\varsigma$ and $\mathfrak{D}_{\mathcal{N}_{\text{as}}}$ two-tensor $\theta$, we define the divergence operator 
\begin{align*}
\slashed{\text{div}}\,\varsigma&:=\slashed{g}{}^{AB}(\nablasl_{A}\varsigma)_{B}  \, , &
(\slashed{\text{div}}\,\theta)_{A}&:=\slashed{g}{}^{CB}(\nablasl_C\theta)_{AB} \, ,
\end{align*}
the curl operator 
\begin{equation*}
\slashed{\text{curl}}\,\varsigma :=\slashed{\varepsilon}{}^{AB}(\nablasl_A\varsigma)_{B} 
\end{equation*}
and the differential operator 
\begin{equation*}
\slashed{\mathcal{D}}_2^{\star}\,\varsigma :=-\frac{1}{2}((\nablasl\varsigma)+(\nablasl\varsigma)^{\mathsf{T}}-(\slashed{\text{div}}\,\varsigma)\,\slashed{g})  \, .
\end{equation*}
The $\mathfrak{D}_{\mathcal{N}_{\text{as}}}$ two-tensor $\slashed{\mathcal{D}}_2^{\star}\,\varsigma$ is symmetric and traceless relative to $\slashed{g}$.

\medskip

For any smooth scalar functions $f$ and $h$, we define the differential operator
\begin{equation*}
\slashed{\mathcal{D}}_1^{\star}(f,h):=-\nablasl f+{}^{\star}\nablasl h \, .
\end{equation*}
We also define the differential operator $\slashed{\mathcal{D}}_2$, which takes any symmetric traceless $\mathfrak{D}_{\mathcal{N}_{\text{as}}}$ two-tensor $\theta$ into the $\mathfrak{D}_{\mathcal{N}_{\text{as}}}$ one-form $\slashed{\text{div}}\,\theta$, and the differential operator $\slashed{\mathcal{D}}_1$, which takes any $\mathfrak{D}_{\mathcal{N}_{\text{as}}}$ one-form $\varsigma$ into the pair of smooth scalar functions $(\slashed{\text{div}}\,\varsigma,\slashed{\text{curl}}\,\varsigma)$.

\medskip

For any smooth scalar function $f$ and $\mathfrak{D}_{\mathcal{N}_{\text{as}}}$ one-form $\varsigma$, we define the horizontal scalar and covariant Laplacian
\begin{align*}
\slashed{\Delta}f&:=\slashed{\text{div}}\nablasl f \, , & \slashed{\Delta}\varsigma&:=\slashed{\text{div}}\nablasl \varsigma \, .
\end{align*}
We have the identities $\slashed{\Delta}f=-\slashed{\text{curl}}^{\star}\nablasl f$ and $\slashed{\Delta}\varsigma=-\slashed{\text{curl}}^{\star}\nablasl \varsigma$.

\subsection{Commutation formulae}

For any $\mathfrak{D}_{\mathcal{N}_{\text{as}}}$ one-form $\varsigma$ and $\mathfrak{D}_{\mathcal{N}_{\text{as}}}$ two-tensor $\theta$, we have the following commutation formulae
\begin{align} 
[\nablasl_4,\nablasl] \, \varsigma =& \,  -\chi \times ( \nablasl\varsigma )+(\etab+\zeta)\otimes(\nablasl_{4}\varsigma) \label{kerr_comm_form_1}\\ & +(\etab,\varsigma)\, \chi -(\chi^{\sharp_2}\cdot\varsigma)\otimes \etab  \, ,  \nonumber \\[5pt] 
[\nablasl_3,\nablasl]\, \varsigma =& \, -\chib \times (\nablasl\varsigma) \label{kerr_comm_form_2}\\ & +(\eta,\varsigma)\,\chib-(\chib^{\sharp_2}\cdot\,\varsigma)\otimes \eta \, ,  \nonumber \\[5pt]
[\nablasl_3,\nablasl_4] \, \varsigma  =& \, \hat{\omega} \,(\nablasl_3 \varsigma)+ 2\,(\nablasl\varsigma)^{\sharp_1} \cdot (\eta-\etab) \label{kerr_comm_form_3}\\
& +2\,(\eta,\varsigma)\,\etab-2\,(\etab,\varsigma)\,\eta  +2\,\sigma({}^{\star}\varsigma) \nonumber
\end{align}
and 
\begin{align}
([\nablasl_4,\nablasl]\,\theta)_{ABC}=& \, -\chi^{\sharp_2}{}_A^D(\nablasl_D\theta)_{BC}+(\etab+\zeta)_A(\nablasl_{4}\theta)_{BC} \label{kerr_comm_form_4} \\ & +\chi{}_{AB}\etab^D\theta_{DC}-\etab{}_B \chi^{\sharp_2}{}_A^D \theta_{DC} \nonumber \\ &+ \chi{}_{AC}\etab^D \theta_{BD} - \etab{}_C  \chi^{\sharp_2 }{}_A^D  \theta_{BD}  \, , \nonumber \\[5pt]
([\nablasl_3,\nablasl]\theta)_{ABC}=& \, -\chib^{\sharp_2 }{}_A^D(\nablasl_D\theta)_{BC} \label{kerr_comm_form_5}\\ & +\chib{}_{AB}\eta^D\theta_{DC}-\eta{}_B \chib^{\sharp_2 }{}_A^D \theta_{DC}  \nonumber\\ & +\chib{}_{AC}\eta^D\theta_{BD}-\eta{}_C \chib^{\sharp_2 }{}_A^D \theta_{BD} \, ,  \nonumber \\[5pt]
([\nablasl_3,\nablasl_4]\,\theta)_{AB} =& \, \hat{\omega}\,(\nablasl_3 \theta)_{AB}+ 2\,{(\nablasl\theta)^{\sharp_1}}{}^{C}_{AB}(\eta-\etab)_{C} \label{kerr_comm_form_6}\\
&+2\,\eta^{C} \theta_{CB}\,\etab{}_{A}-2\,\etab^{C} \theta_{CB}\,\eta{}_{A} +2\,\sigma\,\slashed{\varepsilon}^{\sharp_2}{}_A^C \theta_{CB} \nonumber\\
&+2\,\eta^{C} \theta_{AC}\,\etab{}_{B}-2\,\etab^{C}\theta_{AC}\,\eta{}_{B} +2\,\sigma\,\slashed{\varepsilon}^{\sharp_2}{}_B^C\theta_{AC} \, . \nonumber
\end{align}
See Section 4.9 of \cite{benomio_kerr_system_2} for a proof of the formulae.

\subsection{The \texorpdfstring{$|a|=0$}{a=0} case}

For $|a|=0$, the Kerr exterior manifold reduces to the \emph{Schwarzschild exterior manifold}.~The algebraically special frame $\mathcal{N}_{\text{as}}$ becomes \emph{integrable} and such that
\begin{equation*}
\mathfrak{D}_{\mathcal{N}_{\text{as}}} = T\mathbb{S}^2_{t,r} \, . 
\end{equation*} 
The induced metric, connection coefficients and curvature components of $g$ relative to $\mathcal{N}_{\text{as}}$ are $\mathbb{S}^2_{t,r}$ tensors in the sense of \cite{StabMink} and are such that, contrary to the $|a|\neq 0$ case,
\begin{align*}
(\slashed{\varepsilon}\cdot\chi)=(\slashed{\varepsilon}\cdot\chib)&=0 \, , & \eta=\etab=\zeta&=0 \, ,  &  \sigma &=0 
\end{align*}
on $\mathcal{M}$.

\section{The linearised system of equations} \label{sec_lin_system}

In this section, we report the system of linearised vacuum Einstein equations around the Kerr exterior manifold derived in \cite{Benomio_thesis_doi_2} (see also \cite{benomio_kerr_system_2}).~The unknowns of the system are global, regular $\mathfrak{D}_{\mathcal{N}_{\text{as}}}$ tensors.~A complete list of the unknowns is presented in Table \ref{table:linearised_unknowns}.~A \emph{solution} to the linearised system of equations is a collection of all the linearised quantities in Table \ref{table:linearised_unknowns}.

\medskip

\begin{table}[H]
\centering
\begin{tabular}{ |c||c|c|c| } 
 \hline
 {} & Scalar function & $\mathfrak{D}_{\mathcal{N}_{\text{as}}}$ one-form & $\mathfrak{D}_{\mathcal{N}_{\text{as}}}$ two-tensor \\
 \hline\hline
Metric/Frame & $\overset{\text{\scalebox{.6}{$(1)$}}}{\mathfrak{\underline{\mathfrak{f}}}}_4,\overset{\text{\scalebox{.6}{$(1)$}}}{\mathfrak{\underline{\mathfrak{f}}}}_3,(\text{tr}\overset{\text{\scalebox{.6}{$(1)$}}}{\slashed{g}})$ &  $\overset{\text{\scalebox{.6}{$(1)$}}}{\mathfrak{\underline{\mathfrak{f}}}},\overset{\text{\scalebox{.6}{$(1)$}}}{\mathfrak{\slashed{\mathfrak{f}}}}_{3},\overset{\text{\scalebox{.6}{$(1)$}}}{\mathfrak{\slashed{\mathfrak{f}}}}_{4}$ & $\overset{\text{\scalebox{.6}{$(1)$}}}{\widehat{\slashed{g}}}$ \\
\hline
Connection & $(\overset{\text{\scalebox{.6}{$(1)$}}}{\text{tr}\chi}),(\overset{\text{\scalebox{.6}{$(1)$}}}{\slashed{\varepsilon}\cdot\chi}),(\overset{\text{\scalebox{.6}{$(1)$}}}{\text{tr}\chib}),(\overset{\text{\scalebox{.6}{$(1)$}}}{\slashed{\varepsilon}\cdot\chib}),\overset{\text{\scalebox{.6}{$(1)$}}}{\omegab}$ & $\overset{\text{\scalebox{.6}{$(1)$}}}{\yb},\overset{\text{\scalebox{.6}{$(1)$}}}{\eta},\overset{\text{\scalebox{.6}{$(1)$}}}{\zeta}$ & $\overset{\text{\scalebox{.6}{$(1)$}}}{\chih},\overset{\text{\scalebox{.6}{$(1)$}}}{\chibh}$  \\
\hline
Curvature & $\overset{\text{\scalebox{.6}{$(1)$}}}{\rho},\overset{\text{\scalebox{.6}{$(1)$}}}{\sigma}$ & $\overset{\text{\scalebox{.6}{$(1)$}}}{\beta},\overset{\text{\scalebox{.6}{$(1)$}}}{\betab}$  & $\overset{\text{\scalebox{.6}{$(1)$}}}{\alpha},\overset{\text{\scalebox{.6}{$(1)$}}}{\alphab}$ \\
\hline
\end{tabular}
\caption{Unknowns of the linearised system of equations.}
\label{table:linearised_unknowns}
\end{table}

\subsection{Equations for the linearised frame coefficients}  \label{sec_lin_eqns_frame_coeff}

The equations for the linearised frame coefficients read as follows.~We have the transport equations
\begin{align*}
\nablasl_4 \overset{\text{\scalebox{.6}{$(1)$}}}{\mathfrak{\slashed{\mathfrak{f}}}}_{4}+\chi^{\sharp_2}\cdot \overset{\text{\scalebox{.6}{$(1)$}}}{\mathfrak{\slashed{\mathfrak{f}}}}_{4} -\omegah \overset{\text{\scalebox{.6}{$(1)$}}}{\mathfrak{\slashed{\mathfrak{f}}}}_{4}  &= 0 \, , \\[6pt]
\nablasl_3 \overset{\text{\scalebox{.6}{$(1)$}}}{\mathfrak{\slashed{\mathfrak{f}}}}_{3} +\chib^{\sharp_2} \cdot\, \overset{\text{\scalebox{.6}{$(1)$}}}{\mathfrak{\slashed{\mathfrak{f}}}}_{3} &= \nablasl \overset{\text{\scalebox{.6}{$(1)$}}}{\mathfrak{\underline{\mathfrak{f}}}}_{3}- \overset{\text{\scalebox{.6}{$(1)$}}}{\mathfrak{\underline{\mathfrak{f}}}}_{3}\,(\eta+\etab)+\frac{1}{2}\,(\slashed{\varepsilon}\cdot\chib)\,{}^{\star}\overset{\text{\scalebox{.6}{$(1)$}}}{\mathfrak{\underline{\mathfrak{f}}}} + \overset{\text{\scalebox{.6}{$(1)$}}}{\yb}  \, , 
\end{align*}
the mixed transport equations
\begin{align*}
\nablasl_4 \overset{\text{\scalebox{.6}{$(1)$}}}{\mathfrak{\slashed{\mathfrak{f}}}}_{3} +\chi^{\sharp_2} \cdot \overset{\text{\scalebox{.6}{$(1)$}}}{\mathfrak{\slashed{\mathfrak{f}}}}_{3}  &=   \zetao \, , \\[6pt]
\nablasl_3 \overset{\text{\scalebox{.6}{$(1)$}}}{\mathfrak{\slashed{\mathfrak{f}}}}_{4} +\chib^{\sharp_2} \cdot\, \overset{\text{\scalebox{.6}{$(1)$}}}{\mathfrak{\slashed{\mathfrak{f}}}}_{4}    &=  \nablasl\overset{\text{\scalebox{.6}{$(1)$}}}{\mathfrak{\underline{\mathfrak{f}}}}_{4} +\frac{1}{2}\,(\slashed{\varepsilon}\cdot\chi)\, {}^{\star}\overset{\text{\scalebox{.6}{$(1)$}}}{\mathfrak{\underline{\mathfrak{f}}}} -\omegah \overset{\text{\scalebox{.6}{$(1)$}}}{\mathfrak{\slashed{\mathfrak{f}}}}_{3} + \etao -\zetao   
\end{align*}
and the transport equations
\begin{align*}
\nablasl_{4}\overset{\text{\scalebox{.6}{$(1)$}}}{\mathfrak{\underline{\mathfrak{f}}}}_{4} &= -2\,(\eta-\etab,\overset{\text{\scalebox{.6}{$(1)$}}}{\mathfrak{\slashed{\mathfrak{f}}}}_{4}) \, , \\[6pt]
\nablasl_4 \overset{\text{\scalebox{.6}{$(1)$}}}{\mathfrak{\underline{\mathfrak{f}}}}_{3}+\omegah \overset{\text{\scalebox{.6}{$(1)$}}}{\mathfrak{\underline{\mathfrak{f}}}}_{3}&=\omegabo   -2\,(\eta-\etab,\overset{\text{\scalebox{.6}{$(1)$}}}{\mathfrak{\slashed{\mathfrak{f}}}}_{3})-(\eta+\etab,\overset{\text{\scalebox{.6}{$(1)$}}}{\mathfrak{\underline{\mathfrak{f}}}}) \, , \\[6pt]
\nablasl_4\overset{\text{\scalebox{.6}{$(1)$}}}{\mathfrak{\underline{\mathfrak{f}}}}- \chi^{\sharp_1}\cdot \overset{\text{\scalebox{.6}{$(1)$}}}{\mathfrak{\underline{\mathfrak{f}}}}  +\omegah\overset{\text{\scalebox{.6}{$(1)$}}}{\mathfrak{\underline{\mathfrak{f}}}} &= -2\etao +2\,\overset{\text{\scalebox{.6}{$(1)$}}}{\mathfrak{\underline{\mathfrak{f}}}}_{4}(\eta-\etab)+ 2\,\overset{\text{\scalebox{.6}{$(1)$}}}{\widehat{\slashed{g}}}{}^{\sharp}\cdot(\eta-\etab)+(\text{tr}\overset{\text{\scalebox{.6}{$(1)$}}}{\slashed{g}})(\eta-\etab) \, .   
\end{align*}
We have the elliptic equations
\begin{align*}
\slashed{\text{curl}}\overset{\text{\scalebox{.6}{$(1)$}}}{\mathfrak{\slashed{\mathfrak{f}}}}_{4} &= \frac{1}{2} \,(\overset{\text{\scalebox{.6}{$(1)$}}}{\slashed{\varepsilon}\cdot\chi}) +\frac{1}{2}\,(\slashed{\varepsilon}\cdot\chi)\overset{\text{\scalebox{.6}{$(1)$}}}{\mathfrak{\underline{\mathfrak{f}}}}_4   +\frac{1}{4}\,(\slashed{\varepsilon}\cdot\chi)(\text{tr}\overset{\text{\scalebox{.6}{$(1)$}}}{\slashed{g}})   \, ,  \\[6pt] 
\slashed{\text{curl}}\overset{\text{\scalebox{.6}{$(1)$}}}{\mathfrak{\slashed{\mathfrak{f}}}}_{3}-(\eta+\etab) \wedge \overset{\text{\scalebox{.6}{$(1)$}}}{\mathfrak{\slashed{\mathfrak{f}}}}_{3} &= \frac{1}{2}\, (\overset{\text{\scalebox{.6}{$(1)$}}}{\slashed{\varepsilon}\cdot\chib}) +\frac{1}{2}\,(\slashed{\varepsilon}\cdot\chi)\overset{\text{\scalebox{.6}{$(1)$}}}{\mathfrak{\underline{\mathfrak{f}}}}_3 +\frac{1}{4}\,(\slashed{\varepsilon}\cdot\chib)(\text{tr}\overset{\text{\scalebox{.6}{$(1)$}}}{\slashed{g}})  \, .  
\end{align*}

\subsection{Linearised null structure equations}  \label{sec_lin_eqns_null_structure}

The linearised null structure equations read as follows. We have the linearised first variational formulae
\begin{align*}
\nablasl_4\overset{\text{\scalebox{.6}{$(1)$}}}{\widehat{\slashed{g}}}  +(\slashed{\varepsilon}\cdot\chi)\,{}^{\star}\overset{\text{\scalebox{.6}{$(1)$}}}{\widehat{\slashed{g}}} &= 2\,\chiho +2\,(\etab-\eta)\,\widehat{\otimes}\overset{\text{\scalebox{.6}{$(1)$}}}{\mathfrak{\slashed{\mathfrak{f}}}}_{4} \, , \\[6pt]
\nablasl_3\overset{\text{\scalebox{.6}{$(1)$}}}{\widehat{\slashed{g}}} +(\slashed{\varepsilon}\cdot\chib)\,{}^{\star}\overset{\text{\scalebox{.6}{$(1)$}}}{\widehat{\slashed{g}}}   &=2\,\chibho +2\,(\slashed{\mathcal{D}}_2^{\star}\overset{\text{\scalebox{.6}{$(1)$}}}{\underline{\mathfrak{f}}})-2\,(\etab-\eta)\,\widehat{\otimes}\overset{\text{\scalebox{.6}{$(1)$}}}{\mathfrak{\slashed{\mathfrak{f}}}}_{3} 
\end{align*}
and
\begin{align*}
\nablasl_4(\textup{tr}\overset{\text{\scalebox{.6}{$(1)$}}}{\slashed{g}})&= 2\,(\trchio) +4\,(\etab-\eta,\overset{\text{\scalebox{.6}{$(1)$}}}{\mathfrak{\slashed{\mathfrak{f}}}}_{4})   \, , \\[6pt]
\nablasl_3(\textup{tr}\overset{\text{\scalebox{.6}{$(1)$}}}{\slashed{g}})  &=  2\, (\trchibo)-4\,(\etab-\eta,\overset{\text{\scalebox{.6}{$(1)$}}}{\mathfrak{\slashed{\mathfrak{f}}}}_{3}) -2\,(\slashed{\textup{div}}\overset{\text{\scalebox{.6}{$(1)$}}}{\underline{\mathfrak{f}}}) -2\,(\textup{tr}\chi)\overset{\text{\scalebox{.6}{$(1)$}}}{\underline{\mathfrak{f}}}_3 - 2\,(\textup{tr}\chib)\overset{\text{\scalebox{.6}{$(1)$}}}{\underline{\mathfrak{f}}}_4  \, .
\end{align*}
We have the linearised second variational formulae
\begin{align*}
\nablasl_4 \chiho +(\text{tr}\chi)\chiho-\omegah \,\chiho &= -\alphao \, , \\[6pt]
\nablasl_3 \chibho+(\text{tr}\chib)\chibho &= -2\,\slashed{\mathcal{D}}_2^{\star} \ybo+(\etab-\eta)\,\widehat{\otimes}\ybo    -\alphabo \, .
\end{align*}
We have the linearised Raychaudhuri equations
\begin{align*}
\nablasl_4 (\trchio)+ (\text{tr}\chi)(\trchio)-\omegah \, (\trchio) =& \,  (\slashed{\varepsilon}\cdot\chi)(\overset{\text{\scalebox{.6}{$(1)$}}}{\slashed{\varepsilon}\cdot\chi}) \, , \\[6pt]
\nablasl_3 (\trchibo)+ (\text{tr}\chib)(\trchibo) =& \, (\slashed{\varepsilon}\cdot\chib)(\overset{\text{\scalebox{.6}{$(1)$}}}{\slashed{\varepsilon}\cdot\chib})+(\text{tr}\chib)\omegabo +2\, \slashed{\text{div}} \ybo   +2\,(\etab-\eta,\ybo) \\ &- (\nablasl(\text{tr}\chib),\overset{\text{\scalebox{.6}{$(1)$}}}{\mathfrak{\underline{\mathfrak{f}}}})-\nablasl_4(\text{tr}\chib)\overset{\text{\scalebox{.6}{$(1)$}}}{\mathfrak{\underline{\mathfrak{f}}}}_3  -\nablasl_3(\text{tr}\chib)\overset{\text{\scalebox{.6}{$(1)$}}}{\mathfrak{\underline{\mathfrak{f}}}}_4  \, . \nonumber
\end{align*} 
We have the linearised mixed transport equations
\begin{align*}
\nablasl_4 \chibho+\chi\times\chibho+\omegah\,\chibho =& \, -\frac{1}{2}\,(\text{tr}\chib)\chiho+\frac{1}{2}\,(\slashed{\varepsilon}\cdot\chib){^{\star}\chiho}  \\  &  - \widehat{(\nablasl\overset{\text{\scalebox{.6}{$(1)$}}}{\widehat{\slashed{g}}})^{\sharp_3}\cdot\etab}+ (\nablasl\overset{\text{\scalebox{.6}{$(1)$}}}{\widehat{\slashed{g}}})^{\sharp_1}\cdot\etab  -\frac{1}{2}\,\etab\,\widehat{\otimes}\,(\nablasl(\textup{tr}\overset{\text{\scalebox{.6}{$(1)$}}}{\slashed{g}}))-(\slashed{\textup{div}}\,\etab +(\etab,\etab))\overset{\text{\scalebox{.6}{$(1)$}}}{\widehat{\slashed{g}}} \nonumber\\
&   -\frac{1}{2}\,(\slashed{\varepsilon}\cdot\chi)\,({}^{\star}\etab)\,\widehat{\otimes}\,(-2\overset{\text{\scalebox{.6}{$(1)$}}}{\mathfrak{\slashed{\mathfrak{f}}}}_{3}+\overset{\text{\scalebox{.6}{$(1)$}}}{\mathfrak{\underline{\mathfrak{f}}}})  + (\slashed{\varepsilon}\cdot\chib)\,({}^{\star}\etab)\,\widehat{\otimes}\, \overset{\text{\scalebox{.6}{$(1)$}}}{\mathfrak{\slashed{\mathfrak{f}}}}_{4} \nonumber\\
&+(\nablasl_3\etab-\chib^{\sharp_2}\cdot\etab)\,\widehat{\otimes}\overset{\text{\scalebox{.6}{$(1)$}}}{\mathfrak{\slashed{\mathfrak{f}}}}_{4}+(\nablasl_4\etab-\chi^{\sharp_2}\cdot\etab)\,\widehat{\otimes}\overset{\text{\scalebox{.6}{$(1)$}}}{\mathfrak{\slashed{\mathfrak{f}}}}_{3} \, ,  \nonumber \\[6pt]
\nablasl_3 \chiho+\chib\times\chiho =& \, -\frac{1}{2}\,(\text{tr}\chi)\chibho+\frac{1}{2}\,(\slashed{\varepsilon}\cdot\chi){^{\star}\chibho} -2\, \slashed{\mathcal{D}}_2^{\star}\etao +2\, \eta \,\widehat{\otimes}\,\etao   \\ 
&- \widehat{(\nablasl\overset{\text{\scalebox{.6}{$(1)$}}}{\widehat{\slashed{g}}})^{\sharp_3}\cdot\eta}+ (\nablasl\overset{\text{\scalebox{.6}{$(1)$}}}{\widehat{\slashed{g}}})^{\sharp_1}\cdot\eta  -\frac{1}{2}\,\eta\,\widehat{\otimes}\,(\nablasl(\textup{tr}\overset{\text{\scalebox{.6}{$(1)$}}}{\slashed{g}}))-(\slashed{\textup{div}}\,\eta +(\eta,\eta))\overset{\text{\scalebox{.6}{$(1)$}}}{\widehat{\slashed{g}}}  \nonumber\\
& -\frac{1}{2}\,(\slashed{\varepsilon}\cdot\chi)\,({}^{\star}\eta)\,\widehat{\otimes}\,(-2\overset{\text{\scalebox{.6}{$(1)$}}}{\mathfrak{\slashed{\mathfrak{f}}}}_{3}+\overset{\text{\scalebox{.6}{$(1)$}}}{\mathfrak{\underline{\mathfrak{f}}}})  + (\slashed{\varepsilon}\cdot\chib)\,({}^{\star}\eta)\,\widehat{\otimes}\, \overset{\text{\scalebox{.6}{$(1)$}}}{\mathfrak{\slashed{\mathfrak{f}}}}_{4} \nonumber\\
&+(\nablasl_3\eta-\chib^{\sharp_2}\cdot\,\eta)\,\widehat{\otimes}\overset{\text{\scalebox{.6}{$(1)$}}}{\mathfrak{\slashed{\mathfrak{f}}}}_{4}+(\nablasl_4\eta-\chi^{\sharp_2}\cdot\eta)\,\widehat{\otimes}\overset{\text{\scalebox{.6}{$(1)$}}}{\mathfrak{\slashed{\mathfrak{f}}}}_{3}   \nonumber
\end{align*}
and
\begin{align*}
\nablasl_4( \trchibo)+ \frac{1}{2}\,(\text{tr}\chi)(\trchibo)+\omegah \, (\trchibo) =& \, - \frac{1}{2}\,(\text{tr}\chib)(\trchio)+\frac{1}{2}\,(\slashed{\varepsilon}\cdot\chib)(\overset{\text{\scalebox{.6}{$(1)$}}}{\slashed{\varepsilon}\cdot\chi})+\frac{1}{2}\,(\slashed{\varepsilon}\cdot\chi)(\overset{\text{\scalebox{.6}{$(1)$}}}{\slashed{\varepsilon}\cdot\chib}) \\ 
 &   - 2\,(\slashed{\textup{div}}\overset{\text{\scalebox{.6}{$(1)$}}}{\widehat{\slashed{g}}})^{\sharp}\cdot\etab -(\slashed{\textup{div}}\,\etab+(\etab,\etab))(\textup{tr}\overset{\text{\scalebox{.6}{$(1)$}}}{\slashed{g}}) \nonumber \\ & 
-(\textup{tr}\chi)\,(-2\overset{\text{\scalebox{.6}{$(1)$}}}{\mathfrak{\slashed{\mathfrak{f}}}}_{3}+\overset{\text{\scalebox{.6}{$(1)$}}}{\mathfrak{\underline{\mathfrak{f}}}},\etab) -(\slashed{\varepsilon}\cdot\chi)\,(-2\overset{\text{\scalebox{.6}{$(1)$}}}{\mathfrak{\slashed{\mathfrak{f}}}}_{3}+\overset{\text{\scalebox{.6}{$(1)$}}}{\mathfrak{\underline{\mathfrak{f}}}},{}^{\star}\etab)   \nonumber\\
&  +2\,(\textup{tr}\chib)(\overset{\text{\scalebox{.6}{$(1)$}}}{\mathfrak{\slashed{\mathfrak{f}}}}_{4},\etab)  
+2\,(\slashed{\varepsilon}\cdot\chib)(\overset{\text{\scalebox{.6}{$(1)$}}}{\mathfrak{\slashed{\mathfrak{f}}}}_{4}, {}^{\star}\etab) \nonumber \\
&  +2\,(\nablasl_3\etab-\chib^{\sharp_2}\cdot\etab,\overset{\text{\scalebox{.6}{$(1)$}}}{\mathfrak{\slashed{\mathfrak{f}}}}_{4}) +2\,(\nablasl_4\etab-\chi^{\sharp_2}\cdot\etab,\overset{\text{\scalebox{.6}{$(1)$}}}{\mathfrak{\slashed{\mathfrak{f}}}}_{3}) \nonumber\\ &  +(\text{tr}\chi)(\etab,\overset{\text{\scalebox{.6}{$(1)$}}}{\mathfrak{\underline{\mathfrak{f}}}})+2\rhoo \, , \nonumber \\[6pt]
\nablasl_3 (\trchio)+\frac{1}{2}\,(\text{tr}\chib)(\trchio) =& \, - \frac{1}{2}\,(\text{tr}\chi)(\trchibo)+\frac{1}{2}\,(\slashed{\varepsilon}\cdot\chi)(\overset{\text{\scalebox{.6}{$(1)$}}}{\slashed{\varepsilon}\cdot\chib})+\frac{1}{2}\,(\slashed{\varepsilon}\cdot\chib)(\overset{\text{\scalebox{.6}{$(1)$}}}{\slashed{\varepsilon}\cdot\chi})   \\ 
& -(\text{tr}\chi)\omegabo +2\,\slashed{\textup{div}}\etao+4\,(\eta,\etao) \nonumber\\
& -2 \, (\slashed{\textup{div}}\overset{\text{\scalebox{.6}{$(1)$}}}{\widehat{\slashed{g}}})^{\sharp}\cdot\eta - (\slashed{\textup{div}}\,\eta+(\eta,\eta))(\textup{tr}\overset{\text{\scalebox{.6}{$(1)$}}}{\slashed{g}}) \nonumber\\ & - (\textup{tr}\chi)\,(-2\overset{\text{\scalebox{.6}{$(1)$}}}{\mathfrak{\slashed{\mathfrak{f}}}}_{3}+\overset{\text{\scalebox{.6}{$(1)$}}}{\mathfrak{\underline{\mathfrak{f}}}},\eta) - (\slashed{\varepsilon}\cdot\chi)\,(-2\overset{\text{\scalebox{.6}{$(1)$}}}{\mathfrak{\slashed{\mathfrak{f}}}}_{3}+\overset{\text{\scalebox{.6}{$(1)$}}}{\mathfrak{\underline{\mathfrak{f}}}},{}^{\star}\eta)   \nonumber\\
&  +2\,(\textup{tr}\chib)(\overset{\text{\scalebox{.6}{$(1)$}}}{\mathfrak{\slashed{\mathfrak{f}}}}_{4},\eta)  
+2\,(\slashed{\varepsilon}\cdot\chib)(\overset{\text{\scalebox{.6}{$(1)$}}}{\mathfrak{\slashed{\mathfrak{f}}}}_{4}, {}^{\star}\eta) \nonumber\\
&+2\,(\nablasl_3\eta-\chib^{\sharp_2}\cdot\eta,\overset{\text{\scalebox{.6}{$(1)$}}}{\mathfrak{\slashed{\mathfrak{f}}}}_{4}) +2\,(\nablasl_4\eta-\chi^{\sharp_2}\cdot\eta,\overset{\text{\scalebox{.6}{$(1)$}}}{\mathfrak{\slashed{\mathfrak{f}}}}_{3}) \nonumber\\ 
&-(\nablasl(\text{tr}\chi),\overset{\text{\scalebox{.6}{$(1)$}}}{\mathfrak{\underline{\mathfrak{f}}}})-\nablasl_4(\text{tr}\chi)\overset{\text{\scalebox{.6}{$(1)$}}}{\mathfrak{\underline{\mathfrak{f}}}}_3  -\nablasl_3(\text{tr}\chi)\overset{\text{\scalebox{.6}{$(1)$}}}{\mathfrak{\underline{\mathfrak{f}}}}_4    \nonumber \\
& +(\text{tr}\chi)(\eta,\overset{\text{\scalebox{.6}{$(1)$}}}{\mathfrak{\underline{\mathfrak{f}}}})  +2\rhoo  \, . \nonumber
\end{align*}
We have the linearised transport equations for the antitraces
\begin{align*}
\nablasl_4 (\overset{\text{\scalebox{.6}{$(1)$}}}{\slashed{\varepsilon}\cdot\chi})+(\text{tr}\chi)(\overset{\text{\scalebox{.6}{$(1)$}}}{\slashed{\varepsilon}\cdot\chi})-\omegah \, (\overset{\text{\scalebox{.6}{$(1)$}}}{\slashed{\varepsilon}\cdot\chi})   =& \, -(\slashed{\varepsilon}\cdot\chi)(\trchio) \, , \\[6pt]
\nablasl_3 (\overset{\text{\scalebox{.6}{$(1)$}}}{\slashed{\varepsilon}\cdot\chib})+(\text{tr}\chib)(\overset{\text{\scalebox{.6}{$(1)$}}}{\slashed{\varepsilon}\cdot\chib})  =& \, -(\slashed{\varepsilon}\cdot\chib)(\trchibo)+ (\slashed{\varepsilon}\cdot\chib)\omegabo - 2\,(\eta+\etab)\wedge \ybo + 2\, \slashed{\text{curl}} \, \ybo \\ & -(\nablasl(\slashed{\varepsilon}\cdot\chib),\overset{\text{\scalebox{.6}{$(1)$}}}{\mathfrak{\underline{\mathfrak{f}}}})  -\nablasl_4(\slashed{\varepsilon}\cdot\chib)\overset{\text{\scalebox{.6}{$(1)$}}}{\mathfrak{\underline{\mathfrak{f}}}}_3  -\nablasl_3(\slashed{\varepsilon}\cdot\chib)\overset{\text{\scalebox{.6}{$(1)$}}}{\mathfrak{\underline{\mathfrak{f}}}}_4  \nonumber
\end{align*}
and the mixed transport equations
\begin{align*}
\nablasl_4 (\overset{\text{\scalebox{.6}{$(1)$}}}{\slashed{\varepsilon}\cdot\chib}) +\frac{1}{2}\,(\text{tr}\chi)(\overset{\text{\scalebox{.6}{$(1)$}}}{\slashed{\varepsilon}\cdot\chib})+\omegah \, (\overset{\text{\scalebox{.6}{$(1)$}}}{\slashed{\varepsilon}\cdot\chib})  =& \, -\frac{1}{2}\,(\slashed{\varepsilon}\cdot\chib)(\trchio)-\frac{1}{2}\,(\text{tr}\chib)(\overset{\text{\scalebox{.6}{$(1)$}}}{\slashed{\varepsilon}\cdot \chi})-\frac{1}{2}\,(\slashed{\varepsilon}\cdot \chi)(\trchibo) \\ 
&   -2\,(\nablasl_3\etab-\chib^{\sharp_2}\cdot\etab)\wedge\overset{\text{\scalebox{.6}{$(1)$}}}{\mathfrak{\slashed{\mathfrak{f}}}}_{4} -2\,(\nablasl_4\etab-\chi^{\sharp_2}\cdot\etab)\wedge\overset{\text{\scalebox{.6}{$(1)$}}}{\mathfrak{\slashed{\mathfrak{f}}}}_{3}  \nonumber \\
&+(\slashed{\varepsilon}\cdot\chi)(\etab,\overset{\text{\scalebox{.6}{$(1)$}}}{\mathfrak{\underline{\mathfrak{f}}}}) - (\slashed{\textup{curl}}\etab)(\textup{tr}\overset{\text{\scalebox{.6}{$(1)$}}}{\slashed{g}})+2\sigmao \, , \nonumber \\[6pt]
\nablasl_3 (\overset{\text{\scalebox{.6}{$(1)$}}}{\slashed{\varepsilon}\cdot\chi}) +\frac{1}{2}\,(\text{tr}\chib)(\overset{\text{\scalebox{.6}{$(1)$}}}{\slashed{\varepsilon}\cdot\chi}) =& \,-\frac{1}{2}\,(\slashed{\varepsilon}\cdot\chi)(\trchibo) -\frac{1}{2}\,(\text{tr}\chi)(\overset{\text{\scalebox{.6}{$(1)$}}}{\slashed{\varepsilon}\cdot \chib})-\frac{1}{2}\,(\slashed{\varepsilon}\cdot \chib)(\trchio)   \\
&-(\slashed{\varepsilon}\cdot\chi)\omegabo +2\,\slashed{\textup{curl}}\etao \nonumber \\
&-2\,(\nablasl_3\eta-\chib^{\sharp_2}\cdot\eta)\wedge\overset{\text{\scalebox{.6}{$(1)$}}}{\mathfrak{\slashed{\mathfrak{f}}}}_{4} -2\,(\nablasl_4\eta-\chi^{\sharp_2}\cdot\eta)\wedge\overset{\text{\scalebox{.6}{$(1)$}}}{\mathfrak{\slashed{\mathfrak{f}}}}_{3}  \nonumber \\
&-(\nablasl(\slashed{\varepsilon}\cdot\chi),\overset{\text{\scalebox{.6}{$(1)$}}}{\mathfrak{\underline{\mathfrak{f}}}})  -\nablasl_4(\slashed{\varepsilon}\cdot\chi)\overset{\text{\scalebox{.6}{$(1)$}}}{\mathfrak{\underline{\mathfrak{f}}}}_3  -\nablasl_3(\slashed{\varepsilon}\cdot\chi)\overset{\text{\scalebox{.6}{$(1)$}}}{\mathfrak{\underline{\mathfrak{f}}}}_4  \nonumber \\
&+(\slashed{\varepsilon}\cdot\chi)(\eta,\overset{\text{\scalebox{.6}{$(1)$}}}{\mathfrak{\underline{\mathfrak{f}}}})  - (\slashed{\textup{curl}}\,\eta)(\textup{tr}\overset{\text{\scalebox{.6}{$(1)$}}}{\slashed{g}})  -2\sigmao  \, . \nonumber
\end{align*}
We have the linearised transport equations
\begin{align*}
\nablasl_4 \etao+\frac{1}{2}\,(\text{tr}\chi)\etao -\frac{1}{2}\,(\slashed{\varepsilon}\cdot\chi){}^{\star}\etao =& \, \chiho{}^{\sharp}\cdot\etab+\frac{1}{2}\,(\trchio)\etab-\frac{1}{2}\,(\overset{\text{\scalebox{.6}{$(1)$}}}{\slashed{\varepsilon}\cdot\chi})({}^{\star}\etab-2\,{}^{\star}\eta)-\betao \\
&+2\,(\etab-\eta,\eta)\overset{\text{\scalebox{.6}{$(1)$}}}{\slashed{\mathfrak{f}}}_4+\frac{1}{2}\,(\slashed{\varepsilon}\cdot\chi)({}^{\star}\overset{\text{\scalebox{.6}{$(1)$}}}{\widehat{\slashed{g}}})^{\sharp_1}\cdot(\etab -2\,\eta)  \, , \nonumber \\[6pt]
\nablasl_4\ybo +2\,\omegah\ybo =& \,  -\chibho{}^{\sharp}\cdot\eta -\frac{1}{2}\,(\trchibo)\,\eta+\frac{1}{2}\,(\overset{\text{\scalebox{.6}{$(1)$}}}{\slashed{\varepsilon}\cdot\chib})({}^{\star}\eta-2\,{}^{\star}\etab)-\betabo \\
&-\frac{1}{2}\,(\text{tr}\chib)\etao  +\frac{1}{2}\,(\slashed{\varepsilon}\cdot\chib)\,{}^{\star}\etao -\frac{1}{2}\,(\slashed{\varepsilon}\cdot\chib)({}^{\star}\overset{\text{\scalebox{.6}{$(1)$}}}{\widehat{\slashed{g}}})^{\sharp_1}\cdot(\eta -2\,\etab) \nonumber\\
&+(\nablasl\etab)^{\sharp_1}\cdot \overset{\text{\scalebox{.6}{$(1)$}}}{\underline{\mathfrak{f}}}  -(\slashed{\mathcal{D}}_2^{\star}\overset{\text{\scalebox{.6}{$(1)$}}}{\underline{\mathfrak{f}}})^{\sharp_2}\cdot \etab+\frac{1}{2}\,(\slashed{\textup{div}}\overset{\text{\scalebox{.6}{$(1)$}}}{\underline{\mathfrak{f}}})\,\etab+\frac{1}{2}\,(\slashed{\textup{curl}}\overset{\text{\scalebox{.6}{$(1)$}}}{\underline{\mathfrak{f}}})\,{}^{\star}\etab \nonumber \\
&  -2\,(\eta-\etab,\etab)\overset{\text{\scalebox{.6}{$(1)$}}}{\slashed{\mathfrak{f}}}_3 +(\nablasl_4\etab+\chi^{\sharp_2}\cdot\etab)\overset{\text{\scalebox{.6}{$(1)$}}}{\underline{\mathfrak{f}}}_3+ (\nablasl_3\etab+\chib^{\sharp_2}\cdot\etab)\overset{\text{\scalebox{.6}{$(1)$}}}{\underline{\mathfrak{f}}}_4 \, , \nonumber  
\end{align*}
\begin{align*}
\nablasl_4\omegabo+2\,\omegah\omegabo  =& \,   -2\,(\eta-\etab,\etao)-2\,(\eta-\etab,\zetao) -2\rhoo   \\
&+((\eta-2\,\etab)\,\widehat{\otimes}\,\eta, \overset{\text{\scalebox{.6}{$(1)$}}}{\widehat{\slashed{g}}})+(\textup{tr}\overset{\text{\scalebox{.6}{$(1)$}}}{\slashed{g}})\,(\eta-2\,\etab,\eta) \nonumber \\
& -(\nablasl\omegah,\overset{\text{\scalebox{.6}{$(1)$}}}{\mathfrak{\underline{\mathfrak{f}}}})  -(\nablasl_4\omegah)\overset{\text{\scalebox{.6}{$(1)$}}}{\mathfrak{\underline{\mathfrak{f}}}}_3  -(\nablasl_3\omegah)\overset{\text{\scalebox{.6}{$(1)$}}}{\mathfrak{\underline{\mathfrak{f}}}}_4  \, , \nonumber
\end{align*}
\begin{align*}
\nablasl_4 \zetao +\chi^{\sharp_2}\cdot \zetao +\omegah \zetao    =& \,  \chiho {}^{\sharp}\cdot \etab +\frac{1}{2}\,(\trchio) \, \etab +\frac{1}{2}\,(\overset{\text{\scalebox{.6}{$(1)$}}}{\slashed{\varepsilon}\cdot\chi}) \, {^{\star}\etab} -\betao  \\
& +2\,(\etab-\eta,\eta)\overset{\text{\scalebox{.6}{$(1)$}}}{\slashed{\mathfrak{f}}}_4-(\nablasl_3\omegah)\overset{\text{\scalebox{.6}{$(1)$}}}{\slashed{\mathfrak{f}}}_4-(\nablasl_4\omegah)\overset{\text{\scalebox{.6}{$(1)$}}}{\slashed{\mathfrak{f}}}_3 \nonumber \\
&-\frac{1}{2}\,(\slashed{\varepsilon}\cdot\chi)({}^{\star}\overset{\text{\scalebox{.6}{$(1)$}}}{\widehat{\slashed{g}}})^{\sharp_1}\cdot\etab \, , \nonumber \\[6pt]
\nablasl_3\zetao +\chib^{\sharp_2}\cdot\zetao  =& \,  \nablasl \omegabo  +\chi^{\sharp_2}\cdot \ybo-\omegah \ybo -\chib^{\sharp_2}\cdot \etao -\betabo \\
&-  \,\chibho {}^{\sharp}\cdot \eta -\frac{1}{2}\,(\trchibo)\,\eta  -\frac{1}{2}\,(\overset{\text{\scalebox{.6}{$(1)$}}}{\slashed{\varepsilon}\cdot\chib}){^{\star}\eta}     \nonumber\\
&  +2\,(\eta-\etab,\eta)\overset{\text{\scalebox{.6}{$(1)$}}}{\slashed{\mathfrak{f}}}_3 -(\nablasl_4\eta+\chi^{\sharp_2}\cdot\eta)\overset{\text{\scalebox{.6}{$(1)$}}}{\underline{\mathfrak{f}}}_3 - (\nablasl_3\eta+\chib^{\sharp_2}\cdot\eta)\overset{\text{\scalebox{.6}{$(1)$}}}{\underline{\mathfrak{f}}}_4 \nonumber \\
&-(\nablasl\eta)^{\sharp_1}\cdot \overset{\text{\scalebox{.6}{$(1)$}}}{\underline{\mathfrak{f}}}+(\slashed{\mathcal{D}}_2^{\star}\overset{\text{\scalebox{.6}{$(1)$}}}{\underline{\mathfrak{f}}})^{\sharp_2}\cdot \eta -\frac{1}{2}\,(\slashed{\textup{div}}\overset{\text{\scalebox{.6}{$(1)$}}}{\underline{\mathfrak{f}}})\,\eta -\frac{1}{2}\,(\slashed{\textup{curl}}\overset{\text{\scalebox{.6}{$(1)$}}}{\underline{\mathfrak{f}}})\,{}^{\star}\eta  \nonumber \\
&+\frac{1}{2}\,(\slashed{\varepsilon}\cdot\chib)({}^{\star}\overset{\text{\scalebox{.6}{$(1)$}}}{\widehat{\slashed{g}}})^{\sharp_1}\cdot\eta   \nonumber
\end{align*}
and the linearised elliptic equation
\begin{align*}
\slashed{\textup{curl}}\zetao   =& \,  \frac{1}{4}\,(\slashed{\varepsilon}\cdot\chib)(\trchio)-\frac{1}{4}\,(\text{tr}\chib)(\overset{\text{\scalebox{.6}{$(1)$}}}{\slashed{\varepsilon}\cdot\chi})+\frac{1}{4}\,((\text{tr}\chi)-2\,\omegah)(\overset{\text{\scalebox{.6}{$(1)$}}}{\slashed{\varepsilon}\cdot\chib}) -\frac{1}{4}\,(\slashed{\varepsilon}\cdot\chi)(\trchibo)  \\ & +\frac{1}{2}\,(\slashed{\varepsilon}\cdot\chi)\omegabo  +\sigmao \nonumber \\
&+(\nablasl_3\eta-\chib^{\sharp_2}\cdot\eta)\wedge \overset{\text{\scalebox{.6}{$(1)$}}}{\slashed{\mathfrak{f}}}_4 +(\nablasl_4\eta-\chi^{\sharp_2}\cdot\eta)\wedge \overset{\text{\scalebox{.6}{$(1)$}}}{\slashed{\mathfrak{f}}}_3-\frac{1}{2}\,(\slashed{\varepsilon}\cdot\chi) (\eta,\overset{\text{\scalebox{.6}{$(1)$}}}{\underline{\mathfrak{f}}}) \nonumber \\
&-\frac{1}{2}\,(\slashed{\textup{curl}}\,\eta)(\textup{tr}\overset{\text{\scalebox{.6}{$(1)$}}}{\slashed{g}})    \, . \nonumber
\end{align*}
We have the linearised Codazzi equations
\begin{align*}
\slashed{\text{div}}\chiho+\chiho{}^{\sharp}\cdot\eta =& \, \frac{1}{2}\,\nablasl(\trchio)-\frac{1}{2}\,{}^{\star}\nablasl(\overset{\text{\scalebox{.6}{$(1)$}}}{\slashed{\varepsilon}\cdot\chi}) -\betao \\
& +\frac{1}{2}\,(\trchio)\,\eta -\frac{3}{2}\,(\overset{\text{\scalebox{.6}{$(1)$}}}{\slashed{\varepsilon}\cdot\chi})\,{}^{\star}\eta-\frac{1}{2}\,(\slashed{\varepsilon}\cdot\chi)\,{}^{\star}\zetao  +\frac{1}{2}\,(\text{tr}\chi)\zetao -(\slashed{\varepsilon}\cdot\chi)\,{}^{\star}\etao  \nonumber\\
& +\frac{3}{2}\,(\slashed{\varepsilon}\cdot\chi)({}^{\star}\overset{\text{\scalebox{.6}{$(1)$}}}{\widehat{\slashed{g}}})^{\sharp_1}\cdot\eta+\frac{1}{2}\,({}^{\star}\overset{\text{\scalebox{.6}{$(1)$}}}{\widehat{\slashed{g}}})^{\sharp_1}\cdot(\nablasl(\slashed{\varepsilon}\cdot\chi))  \nonumber\\
&+\frac{1}{2}\,(\nablasl_3(\text{tr}\chi))\overset{\text{\scalebox{.6}{$(1)$}}}{\slashed{\mathfrak{f}}}_4+\frac{1}{2}\,(\nablasl_4(\text{tr}\chi))\overset{\text{\scalebox{.6}{$(1)$}}}{\slashed{\mathfrak{f}}}_3-\frac{1}{2}\,(\nablasl_3(\slashed{\varepsilon}\cdot\chi))\,{}^{\star}\overset{\text{\scalebox{.6}{$(1)$}}}{\slashed{\mathfrak{f}}}_4-\frac{1}{2}\,(\nablasl_4(\slashed{\varepsilon}\cdot\chi))\,{}^{\star}\overset{\text{\scalebox{.6}{$(1)$}}}{\slashed{\mathfrak{f}}}_3 \, , \nonumber \\[6pt]
\slashed{\text{div}}\chibho-\chibho{}^{\sharp}\cdot\eta =& \,\frac{1}{2}\,\nablasl(\trchibo)-\frac{1}{2}\,{}^{\star}\nablasl(\overset{\text{\scalebox{.6}{$(1)$}}}{\slashed{\varepsilon}\cdot\chib}) +\betabo \\
&-\frac{1}{2}\,(\trchibo)\,\eta +\frac{1}{2}\,(\overset{\text{\scalebox{.6}{$(1)$}}}{\slashed{\varepsilon}\cdot\chib})({}^{\star}\eta-2\,{}^{\star}\etab)+\frac{1}{2}\,(\slashed{\varepsilon}\cdot\chib)\,{}^{\star}\zetao   -\frac{1}{2}\,(\text{tr}\chib)\zetao -(\slashed{\varepsilon}\cdot\chi)\,{}^{\star}\ybo  \nonumber \\
& -\frac{1}{2}\,(\slashed{\varepsilon}\cdot\chib)({}^{\star}\overset{\text{\scalebox{.6}{$(1)$}}}{\widehat{\slashed{g}}})^{\sharp_1}\cdot(\eta-2\,\etab)+\frac{1}{2}\,({}^{\star}\overset{\text{\scalebox{.6}{$(1)$}}}{\widehat{\slashed{g}}})^{\sharp_1}\cdot(\nablasl(\slashed{\varepsilon}\cdot\chib)) \nonumber \\
&+\frac{1}{2}\,(\nablasl_3(\text{tr}\chib))\overset{\text{\scalebox{.6}{$(1)$}}}{\slashed{\mathfrak{f}}}_4+\frac{1}{2}\,(\nablasl_4(\text{tr}\chib))\overset{\text{\scalebox{.6}{$(1)$}}}{\slashed{\mathfrak{f}}}_3-\frac{1}{2}\,(\nablasl_3(\slashed{\varepsilon}\cdot\chib))\,{}^{\star}\overset{\text{\scalebox{.6}{$(1)$}}}{\slashed{\mathfrak{f}}}_4-\frac{1}{2}\,(\nablasl_4(\slashed{\varepsilon}\cdot\chib))\,{}^{\star}\overset{\text{\scalebox{.6}{$(1)$}}}{\slashed{\mathfrak{f}}}_3  \nonumber
\end{align*}
and the linearised Gauss equation
\begin{align*}
\overset{\text{\scalebox{.6}{$(1)$}}}{\widetilde{\slashed{K}}} &= -\frac{1}{4}\,(\text{tr}\chib)(\trchio)-\frac{1}{4}\,(\text{tr}\chi)(\trchibo)-\frac{1}{4}\,(\slashed{\varepsilon}\cdot\chib)(\overset{\text{\scalebox{.6}{$(1)$}}}{\slashed{\varepsilon}\cdot\chi})-\frac{1}{4}\,(\slashed{\varepsilon}\cdot\chi)(\overset{\text{\scalebox{.6}{$(1)$}}}{\slashed{\varepsilon}\cdot\chib})-\rhoo \, .
\end{align*}

\subsection{Linearised Bianchi equations}  \label{sec_lin_eqns_Bianchi}

The linearised Bianchi equations read
\begin{gather}
\nablasl_3\alphao+\frac{1}{2}\,(\text{tr}\chib)\alphao +\frac{1}{2}(\slashed{\varepsilon}\cdot\chib){}^{\star}\alphao =   -2\,\slashed{\mathcal{D}}_2^{\star}\betao-3\,\rho\,\chiho -3\,\sigma\,{}^{\star}\chiho+5\,\eta\,\widehat{\otimes}\betao  \, ,\\[6pt]
\nablasl_4\betao+2 \, (\text{tr}\chi)\betao-2\, (\slashed{\varepsilon}\cdot\chi){}^{\star}\betao-\omegah\betao =  \slashed{\text{div}}\alphao+(\etab^{\sharp}+2\,\eta^{\sharp})\cdot\alphao \, , 
\end{gather}
\begin{align}
\nablasl_3\betao+(\text{tr} \chib)\betao+(\slashed{\varepsilon}\cdot\chi){}^{\star}\betao  =& \,\slashed{\mathcal{D}}_1^{\star}(-\rhoo,\sigmao)+3\, \rho\etao +3\rhoo\eta  +3\,\sigma\,{}^{\star}\etao +3\sigmao {}^{\star}\eta \\
&+(\nablasl_3\rho)\overset{\text{\scalebox{.6}{$(1)$}}}{\mathfrak{\slashed{\mathfrak{f}}}}_{4}+(\nablasl_3\sigma){}^{\star}\overset{\text{\scalebox{.6}{$(1)$}}}{\mathfrak{\slashed{\mathfrak{f}}}}_{4}+ (\nablasl_4\rho)\overset{\text{\scalebox{.6}{$(1)$}}}{\mathfrak{\slashed{\mathfrak{f}}}}_{3}+ (\nablasl_4\sigma){}^{\star}\overset{\text{\scalebox{.6}{$(1)$}}}{\mathfrak{\slashed{\mathfrak{f}}}}_{3}  \nonumber \\
&-({}^{\star}\overset{\text{\scalebox{.6}{$(1)$}}}{\widehat{\slashed{g}}})^{\sharp_1}\cdot(\nablasl\sigma +3\,\sigma\,\eta) \nonumber \, ,
\end{align}
\begin{gather}
\nablasl_4 \rhoo+\frac{3}{2}\,(\text{tr}\chi)\rhoo  = \slashed{\text{div}}\betao+(2\etab+\eta,\betao)-\frac{3}{2}\,\rho\,(\trchio)-\frac{3}{2}\,\sigma\,(\overset{\text{\scalebox{.6}{$(1)$}}}{\slashed{\varepsilon}\cdot\chi}) -\frac{3}{2}\,(\slashed{\varepsilon}\cdot\chi)\sigmao \, , \\[6pt]
\nablasl_4 \sigmao+\frac{3}{2}\,(\text{tr}\chi)\sigmao =  -\slashed{\text{curl}}\betao-(2\etab+\eta)\wedge\betao-\frac{3}{2}\,\sigma\,(\trchio) +\frac{3}{2}\,\rho\,(\overset{\text{\scalebox{.6}{$(1)$}}}{\slashed{\varepsilon}\cdot\chi})+\frac{3}{2}\,(\slashed{\varepsilon}\cdot\chi)\rhoo \, ,  
\end{gather}
\begin{align}
\nablasl_3 \rhoo+\frac{3}{2}\,(\text{tr}\chib)\rhoo  =& \, -\slashed{\text{div}}\betabo-(\eta,\betabo)-\frac{3}{2}\,\rho\,(\trchibo)+\frac{3}{2}\,\sigma\,(\overset{\text{\scalebox{.6}{$(1)$}}}{\slashed{\varepsilon}\cdot\chib}) +\frac{3}{2}\,(\slashed{\varepsilon}\cdot\chib)\sigmao \\
& -(\nablasl \rho,\overset{\text{\scalebox{.6}{$(1)$}}}{\mathfrak{\underline{\mathfrak{f}}}}) -(\nablasl_4\rho)\overset{\text{\scalebox{.6}{$(1)$}}}{\mathfrak{\underline{\mathfrak{f}}}}_3  -(\nablasl_3\rho)\overset{\text{\scalebox{.6}{$(1)$}}}{\mathfrak{\underline{\mathfrak{f}}}}_4 \, , \nonumber \\[6pt]
\nablasl_3 \sigmao+\frac{3}{2}\,(\text{tr}\chib)\sigmao  =& \, -\slashed{\text{curl}}\betabo-\eta\wedge\betabo-\frac{3}{2}\,\sigma\,(\trchibo)-\frac{3}{2}\,\rho\,(\overset{\text{\scalebox{.6}{$(1)$}}}{\slashed{\varepsilon}\cdot\chib})-\frac{3}{2}\,(\slashed{\varepsilon}\cdot\chib)\rhoo \\
& -(\nablasl \sigma,\overset{\text{\scalebox{.6}{$(1)$}}}{\mathfrak{\underline{\mathfrak{f}}}})  -(\nablasl_4\sigma)\overset{\text{\scalebox{.6}{$(1)$}}}{\mathfrak{\underline{\mathfrak{f}}}}_3  -(\nablasl_3\sigma)\overset{\text{\scalebox{.6}{$(1)$}}}{\mathfrak{\underline{\mathfrak{f}}}}_4 \, , \nonumber
\end{align}
\begin{align}
\nablasl_4\betabo+(\text{tr} \chi)\betabo+(\slashed{\varepsilon}\cdot\chib){}^{\star}\betabo+\omegah\betabo  =& \,\slashed{\mathcal{D}}_1^{\star}(\rhoo,\sigmao)-3\,\rhoo\etab  +3\,\sigmao {}^{\star}\etab  \\
&-(\nablasl_3\rho)\overset{\text{\scalebox{.6}{$(1)$}}}{\mathfrak{\slashed{\mathfrak{f}}}}_{4}+ (\nablasl_3\sigma){}^{\star}\overset{\text{\scalebox{.6}{$(1)$}}}{\mathfrak{\slashed{\mathfrak{f}}}}_{4}- (\nablasl_4\rho)\overset{\text{\scalebox{.6}{$(1)$}}}{\mathfrak{\slashed{\mathfrak{f}}}}_{3}+ (\nablasl_4\sigma){}^{\star}\overset{\text{\scalebox{.6}{$(1)$}}}{\mathfrak{\slashed{\mathfrak{f}}}}_{3} \nonumber \\
&-({}^{\star}\overset{\text{\scalebox{.6}{$(1)$}}}{\widehat{\slashed{g}}})^{\sharp_1}\cdot(\nablasl\sigma +3\,\sigma\,\etab)  \, , \nonumber 
\end{align}
\begin{gather}
\nablasl_3\betabo+2\,(\text{tr}\chib)\betabo-2\,(\slashed{\varepsilon}\cdot\chib){}^{\star}\betabo =  -\slashed{\text{div}}\alphabo+\eta^{\sharp}\cdot\alphabo -3\,\rho\ybo+3\,\sigma\, {}^{\star}\ybo \, , \\[6pt] 
\nablasl_4\alphabo+\frac{1}{2}\,(\text{tr}\chi)\alphabo-\frac{1}{2}\,(\slashed{\varepsilon}\cdot\chi){}^{\star}\alphabo+2\,\omegah\alphabo =    2\,\slashed{\mathcal{D}}_2^{\star}\betabo-3\,\rho\chibho +3\,\sigma\, {}^{\star}\chibho-(4\etab-\eta)\,\widehat{\otimes}\betabo \, .  
\end{gather}

\section{The main theorem} \label{sec_main_theorem}

For any $\mathfrak{D}_{\mathcal{N}_{\text{as}}}$ covariant tensor $\varsigma$, we define the pointwise norm 
\begin{equation*}
|\varsigma|^2_{\slashed{g}}:=\varsigma\cdot_{\slashed{g}}\varsigma
\end{equation*}
and the $L^2(\mathbb{S}^2)$-norm
\begin{equation}
\| \varsigma\|_{L^2(\mathbb{S}^2_{t,r},\slashed{g})}^2 := \int_{\mathbb{S}^2_{t,r}} |\varsigma|^2_{\slashed{g}} \, \, \slashed{\varepsilon}_{\slashed{\gamma}} \, , \label{def_L2_norm_non_integr}
\end{equation}
with 
\begin{equation*}
    \slashed{\gamma}=\Sigma \, d\theta^2+\frac{(r^2+a^2)^2-a^2\Delta\sin^2\theta}{\Sigma}\sin^2\theta \, d\phi^2
\end{equation*}
the metric induced by $g$ over the $\mathbb{S}^2_{t,r}$-spheres and $\slashed{\varepsilon}_{\slashed{\gamma}}$ the standard volume form of $\slashed{\gamma}$.~For any $\mathfrak{D}_{\mathcal{N}_{\text{as}}}$ covariant tensors $\varsigma_{j_1},\ldots,\varsigma_{j_n}$, we introduce the schematic notation
\begin{equation}
\|\nablasl^{i_1}_4\nablasl^{i_2}_3\nablasl^{i_3}(\varsigma_{j_1},\ldots,\varsigma_{j_n})\|_{L^2(\mathbb{S}^2_{t,r},\slashed{g})}^2= \|\nablasl^{i_1}_4\nablasl^{i_2}_3\nablasl^{i_3}\varsigma_{j_1}\|_{L^2(\mathbb{S}^2_{t,r},\slashed{g})}^2+\ldots +\|\nablasl^{i_1}_4\nablasl^{i_2}_3\nablasl^{i_3} \varsigma_{j_n}\|_{L^2(\mathbb{S}^2_{t,r},\slashed{g})}^2 \, . \label{schematic_notation_th}
\end{equation}
and also, for $k\in\mathbb N$,
\begin{align}
    \|\mathfrak{\partial}^{\leq k}(\varsigma_{j_1},\dots \varsigma_{j_n})\|_{L^{2}(\mathbb{S}^2_{t,r},\slashed{g})}&= \sum_{0\leq i_1+i_2+i_3\leq k}\|\nablasl^{i_1}_4\nablasl^{i_2}_3\nablasl^{i_3}(\varsigma_{j_1},\dots \varsigma_{j_n})\|_{L^2(\mathbb{S}^2_{t,r},\slashed{g})} \, , \label{schematic_notation_th_a}\\
    \|\mathfrak{\partial}^{\leq k}(\varsigma_{j_1},\dots \varsigma_{j_n})\|_{L^{2}_{w}(\mathbb{S}^2_{t,r},\slashed{g})}&= r^{-i_1-i_2}\left(\Delta r^{-2}\right)^{i_2}\sum_{0\leq i_1+i_2+i_3\leq k}\|\nablasl^{i_1}_4\nablasl^{i_2}_3\nablasl^{i_3}(\varsigma_{j_1},\dots \varsigma_{j_n})\|_{L^2(\mathbb{S}^2_{t,r},\slashed{g})} \, . \label{schematic_notation_th_b}
\end{align}

\medskip

The following is our main theorem.

\medskip

\begin{theorem}[Elliptic $L^2(\mathbb{S}^2)$-estimates for linearised curvature components] \label{main_th}
Let $0\leq |a|<M$, $k\in\mathbb{N}$ with $k\geq 3$, and finite constant $R>r_+$.~Then, there exists a constant $C_{k,R}>0$ such that, for any solution to the linearised system of equations, with notation
\begin{align*}
\psio&= \left\lbrace \betao,\betabo,(\rhoo,\sigmao) \right\rbrace \, , \\
\Gammao&=\left\lbrace (\overset{\text{\scalebox{.6}{$(1)$}}}{\textup{tr}\chi}),(\overset{\text{\scalebox{.6}{$(1)$}}}{\slashed{\varepsilon}\cdot\chi}),(\overset{\text{\scalebox{.6}{$(1)$}}}{\textup{tr}\chib}),(\overset{\text{\scalebox{.6}{$(1)$}}}{\slashed{\varepsilon}\cdot\chib}),\overset{\text{\scalebox{.6}{$(1)$}}}{\omegab},\overset{\text{\scalebox{.6}{$(1)$}}}{\yb},\overset{\text{\scalebox{.6}{$(1)$}}}{\eta},\overset{\text{\scalebox{.6}{$(1)$}}}{\zeta},\overset{\text{\scalebox{.6}{$(1)$}}}{\chih},\overset{\text{\scalebox{.6}{$(1)$}}}{\chibh} \right\rbrace \, , \\
\fo&=\left\lbrace \overset{\text{\scalebox{.6}{$(1)$}}}{\mathfrak{\underline{\mathfrak{f}}}}_4,\overset{\text{\scalebox{.6}{$(1)$}}}{\mathfrak{\underline{\mathfrak{f}}}}_3,(\textup{tr}\overset{\text{\scalebox{.6}{$(1)$}}}{\slashed{g}}),\overset{\text{\scalebox{.6}{$(1)$}}}{\mathfrak{\underline{\mathfrak{f}}}},\overset{\text{\scalebox{.6}{$(1)$}}}{\mathfrak{\slashed{\mathfrak{f}}}}_{3},\overset{\text{\scalebox{.6}{$(1)$}}}{\mathfrak{\slashed{\mathfrak{f}}}}_{4},\overset{\text{\scalebox{.6}{$(1)$}}}{\widehat{\slashed{g}}} \right\rbrace \, ,
\end{align*}
the estimates
\begin{equation*}
\sum_{0\leq i_1+i_2+i_3\leq k}\|\nablasl^{i_1}_4\nablasl^{i_2}_3\nablasl^{i_3}\psio\|_{L^2(\mathbb{S}^2_{t,r},\slashed{g})} \leq C_{k,R}\left[\|\mathfrak{\partial}^{\leq k}(\alphao,\alphabo)\|_{L^2(\mathbb{S}^2_{t,r},\slashed{g})} +\|\mathfrak{\partial}^{\leq k-1}(\fo,\Gammao,\psio)\|_{L^2(\mathbb{S}^2_{t,r},\slashed{g})}\right]
\end{equation*}
hold for all $\mathbb{S}^2_{t,r}$-spheres with $r_+\leq r \leq R$.
\end{theorem}

\medskip

\begin{remark}
From our proof of Theorem \ref{main_th}, one can check that the constant $C_{k,R}$ in the theorem is, in fact, uniform in $0\leq |a|\leq M$. From our proof, one can also sharpen the norms of the extremal linearised curvature components on the right hand side of the inequalities (see, for instance, Proposition \ref{main_th_angular_version}).
\end{remark}

\medskip

\begin{remark}
In our proof of Theorem \ref{main_th}, we do not keep track of the lower order terms. By computing the lower order terms, it is possible to sharpen the inequalities in the theorem by only including the relevant lower order terms on the right hand side.
\end{remark}

\medskip

We will prove Theorem \ref{main_th} in the case $k=3$.~Higher-order estimates follow by iterating the scheme that is presented in the proof (see the related Remark \ref{rmk_high_order_th} at the end of Section \ref{sec_proof_main_th}).

\section{The \texorpdfstring{$\mathbb{S}^2$}{S2}-projection procedure} 
\label{sec_proj_formulae}

In this section, we employ the projection procedure of \cite{Benomio_thesis_doi_2}, which we recall here from Section 7 of \cite{benomio_kerr_system_2}, to map $\mathfrak{D}_{\mathcal{N}_{\text{as}}}$ tensors to $\mathbb{S}^2_{t,r}$ tensors.\footnote{The projection procedure from Section 7 of \cite{benomio_kerr_system_2} is, in fact, more general than its application here, in that it allows to map tensors over a non-integrable distribution to tensors over another non-integrable distribution.~The fact that here the landing distribution is integrable (i.e.~$T\mathbb{S}^2_{t,r}$) introduces various simplifications.}~In summary, the procedure starts by extending $\mathfrak{D}_{\mathcal{N}_{\text{as}}}$ tensors to spacetime tensors which identically vanish when evaluated on either $e_4^{\text{as}}$ or $e_3^{\text{as}}$.~It then projects the obtained spacetime tensors to the $\mathbb{S}^2_{t,r}$-spheres in the natural way.~We now introduce the basic objects involved in the procedure.

\medskip

We recall that, in Section \ref{sec_kerr_exterior_manifold} above, we denoted by $g$ the spacetime Kerr metric and by $\slashed{g}$ the $\mathfrak{D}_{\mathcal{N}_{\text{as}}}$ metric tensor induced by $g$ over $\mathfrak{D}_{\mathcal{N}_{\text{as}}}$.~We also recall that we denoted by $\slashed{\gamma}$ the metric induced by $g$ on the $\mathbb{S}^2_{t,r}$-spheres.~Let now 
\begin{equation*}
    \check{\slashed{g}}=\Sigma \, d\theta^2+\frac{(r^2+a^2)^2}{\Sigma}\sin^2\theta \, d\phi^2 
\end{equation*}
be the metric induced by $\slashed{g}$ on the foliation spheres $\mathbb{S}^2_{t,r}$ via the projection procedure from Section 7 of \cite{benomio_kerr_system_2}, with associated inverse metric $\check{\slashed{g}}{}^{-1}$ and standard volume form $\slashed{\varepsilon}_{\check{\slashed{g}}}$.~We remark that $\check{\slashed{g}}=\slashed{\gamma}$ on $\mathbb{S}^2_{t,r_+}$, i.e. 
\begin{equation} \label{aux_0_b}
\check{\slashed{g}}|_{\mathcal{H}^+}=\slashed{\gamma}|_{\mathcal{H}^+} \, ,
\end{equation}
where the property essentially follows from the properties \eqref{intro_e4_horizon_killing} and \eqref{e4_distribution_horizon} of the frame $\mathcal{N}_{\text{as}}$ on $\mathcal{H}^+$.

\medskip

We recall that, in Section \ref{sec_kerr_exterior_manifold} above, we denoted by $\nablasl$ the connection induced by $\nabla$ over $\mathfrak{D}_{\mathcal{N}_{\text{as}}}$.~Let $\check{\nablasl}$ be the Levi-Civita connection of $\check{\slashed{g}}$ over the $\mathbb{S}^2_{t,r}$-spheres.~Let $\nablasl^{\slashed{\gamma}}$ be the Levi-Civita connection of $\slashed{\gamma}$ over the $\mathbb{S}^2_{t,r}$-spheres (also coinciding with the connection induced by $\nabla$ over the $\mathbb{S}^2_{t,r}$-spheres).~For any $\mathbb{S}^2_{t,r}$ covariant tensor $\varsigma$, one has (in schematic form)
\begin{equation} \label{aux_difference_connections}
(\check{\nablasl}-\nablasl^{\slashed{\gamma}})\,\varsigma\sim \slashed{\Gamma}_{h}\cdot \varsigma 
\end{equation}
for some non-trivial $\mathbb{S}^2_{t,r}$ $(1,2)$-tensors $\slashed{\Gamma}_{h}$ such that 
\begin{equation} \label{aux_0_c}
\slashed{\Gamma}_{h}|_{\mathcal{H}^+} = 0 \, ,
\end{equation}
where the property \eqref{aux_0_c} follows from the identity \eqref{aux_0_b}.~For any smooth scalar function $f$, one has $(\check{\nablasl}-\nablasl^{\slashed{\gamma}})f = 0$.

\subsection{The \texorpdfstring{$\mathbb{S}^2$}{S2}-projection formulae} \label{sec_proj_formulae_2}

In this section, we state the $\mathbb{S}^2$-projection formulae which are necessary for the sequel.~For a proof of the formulae, the reader may refer to Proposition 7.47 (and previous propositions) of \cite{benomio_kerr_system_2}.

\medskip

Let $f$ be any smooth scalar function, $\varsigma,\varpi$ any $\mathfrak{D}_{\mathcal{N}_{\text{as}}}$ one-forms and $\theta$ any $\mathfrak{D}_{\mathcal{N}_{\text{as}}}$ two-tensor.~Their $\mathbb{S}^2$-projections will be denoted as the $\mathbb{S}^2_{t,r}$ one-forms $\widetilde{\varsigma},\widetilde{\varpi}$ and the $\mathbb{S}^2_{t,r}$ two-tensor $\widetilde{\vartheta}$.~One has the identities
\begin{align*}
\widetilde{\varsigma}\cdot_{\check{\slashed{g}}}\widetilde{\varpi}&=\varsigma\cdot_{\slashed{g}}\varpi \, , & \widetilde{\varsigma}\wedge_{\check{\slashed{g}}}\widetilde{\varpi}&=\varsigma\wedge_{\slashed{g}} \varpi \, , & \widetilde{\varsigma}\,\widehat{\otimes}_{\check{\slashed{g}}}\widetilde{\varpi}&=\widetilde{\varsigma\,\widehat{\otimes}_{\slashed{g}}\varpi} \, .
\end{align*}
In particular, we have the pointwise-norm identity
\begin{equation} \label{equality_pointwise_norms}
|\, \widetilde{\varsigma}\, |^2_{\check{\slashed{g}}} =|\, \varsigma\, |^2_{\slashed{g}} \, .
\end{equation}
These identities generalise to higher rank tensors in the natural way.~We also have
\begin{align*}
\text{tr}_{\check{\slashed{g}}}\widetilde{\theta}&= \text{tr}_{\slashed{g}}\theta \, ,   &  \widetilde{\theta}^{\sharp_2}\cdot_{\check{\slashed{g}}}\widetilde{\varsigma}&=\widetilde{\theta^{\sharp_2}\cdot_{\slashed{g}}\varsigma} 
\end{align*}
and analogous identities for contractions of tensors of different rank and Hodge duals.~In particular, from the projection formula for the trace, one sees that if $\theta$ is a symmetric traceless (with respect to $\slashed{g}$) $\mathfrak{D}_{\mathcal{N}_{\text{as}}}$ two-tensor, then $\widetilde{\theta}$ is a symmetric traceless (with respect to $\check{\slashed{g}}$) $\mathbb{S}^2_{t,r}$ two-tensor.

\medskip

For any smooth scalar function $f$ and $\mathfrak{D}_{\mathcal{N}_{\text{as}}}$ covariant tensor $\varsigma$, we have the identities (we denote the lower order terms appearing in the formulae by ``l.o.t.'')
\begin{align*}
\nablasl_4 f &= \nablasl^{\slashed{\gamma}}_4 f \, , &
\nablasl_3 f &= \nablasl^{\slashed{\gamma}}_3 f \, , &
\widetilde{\nablasl f} &=\nablasl^{\slashed{\gamma}}f+ (\nablasl^{\slashed{\gamma}}_4 f) \,\mathfrak{k}+(\nablasl^{\slashed{\gamma}}_3 f)\,\mathfrak{h}+\textup{l.o.t.}
\end{align*}
and
\begin{align*}
\widetilde{\nablasl_{4}\varsigma}&= \nablasl^{\slashed{\gamma}}_{4}\, \widetilde{\varsigma}+\textup{l.o.t.} \, ,  &
\widetilde{\nablasl_{3}\varsigma}&= \nablasl^{\slashed{\gamma}}_{3}\, \widetilde{\varsigma}+\textup{l.o.t.} \, ,   &
\widetilde{\nablasl\varsigma} &=  \nablasl^{\slashed{\gamma}}\, \widetilde{\varsigma}+\mathfrak{k}\otimes \nablasl^{\slashed{\gamma}}_4\, \widetilde{\varsigma} +\mathfrak{h}\otimes \nablasl^{\slashed{\gamma}}_3\, \widetilde{\varsigma}+\textup{l.o.t.}
\end{align*}
with $\mathbb{S}^2_{t,r}$ one-forms $\mathfrak{k}$ and $\mathfrak{h}$ such that, for any local frame $(e_1^{\text{ad}},e_2^{\text{ad}})$ of $\mathbb{S}^2_{t,r}$, we have
\begin{align*}
    \mathfrak{k}(e_A^{\text{ad}})&:=\frac{1}{2}\, g(e_3^{\text{as}},e_A^{\text{ad}}) \, , & \mathfrak{h}(e_A^{\text{ad}})&:=\frac{1}{2}\, g(e_4^{\text{as}},e_A^{\text{ad}}) \, , & A&=\left\lbrace 1,2 \right\rbrace \, .
\end{align*}
See Section \ref{sec_kerr_one_forms} for the explicit coordinate form of $\mathfrak{k}$ and $\mathfrak{h}$ and their relevant properties. 

\medskip

From the formulae above, one can deduce the $\mathbb{S}^2$-projection formulae for angular operators, e.g.
\begin{align*}
\slashed{\text{div}}\, \varsigma &= \check{\slashed{\text{div}}}\, \widetilde{\varsigma} +\mathfrak{k}\cdot_{\check{\slashed{g}}} \check{\nablasl}_4\, \widetilde{\varsigma} +\mathfrak{h}\cdot_{\check{\slashed{g}}} \check{\nablasl}_3\, \widetilde{\varsigma}+\textup{l.o.t.} \, , &
\slashed{\text{curl}}\, \varsigma &= \check{\slashed{\text{curl}}}\, \widetilde{\varsigma} +\mathfrak{k}\wedge_{\check{\slashed{g}}} \check{\nablasl}_4\, \widetilde{\varsigma} +\mathfrak{h}\wedge_{\check{\slashed{g}}} \check{\nablasl}_3\, \widetilde{\varsigma}+\textup{l.o.t.}
\end{align*}
and
\begin{equation*}
-2\, \widetilde{\slashed{\mathcal{D}}_2^{\star} \varsigma} = -2\, \check{\slashed{\mathcal{D}}}{}_2^{\star} \, \widetilde{\varsigma} +\mathfrak{k}\,\widehat{\otimes}_{\check{\slashed{g}}} \check{\nablasl}_4\, \widetilde{\varsigma} +\mathfrak{h}\,\widehat{\otimes}_{\check{\slashed{g}}} \check{\nablasl}_3\, \widetilde{\varsigma}+\textup{l.o.t.} \, .
\end{equation*}

\medskip

\begin{remark}
One can replace the connection $\nablasl^{\slashed{\gamma}}$ by the connection $\check{\nablasl}$ in the $\mathbb{S}^2$-projection formulae for covariant derivatives of tensors, with the difference of the two connections only generating lower order terms (which moreover identically vanish at the event horizon, see \eqref{aux_difference_connections} and \eqref{aux_0_c}).~The replacement of the connection is also possible in the $\mathbb{S}^2$-projection formulae for covariant derivatives of scalar functions, with the difference of the two connections applied to a scalar function being identically zero.~In fact, in the sequel (see the equations of Section \ref{sec_proj_lin_system}), we will apply the $\mathbb{S}^2$-projection formulae by employing the connection $\check{\nablasl}$.~The formulae are stated here with the connection $\nablasl^{\slashed{\gamma}}$ to enable the reader to directly read them off from Proposition 7.47 of \cite{benomio_kerr_system_2}. 
\end{remark}

\subsection{The \texorpdfstring{$\mathbb{S}^2$}{S2} one-forms \texorpdfstring{$\mathfrak{k}$}{k} and \texorpdfstring{$\mathfrak{h}$}{h}} \label{sec_kerr_one_forms}

By adopting the orthonormal (relative to $\check{\slashed{g}}$) frame
\begin{align*}
    e_1^{\text{ad}}&=\frac{1}{\Sigma^{1/2}}\partial_{\theta} \, , &  e_2^{\text{ad}}&=\frac{\Sigma^{1/2}}{(r^2+a^2)\sin\theta}\partial_{\phi}
\end{align*}
of $\mathbb{S}^2_{t,r}$, we compute
\begin{align*}
    \mathfrak{k}(e_1^{\text{ad}})&=0 \, , & \mathfrak{k}(e_2^{\text{ad}})&=\frac{a \sqrt{\Sigma} \sin \theta }{2\left(a^2+r^2\right)} \, , & \mathfrak{h}(e_1^{\text{ad}})&=0 \, , & \mathfrak{h}(e_2^{\text{ad}})&= \frac{ a \Delta \sin \theta}{2\sqrt{\Sigma}\left(a^2+r^2\right)}
\end{align*}
and deduce
\begin{align*}
    |\mathfrak{k}|^2_{\check{\slashed{g}}} &=\frac{a^2 \Sigma \sin^2 \theta}{4\left(a^2+r^2\right)^2} \, , & |\mathfrak{h}|^2_{\check{\slashed{g}}} &=\frac{ a^2 \Delta^2 \sin^2 \theta}{4\Sigma\left(a^2+r^2\right)^2}
\end{align*}
and
\begin{align*}
    \mathfrak{k}\cdot_{\check{\slashed{g}}}\mathfrak{h} &= \frac{a^2\Delta \sin^2 \theta }{4\left(a^2+r^2\right)^2} \, , &  \mathfrak{k}\wedge_{\check{\slashed{g}}}\mathfrak{h} &=0 \, .
\end{align*}
By defining the $L^{\infty}(\mathbb{S}^2)$-norms
\begin{align*}
|f|_{\infty}&:=\sup_{\mathbb{S}^2}|f| \, , & |\varsigma|_{\infty}&:=\sup_{\mathbb{S}^2}|\varsigma|_{\check{\slashed{g}}}
\end{align*}
for any scalar function $f$ and $\mathbb{S}^2_{t,r}$ covariant tensor $\varsigma$, we have
\begin{align*}
|\mathfrak{k}|_{\infty}&=\frac{ar}{2(a^2+r^2)} \, , &
|\mathfrak{h}|_{\infty}&=\frac{a\Delta}{2r(a^2+r^2)} \, , \\
|\mathfrak{k}\cdot_{\check{\slashed{g}}}\mathfrak{h}|_{\infty}&=\frac{a^2\Delta}{4(a^2+r^2)^2} =|\mathfrak{h}|_{\infty}|\mathfrak{k}|_{\infty}\, , & |\mathfrak{h}\cdot_{\check{\slashed{g}}}\mathfrak{h}|_{\infty}&=\frac{a^2\Delta^2}{4r^2(a^2+r^2)^2}=|\mathfrak{h}|_{\infty}^2 \, .
\end{align*}

\medskip

\begin{remark}
We note that
\begin{equation}
\mathfrak{h}|_{\mathcal{H}^+} = 0 \, ,
\end{equation}
which essentially follows from the properties \eqref{intro_e4_horizon_killing}--\eqref{eA_kerr_fermi_horizon} of the frame $\mathcal{N}_{\textup{as}}$ on $\mathcal{H}^+$.
\end{remark}

\medskip

\begin{remark}
We have 
\begin{align*}
&\sup_{i_1,i_2,i_3\geq 0}|\check{\nablasl}{}_4^{i_1}\check{\nablasl}{}_3^{i_2}\check{\nablasl}{}^{i_3}\mathfrak{k}|_{\infty} \rightarrow 0 \, , & &\sup_{i_1,i_2,i_3\geq 0}|\check{\nablasl}{}_4^{i_1}\check{\nablasl}{}_3^{i_2}\check{\nablasl}{}^{i_3}\mathfrak{h}|_{\infty} \rightarrow 0
\end{align*}
as $|a|\rightarrow 0$.~In particular, for $|a|=0$, we have $\mathfrak{k}=\mathfrak{h}= 0$ on $\mathcal{M}$.  
\end{remark}

\medskip

\begin{remark}
For later convenience, we note the identities
    \begin{align}
        \mathfrak{h}&=\frac{\Delta}{\Sigma}\, \mathfrak{k} \, , & |\mathfrak{h}|_{\infty}&=\frac{\Delta}{r^2}\, |\mathfrak{k}|_{\infty} \, ,
    \end{align}
and the inequalities
\begin{align*}
    |\mathfrak{k}|_{\infty}&\leq \frac{|a|}{4M}\,, & |\mathfrak{h}|_{\infty}&<0.07\frac{|a|}{M}\,.
\end{align*}
We also introduce the radial function 
\begin{equation} \label{def_function_l}
    l(r):=\frac{Mr}{2(r^2+a^2)} \, .
\end{equation}
\end{remark}

\subsection{The \texorpdfstring{$\mathbb{S}^2$}{S2}-projected linearised system of equations} \label{sec_proj_lin_system}

In this section, we state the $\mathbb{S}^2$-projected version of the linearised system of equations of Section \ref{sec_lin_system}.~The $\mathbb{S}^2$-projected system is derived by applying the $\mathbb{S}^2$-projection formulae of Section \ref{sec_proj_formulae_2} to the equations of Section \ref{sec_lin_system}.~The unknowns of the $\mathbb{S}^2$-projected system are global, regular $\mathbb{S}^2_{t,r}$ tensors (in the sense of \cite{StabMink}).~A complete list of the unknowns is obtained by taking the $\mathbb{S}^2$-projection of the linearised quantities in Table \ref{table:linearised_unknowns}.

\medskip

In the $\mathbb{S}^2$-projected system and later in the paper, we adopt the schematic notation 
\begin{equation*}
\widetilde{\fo} \, , \, \widetilde{\Gammao} \,  , \,  \widetilde{\psio}
\end{equation*}
to denote a zero order term involving a $\mathbb{S}^2$-projected linearised metric or frame coefficient, a zero order term involving a $\mathbb{S}^2$-projected linearised connection coefficient and a zero order term involving a $\mathbb{S}^2$-projected linearised curvature component (including $\alpha$ and $\alphab$) respectively.~These terms are $\mathbb{S}^2_{t,r}$ tensors.~Depending on the equation, the terms may have different rank.~For instance, the schematic terms in equation \eqref{4_chih_alpha} are symmetric traceless $\mathbb{S}^2_{t,r}$ two-tensor, whereas the schematic terms in equations \eqref{4_eta_beta} are $\mathbb{S}^2_{t,r}$ one-forms.~We denote by $$\check{\nablasl}{}^{i_1}_4\check{\nablasl}{}^{i_2}_3\check{\nablasl}{}^{i_3}\widetilde{\fo} \, , \, \check{\nablasl}{}^{i_1}_4\check{\nablasl}{}^{i_2}_3\check{\nablasl}{}^{i_3}\widetilde{\Gammao} \, , \, \check{\nablasl}{}^{i_1}_4\check{\nablasl}{}^{i_2}_3\check{\nablasl}{}^{i_3}\widetilde{\psio}$$ the terms involving $\check{\nablasl}{}^{i_1}_4\check{\nablasl}{}^{i_2}_3\check{\nablasl}{}^{i_3}$-derivatives of $\mathbb{S}^2$-projected linearised quantities.~These terms are also $\mathbb{S}^2_{t,r}$ tensors whose rank depends on the equation where the terms appear.

\medskip

The $\mathbb{S}^2$-projected linearised null structure equations include the outgoing transport equations
\begin{equation}
\check{\nablasl}_4\widetilde{\chiho}=-\widetilde{\alphao} +\widetilde{\Gammao} \, , \label{4_chih_alpha}
\end{equation}
\begin{align}
\check{\nablasl}_4(\trchibo)&=2\rhoo +\sum_{i=0}^1\check{\nablasl}{}^{i}\widetilde{\fo}+\widetilde{\Gammao} \, , &
\check{\nablasl}_4(\overset{\text{\scalebox{.6}{$(1)$}}}{\slashed{\varepsilon}\cdot\chib})&=2\sigmao +\widetilde{\fo}+\widetilde{\Gammao} \, , \nonumber\\
\check{\nablasl}_4\widetilde{\etao}&=-\widetilde{\betao}+\widetilde{\fo}+\widetilde{\Gammao} \, ,  & 
\check{\nablasl}_4\widetilde{\zetao}&=-\widetilde{\betao}+\widetilde{\fo}+\widetilde{\Gammao} \label{4_eta_beta} \, , \\
\check{\nablasl}_4\widetilde{\ybo}&=-\widetilde{\betabo} +\sum_{i=0}^1\check{\nablasl}{}^{i}\widetilde{\fo}+\widetilde{\Gammao} \, ,  &
\check{\nablasl}_4\omegabo&=-2\rhoo +\widetilde{\fo}+\widetilde{\Gammao} \, , \nonumber 
\end{align}
the ingoing transport equations
\begin{align*}
\check{\nablasl}_3\widetilde{\chibho}&=-2\, \check{\slashed{\mathcal{D}}}{}_2^{\star} \, \widetilde{\ybo} +\mathfrak{k}\widehat{\otimes}_{\check{\slashed{g}}} \check{\nablasl}_4\, \widetilde{\ybo} +\mathfrak{h}\widehat{\otimes}_{\check{\slashed{g}}} \check{\nablasl}_3\, \widetilde{\ybo}-\widetilde{\alphabo} +\widetilde{\Gammao} \, , \\
\check{\nablasl}_3(\trchibo)&=2\check{\slashed{\text{div}}}\, \widetilde{\ybo} +2\mathfrak{k}\cdot_{\check{\slashed{g}}} \check{\nablasl}_4\, \widetilde{\ybo} +2\mathfrak{h}\cdot_{\check{\slashed{g}}} \check{\nablasl}_3\, \widetilde{\ybo} +\widetilde{\fo}+\widetilde{\Gammao} \, , \\
\check{\nablasl}_3(\overset{\text{\scalebox{.6}{$(1)$}}}{\slashed{\varepsilon}\cdot\chib})&=2\check{\slashed{\text{curl}}}\, \widetilde{\ybo} +2\mathfrak{k}\wedge_{\check{\slashed{g}}} \check{\nablasl}_4\, \widetilde{\ybo} +2\mathfrak{h}\wedge_{\check{\slashed{g}}} \check{\nablasl}_3\, \widetilde{\ybo}+\widetilde{\fo}+\widetilde{\Gammao} \, , 
\end{align*}
\begin{align*}
\check{\nablasl}_3\widetilde{\chiho}&=-2\, \check{\slashed{\mathcal{D}}}{}_2^{\star} \, \widetilde{\etao} +\mathfrak{k}\widehat{\otimes}_{\check{\slashed{g}}} \check{\nablasl}_4\, \widetilde{\etao} +\mathfrak{h}\widehat{\otimes}_{\check{\slashed{g}}} \check{\nablasl}_3\, \widetilde{\etao}+\sum_{i=0}^1\check{\nablasl}{}^i\widetilde{\fo}+\widetilde{\Gammao} \, , \\
\check{\nablasl}_3(\trchio)&=2\check{\slashed{\text{div}}}\, \widetilde{\etao} +2\mathfrak{k}\cdot_{\check{\slashed{g}}} \check{\nablasl}_4\, \widetilde{\etao} +2\mathfrak{h}\cdot_{\check{\slashed{g}}} \check{\nablasl}_3\, \widetilde{\etao}+2\rhoo +\sum_{i=0}^1\check{\nablasl}{}^i\widetilde{\fo}+\widetilde{\Gammao} \, , \\
\check{\nablasl}_3(\overset{\text{\scalebox{.6}{$(1)$}}}{\slashed{\varepsilon}\cdot\chi})&=2\check{\slashed{\text{curl}}}\, \widetilde{\etao} +2\mathfrak{k}\wedge_{\check{\slashed{g}}} \check{\nablasl}_4\, \widetilde{\etao} +2\mathfrak{h}\wedge_{\check{\slashed{g}}} \check{\nablasl}_3\, \widetilde{\etao}-2\sigmao +\widetilde{\fo}+\widetilde{\Gammao} \, ,
\end{align*}
\begin{equation*}
\check{\nablasl}_3\widetilde{\zetao}=\check{\nablasl}\omegabo+ (\check{\nablasl}_4 \omegabo) \mathfrak{k}+(\check{\nablasl}_3 \omegabo)\mathfrak{h}-\widetilde{\betabo}+\sum_{i=0}^1\check{\nablasl}{}^i\widetilde{\fo}+\widetilde{\Gammao} 
\end{equation*}
and the elliptic equations
\begin{equation*}
\check{\slashed{\text{curl}}}\, \widetilde{\zetao} =-\mathfrak{k}\wedge_{\check{\slashed{g}}} \check{\nablasl}_4\, \widetilde{\zetao} -\mathfrak{h}\wedge_{\check{\slashed{g}}} \check{\nablasl}_3\, \widetilde{\zetao}+\sigmao +\widetilde{\fo}+\widetilde{\Gammao}  \, , 
\end{equation*}
\begin{align*}
\check{\slashed{\text{div}}}\, \widetilde{\chiho} &=-\mathfrak{k}\cdot_{\check{\slashed{g}}} \check{\nablasl}_4\, \widetilde{\chiho} -\mathfrak{h}\cdot_{\check{\slashed{g}}} \check{\nablasl}_3\, \widetilde{\chiho} \\
&+\frac{1}{2}\check{\nablasl}(\trchio)+ \frac{1}{2}(\check{\nablasl}_4 (\trchio)) \mathfrak{k}+\frac{1}{2}(\check{\nablasl}_3 (\trchio))\mathfrak{h} \nonumber \\
&-\frac{1}{2}{}^{\star}\check{\nablasl}(\overset{\text{\scalebox{.6}{$(1)$}}}{\slashed{\varepsilon}\cdot\chi})- \frac{1}{2}(\check{\nablasl}_4 (\overset{\text{\scalebox{.6}{$(1)$}}}{\slashed{\varepsilon}\cdot\chi})) {}^{\star}\mathfrak{k}-\frac{1}{2}(\check{\nablasl}_3 (\overset{\text{\scalebox{.6}{$(1)$}}}{\slashed{\varepsilon}\cdot\chi})){}^{\star}\mathfrak{h}-\widetilde{\betao} \nonumber\\
&+\widetilde{\fo}+\widetilde{\Gammao} \, , \nonumber\\
\check{\slashed{\text{div}}}\, \widetilde{\chibho} &=-\mathfrak{k}\cdot_{\check{\slashed{g}}} \check{\nablasl}_4\, \widetilde{\chibho} -\mathfrak{h}\cdot_{\check{\slashed{g}}} \check{\nablasl}_3\, \widetilde{\chibho} \\
&+\frac{1}{2}\check{\nablasl}(\trchibo)+ \frac{1}{2}(\check{\nablasl}_4 (\trchibo)) \mathfrak{k}+\frac{1}{2}(\check{\nablasl}_3 (\trchibo))\mathfrak{h} \nonumber\\
&-\frac{1}{2}{}^{\star}\check{\nablasl}(\overset{\text{\scalebox{.6}{$(1)$}}}{\slashed{\varepsilon}\cdot\chib})- \frac{1}{2}(\check{\nablasl}_4 (\overset{\text{\scalebox{.6}{$(1)$}}}{\slashed{\varepsilon}\cdot\chib})) {}^{\star}\mathfrak{k}-\frac{1}{2}(\check{\nablasl}_3 (\overset{\text{\scalebox{.6}{$(1)$}}}{\slashed{\varepsilon}\cdot\chib})){}^{\star}\mathfrak{h}+\widetilde{\betabo} \nonumber\\
&+\widetilde{\fo}+\widetilde{\Gammao} \, . \nonumber
\end{align*}
The $\mathbb{S}^2$-projected linearised Bianchi equations read
\begin{equation}
\check{\nablasl}_3\widetilde{\alphao}=-2\, \check{\slashed{\mathcal{D}}}{}_2^{\star} \, \widetilde{\betao} +\mathfrak{k}\widehat{\otimes}_{\check{\slashed{g}}} \check{\nablasl}_4\, \widetilde{\betao} +\mathfrak{h}\widehat{\otimes}_{\check{\slashed{g}}} \check{\nablasl}_3\, \widetilde{\betao}+\widetilde{\Gammao}+\widetilde{\psio} \, , \label{3_alpha}
\end{equation}
\begin{gather}
\check{\nablasl}_4\widetilde{\betao}=\check{\slashed{\text{div}}}\, \widetilde{\alphao} +\mathfrak{k}\cdot_{\check{\slashed{g}}} \check{\nablasl}_4\, \widetilde{\alphao} +\mathfrak{h}\cdot_{\check{\slashed{g}}} \check{\nablasl}_3\, \widetilde{\alphao}+\widetilde{\psio} \, , \label{4_beta} \\
\check{\nablasl}_3\widetilde{\betao}= \check{\nablasl}\rhoo+ (\check{\nablasl}_4 \rhoo) \mathfrak{k}+(\check{\nablasl}_3 \rhoo)\mathfrak{h} +{}^{\star}\check{\nablasl}\sigmao+ (\check{\nablasl}_4 \sigmao) {}^{\star}\mathfrak{k}+(\check{\nablasl}_3 \sigmao){}^{\star}\mathfrak{h} +\widetilde{\fo}+\widetilde{\Gammao}+\widetilde{\psio} \, , \label{3_beta}
\end{gather}
\begin{align}
\check{\nablasl}_4\rhoo&= \check{\slashed{\text{div}}}\, \widetilde{\betao} +\mathfrak{k}\cdot_{\check{\slashed{g}}} \check{\nablasl}_4\, \widetilde{\betao} +\mathfrak{h}\cdot_{\check{\slashed{g}}} \check{\nablasl}_3\, \widetilde{\betao} +\widetilde{\Gammao}+\widetilde{\psio} \, , \label{4_rho}\\
\check{\nablasl}_4\sigmao&= -\check{\slashed{\text{curl}}}\, \widetilde{\betao} -\mathfrak{k}\wedge_{\check{\slashed{g}}} \check{\nablasl}_4\, \widetilde{\betao} -\mathfrak{h}\wedge_{\check{\slashed{g}}} \check{\nablasl}_3\, \widetilde{\betao}+\widetilde{\Gammao}+\widetilde{\psio} \, , \label{4_sigma}
\end{align}
\begin{align}
\check{\nablasl}_3\rhoo&= -\check{\slashed{\text{div}}}\, \widetilde{\betabo} -\mathfrak{k}\cdot_{\check{\slashed{g}}} \check{\nablasl}_4\, \widetilde{\betabo} -\mathfrak{h}\cdot_{\check{\slashed{g}}} \check{\nablasl}_3\, \widetilde{\betabo}+\widetilde{\fo}+\widetilde{\Gammao}+\widetilde{\psio} \, , \label{3_rho}\\
\check{\nablasl}_3\sigmao&= -\check{\slashed{\text{curl}}}\, \widetilde{\betabo} -\mathfrak{k}\wedge_{\check{\slashed{g}}} \check{\nablasl}_4\, \widetilde{\betabo} -\mathfrak{h}\wedge_{\check{\slashed{g}}} \check{\nablasl}_3\, \widetilde{\betabo}+\widetilde{\fo}+\widetilde{\Gammao}+\widetilde{\psio} \, , \label{3_sigma}
\end{align}
\begin{gather}
\check{\nablasl}_4\widetilde{\betabo}=-\check{\nablasl}\rhoo- (\check{\nablasl}_4 \rhoo) \mathfrak{k}-(\check{\nablasl}_3 \rhoo)\mathfrak{h} +{}^{\star}\check{\nablasl}\sigmao+ (\check{\nablasl}_4 \sigmao) {}^{\star}\mathfrak{k}+(\check{\nablasl}_3 \sigmao){}^{\star}\mathfrak{h} +\widetilde{\fo}+\widetilde{\psio} \, , \label{4_betab}\\
\check{\nablasl}_3\widetilde{\betabo}=-\check{\slashed{\text{div}}}\, \widetilde{\alphabo} -\mathfrak{k}\cdot_{\check{\slashed{g}}} \check{\nablasl}_4\, \widetilde{\alphabo} -\mathfrak{h}\cdot_{\check{\slashed{g}}} \check{\nablasl}_3\, \widetilde{\alphabo} +\widetilde{\Gammao}+\widetilde{\psio} \, , \label{3_betab}
\end{gather}
\begin{equation}
\check{\nablasl}_4\widetilde{\alphabo}=2\, \check{\slashed{\mathcal{D}}}{}_2^{\star} \, \widetilde{\betabo} -\mathfrak{k}\widehat{\otimes}_{\check{\slashed{g}}} \check{\nablasl}_4\, \widetilde{\betabo} -\mathfrak{h}\widehat{\otimes}_{\check{\slashed{g}}} \check{\nablasl}_3\, \widetilde{\betabo}+\widetilde{\Gammao}+\widetilde{\psio} \, . \label{4_alphab}
\end{equation}

\subsection{The \texorpdfstring{$\mathbb{S}^2$}{S2}-projected version of the main theorem} \label{sec_proj_th}

For any $\mathbb{S}^2_{t,r}$ covariant tensor $\widetilde{\varsigma}$, we define the $L^2(\mathbb{S}^2)$-norm
\begin{equation*}
 \| \widetilde{\varsigma}\|_{L^2(\mathbb{S}^2_{t,r},\check{\slashed{g}})}^2 :=\int_{\mathbb{S}^2_{t,r}} |\widetilde{\varsigma}|^2_{\check{\slashed{g}}} \,\, \slashed{\varepsilon}_{\check{\slashed{g}}} \, .
\end{equation*}
If a $\mathfrak{D}_{\mathcal{N}_{\text{as}}}$ covariant tensor $\varsigma$ is mapped, under $\mathbb{S}^2$-projection, to the $\mathbb{S}^2_{t,r}$ covariant tensor $\widetilde{\varsigma}$, then we have the equivalence of $L^2(\mathbb{S}^2)$-norms (recall definition \eqref{def_L2_norm_non_integr})
\begin{equation} \label{equivalence_L2_norms}
\| \varsigma\|_{L^2(\mathbb{S}^2_{t,r},\slashed{g})} \sim_{r} \| \widetilde{\varsigma}\|_{L^2(\mathbb{S}^2_{t,r},\check{\slashed{g}})} \, ,
\end{equation}
which directly follows from the identity \eqref{equality_pointwise_norms}, and we have, for any $k\in\mathbb{N}$, the inequalities
\begin{align} 
\sum_{i_1+i_2+i_3= k}\|\nablasl^{i_1}_4\nablasl^{i_2}_3\nablasl^{i_3}\varsigma\|_{L^2(\mathbb{S}^2_{t,r},\slashed{g})}\lesssim_{k,r} \sum_{i_1+i_2+i_3= k}\|\check{\nablasl}{}^{i_1}_4\check{\nablasl}{}^{i_2}_3\check{\nablasl}{}^{i_3}\,\widetilde{\varsigma}\|_{L^2(\mathbb{S}^2_{t,r},\check{\slashed{g}})} +\|\text{l.o.t.}\|_{L^2(\mathbb{S}^2_{t,r},\check{\slashed{g}})} \, , \label{equivalence_L2_norms_derivatives_1}\\
\sum_{i_1+i_2+i_3= k}\|\check{\nablasl}{}^{i_1}_4\check{\nablasl}{}^{i_2}_3\check{\nablasl}{}^{i_3}\,\widetilde{\varsigma}\|_{L^2(\mathbb{S}^2_{t,r},\check{\slashed{g}})} \lesssim_{k,r} \sum_{i_1+i_2+i_3= k}\|\nablasl^{i_1}_4\nablasl^{i_2}_3\nablasl^{i_3}\varsigma\|_{L^2(\mathbb{S}^2_{t,r},\slashed{g})}+\|\text{l.o.t.}\|_{L^2(\mathbb{S}^2_{t,r},\slashed{g})} \, ,\label{equivalence_L2_norms_derivatives_2}
\end{align}
which can be easily checked by using \eqref{equivalence_L2_norms} and the $\mathbb{S}^2$-projection formulae for covariant derivatives.

\medskip

In this section, we state the $\mathbb{S}^2$-projected version of Theorem \ref{main_th}, which can be obtained by adding appropriate linearised curvature $\|\text{l.o.t.}\|_{L^2(\mathbb{S}^2_{t,r},\slashed{g})}$-terms to the left and right hand side of the inequalities in Theorem \ref{main_th} (for each inequality, the added lower order terms involve the linearised curvature component appearing on the left hand side of the inequality) and using the inequalities \eqref{equivalence_L2_norms_derivatives_1} and \eqref{equivalence_L2_norms_derivatives_2}.~In the statement, we adopt the schematic notation \eqref{schematic_notation_th}--\eqref{schematic_notation_th_b}, for which we introduce the symbol $\check{\mathfrak{\partial}}^{\leq k}$ in the notation \eqref{schematic_notation_th_a} and \eqref{schematic_notation_th_b} with respect to the projected covariant derivatives.

\medskip

\begin{theorem}[Elliptic $L^2(\mathbb{S}^2)$-estimates for linearised curvature components, $\mathbb{S}^2$-projected version] \label{main_th_projected}
Let $0\leq |a|<M$, $k\in\mathbb{N}$ with $k\geq 3$, and finite constant $R>r_+$.~Then, there exists a constant $C_{k,R}>0$ such that, for any solution to the $\mathbb{S}^2$-projected linearised system of equations, with notation
\begin{align}
\widetilde{\psio}&= \left\lbrace \widetilde{\betao},\widetilde{\betabo},(\rhoo,\sigmao) \right\rbrace \, , \label{schematic_lin_q_1}\\
\widetilde{\Gammao}&=\left\lbrace (\overset{\text{\scalebox{.6}{$(1)$}}}{\textup{tr}\chi}),(\overset{\text{\scalebox{.6}{$(1)$}}}{\slashed{\varepsilon}\cdot\chi}),(\overset{\text{\scalebox{.6}{$(1)$}}}{\textup{tr}\chib}),(\overset{\text{\scalebox{.6}{$(1)$}}}{\slashed{\varepsilon}\cdot\chib}),\overset{\text{\scalebox{.6}{$(1)$}}}{\omegab},\widetilde{\overset{\text{\scalebox{.6}{$(1)$}}}{\yb}},\widetilde{\overset{\text{\scalebox{.6}{$(1)$}}}{\eta}},\widetilde{\overset{\text{\scalebox{.6}{$(1)$}}}{\zeta}},\widetilde{\overset{\text{\scalebox{.6}{$(1)$}}}{\chih}},\widetilde{\overset{\text{\scalebox{.6}{$(1)$}}}{\chibh}} \right\rbrace \, , \label{schematic_lin_q_2}\\
\widetilde{\fo}&=\left\lbrace \overset{\text{\scalebox{.6}{$(1)$}}}{\mathfrak{\underline{\mathfrak{f}}}}_4,\overset{\text{\scalebox{.6}{$(1)$}}}{\mathfrak{\underline{\mathfrak{f}}}}_3,(\textup{tr}\overset{\text{\scalebox{.6}{$(1)$}}}{\slashed{g}}),\widetilde{\overset{\text{\scalebox{.6}{$(1)$}}}{\mathfrak{\underline{\mathfrak{f}}}}},\widetilde{\overset{\text{\scalebox{.6}{$(1)$}}}{\mathfrak{\slashed{\mathfrak{f}}}}}_{3},\widetilde{\overset{\text{\scalebox{.6}{$(1)$}}}{\mathfrak{\slashed{\mathfrak{f}}}}}_{4},\widetilde{\overset{\text{\scalebox{.6}{$(1)$}}}{\widehat{\slashed{g}}}} \right\rbrace \,  \label{schematic_lin_q_3}
\end{align}
the estimates
\begin{equation*}
    \sum_{0\leq i_1+i_2+i_3\leq k}\|\check{\nablasl}{}^{i_1}_4\check{\nablasl}{}^{i_2}_3\check{\nablasl}{}^{i_3}\,\widetilde{\psio}\|_{L^2(\mathbb{S}^2_{t,r},\check{\slashed{g}})} \leq C_{k,R}\left[\|\check{\mathfrak{\partial}}^{\leq k}(\widetilde{\alphao},\widetilde{\alphabo})\|_{L^2(\mathbb{S}^2_{t,r},\check{\slashed{g}})} +\|\check{\mathfrak{\partial}}^{\leq k-1}(\widetilde{\fo},\widetilde{\Gammao},\widetilde{\psio})\|_{L^2(\mathbb{S}^2_{t,r},\check{\slashed{g}})} \right]
\end{equation*}
hold for all $\mathbb{S}^2_{t,r}$-spheres with $r_+\leq r \leq R$.
\end{theorem}

\medskip

By a similar logic to the one explained, one can check that the inequalities stated in Theorem \ref{main_th_projected} imply the inequalities stated in Theorem \ref{main_th}.~In Section \ref{sec_proof_main_th}, the proof of Theorem \ref{main_th} will be carried out by proving Theorem \ref{main_th_projected}.

\section{Proof of the main theorem} \label{sec_proof_main_th}

In this section, we prove the following proposition, which coincides with Theorem \ref{main_th_projected} in the case $k=3$.~As already pointed out, Proposition \ref{prop_top_order_curvature_horizon} implies Theorem \ref{main_th} in the case $k=3$.~Higher-order estimates follow by iterating the scheme that is presented in the proof (see Remark \ref{rmk_high_order_th}).

\medskip

\begin{prop} \label{prop_top_order_curvature_horizon}
Let $0\leq |a|<M$ and finite constant $R>r_+$.~Then, there exists a constant $C_{R}>0$ such that, for any solution to the $\mathbb{S}^2$-projected linearised system of equations, the estimates
\begin{equation*}
    \sum_{0\leq i_1+i_2+i_3\leq 3}\|\check{\nablasl}{}^{i_1}_4\check{\nablasl}{}^{i_2}_3\check{\nablasl}{}^{i_3}\,\widetilde{\psio}\|_{L^2(\mathbb{S}^2_{t,r},\check{\slashed{g}})} \leq C_{R}\left[\|\check{\mathfrak{\partial}}^{\leq 3}(\widetilde{\alphao},\widetilde{\alphabo})\|_{L^2(\mathbb{S}^2_{t,r},\check{\slashed{g}})} +\|\check{\mathfrak{\partial}}^{\leq 2}(\widetilde{\fo},\widetilde{\Gammao},\widetilde{\psio})\|_{L^2(\mathbb{S}^2_{t,r},\check{\slashed{g}})} \right]
\end{equation*}
hold for all $\mathbb{S}^2_{t,r}$-spheres with $r_+\leq r \leq R$, with notation \eqref{schematic_lin_q_1}--\eqref{schematic_lin_q_3} for the linearised quantities.
\end{prop}

\subsection{\texorpdfstring{$L^2(\mathbb{S}^2)$}{L2(S2)}-estimates for mixed derivatives}

As a first step towards proving Proposition \ref{prop_top_order_curvature_horizon}, we estimate third order derivatives of the linearised curvature components containing one or two null derivatives in terms of third order \emph{angular} derivatives of linearised curvature components (plus third order mixed derivatives of $\alpha$ and $\alphab$ and additional lower order terms).~This is achieved in Propositions \ref{prop_commutation_1} and \ref{prop_commutation_2}. 

\medskip

We start with a preliminary proposition.~All norms appearing in this and the following propositions are $L^2(\mathbb{S}^2_{t,r},\check{\slashed{g}})$-norms. We also adopt the (projected version of the) notation \eqref{schematic_notation_th}--\eqref{schematic_notation_th_b}, with the $L^2_w(\mathbb{S}^2_{t,r},\check{\slashed{g}})$-norms denoted here by the subscript $w$, and the radial function \eqref{def_function_l}.

\medskip

\begin{prop} \label{prop_preliminary_null_derivatives} 
Let $0\leq |a|<M$. Then, there exists a uniform constant $C>0$ such that the following holds. For any finite constant $R>r_+$, there exists a constant $C_{R}>0$ such that, for any solution to the $\mathbb{S}^2$-projected linearised system of equations, the estimates
\begin{align}
    \|\check\nablasl_4\widetilde{\betao}\| 
    &\leq C\|\check\partial{}^{\leq 1}\widetilde{\alphao}\|_w +C_R\|(\widetilde{\fo},\widetilde{\Gammao},\widetilde{\psio})\| \, , \label{est_4_beta}\\
    \left(1-2\frac{a^2}{M^2}\frac{\Delta }{r^2}l^2\right) \|\check\nablasl_4\widetilde{\betabo}\| 
    &\leq \frac{|a|}{M}l\|\check\nablasl\widetilde{\betao}\|+\|\check\nablasl (\rhoo,\sigmao)\| +\frac{|a|}{M}\frac{\Delta}{r^2}l\|\check\nablasl\widetilde{\betabo}\|\label{est_4_betab}\\
    &\qquad + C\frac{a^2}{M^2}l^2\left(\|\check\partial{}^{\leq 1}\widetilde{\alphao}\|_w+\frac{\Delta^2}{r^4}\|\check\partial{}^{\leq 1}\widetilde{\alphabo}\|_w\right)+C_R\|(\widetilde{\fo},\widetilde{\Gammao},\widetilde{\psio})\| \, ,\nonumber\\
    \left(1-2\frac{a^2}{M^2}\frac{\Delta}{r^2}l^2\right) \|\check\nablasl_4 (\rhoo,\sigmao)\| &\leq \left(1-\frac{a^2}{M^2}\frac{\Delta }{r^2}l^2\right)\|\check\nablasl\widetilde{\betao}\|+\frac{|a|}{M}\frac{\Delta}{r^2}l\|\check\nablasl(\rhoo,\sigmao)\|+\frac{a^2}{M^2}\frac{\Delta^2}{r^4}l^2\|\check\nablasl\widetilde{\betabo}\| \label{est_4_rho_sigma}\\
    &\qquad+ C\frac{|a|}{M}l\left(\|\check\partial{}^{\leq 1}\widetilde{\alphao}\|_w+\frac{\Delta^3}{r^6}l^2\|\check\partial{}^{\leq 1}\widetilde{\alphabo}\|_w\right)+C_R\|(\widetilde{\fo},\widetilde{\Gammao},\widetilde{\psio})\| \nonumber
\end{align}
and
\begin{align}
    \left(1-2\frac{a^2}{M^2}\frac{\Delta }{r^2}l^2\right) \|\check\nablasl_3\widetilde{\betao}\| 
    &\leq \frac{|a|}{M}l\|\check\nablasl\widetilde{\betao}\|+\|\check\nablasl (\rhoo,\sigmao)\| +\frac{|a|}{M}\frac{\Delta}{r^2}l\|\check\nablasl\widetilde{\betabo}\| \label{est_3_beta}\\
    &\qquad + C\frac{a^2}{M^2}l^2\left(\|\check\partial{}^{\leq 1}\widetilde{\alphao}\|_w+\frac{\Delta^2}{r^4}\|\check\partial{}^{\leq 1}\widetilde{\alphabo}\|_w\right)+C_R\|(\widetilde{\fo},\widetilde{\Gammao},\widetilde{\psio})\| \, ,\nonumber\\
    \|\check\nablasl_3\widetilde{\betabo}\| 
    &\leq C\|\check\partial{}^{\leq 1}\widetilde{\alphabo}\|_w+C_R\|(\widetilde{\fo},\widetilde{\Gammao},\widetilde{\psio})\| \, ,\label{est_3_betab} \\
    \left(1-2\frac{a^2}{M^2}\frac{\Delta}{r^2}l^2\right) \|\check\nablasl_3 (\rhoo,\sigmao)\| &\leq \frac{a^2}{M^2}l^2\|\check\nablasl\widetilde{\betao}\|+\frac{|a|}{M}l\|\check\nablasl (\rhoo,\sigmao)\|+\left(1-\frac{a^2}{M^2}\frac{\Delta }{r^2}l^2\right)\|\check\nablasl\widetilde{\betabo}\| \label{est_3_rho_sigma}\\
    &\qquad+ C\frac{|a|}{M}l\left(l^2\|\check\partial{}^{\leq 1}\widetilde{\alphao}\|_w+\frac{\Delta}{r^2}\|\check\partial{}^{\leq 1}\widetilde{\alphabo}\|_w\right)+C_R\|(\widetilde{\fo},\widetilde{\Gammao},\widetilde{\psio})\|  \nonumber
\end{align}
hold for all $\mathbb{S}^2_{t,r}$-spheres with $r_+\leq r \leq R$, with notation \eqref{schematic_lin_q_1}--\eqref{schematic_lin_q_3} for the linearised quantities.
\end{prop}

\medskip

\begin{proof}
All products in the proof are taken relative to $\check{\slashed{g}}$. Recall that, for any scalar functions $f_1,f_2$ and $\mathbb S^2_{t,r}$ one-forms $\varsigma,\xi$, one has the identities
\begin{gather*}
    |f_1\varsigma+f_2{}^\star \varsigma|^2= |\varsigma|^2(f_1^2+f_2^2) \, ,\\
    (\varsigma\cdot \xi)^2+(\varsigma\wedge \xi)^2 = |\varsigma|^2|\xi|^2 \, ,\\
    \|\check\divsl\,\varsigma\|^2+\|\check\curlsl\,\varsigma\|^2=\|\check\nablasl\varsigma\|^2+\text{l.o.t.} \, ,\\
    \|\check{\slashed{\mathcal D}}{}_1^\star(f_1,f_2)\|^2= \|\check\nablasl f_1\|^2+\|\check\nablasl f_2\|^2 \, ,
\end{gather*}
where ``l.o.t.''\ denotes terms which are zeroth order in $\varsigma$.

\medskip

The estimates \eqref{est_4_beta} and \eqref{est_3_betab} are immediate from equations \eqref{4_beta} and \eqref{3_betab} respectively, which imply
\begin{align*}
    \|\check\nablasl_4\widetilde{\betao}\| \leq \| \check{\divsl} \widetilde{\alphao} \| +|\mathfrak k|_\infty \,\|\check{\nablasl}_4\widetilde{\alphao} \| +|\mathfrak h|_\infty\, \|\check{\nablasl}_3\widetilde{\alphao} \|+\|\widetilde{\psio}\| \, ,\\
     \|\check\nablasl_3\widetilde{\betabo}\| \leq \| \check{\divsl} \widetilde{\alphabo} \| +|\mathfrak k|_\infty\, \|\check{\nablasl}_4\widetilde{\alphabo} \| +|\mathfrak h|_\infty\,\|\check{\nablasl}_3\widetilde{\alphabo}\| +\|\widetilde{\psio}\| \, .
\end{align*}
We turn to estimates~\eqref{est_4_betab} and \eqref{est_4_rho_sigma}. First, we use the $\mathbb{S}^2$-projected linearised Bianchi equations \eqref{4_betab}, \eqref{3_rho} and \eqref{3_sigma} to derive the identity
\begin{align}
\check{\nablasl}_4\widetilde{\betabo}=& \,  -\check{\nablasl}\rhoo- (\check{\nablasl}_4 \rhoo) \mathfrak{k}+(\check{\slashed{\text{div}}}\, \widetilde{\betabo})\mathfrak{h} +(\mathfrak{k}\cdot \check{\nablasl}_4\, \widetilde{\betabo})\mathfrak{h}  +(\mathfrak{h}\cdot \check{\nablasl}_3\, \widetilde{\betabo})\mathfrak{h}\nonumber\\
&+{}^{\star}\check{\nablasl}\sigmao+ (\check{\nablasl}_4 \sigmao) {}^{\star}\mathfrak{k}-(\check{\slashed{\text{curl}}}\, \widetilde{\betabo}){}^{\star}\mathfrak{h} -(\mathfrak{k}\wedge \check{\nablasl}_4\, \widetilde{\betabo}){}^{\star}\mathfrak{h}  -(\mathfrak{h}\wedge \check{\nablasl}_3\, \widetilde{\betabo}){}^{\star}\mathfrak{h}\nonumber\\
&+\widetilde{\fo}+\widetilde{\Gammao}+\widetilde{\psio} \nonumber\\
=& \, \check{\slashed{\mathcal{D}}}{}_1^\star(\rhoo,\sigmao)+\Big[(\check{\nablasl}_4 \sigmao) {}^{\star}\mathfrak{k}- (\check{\nablasl}_4 \rhoo) \mathfrak{k}\Big] \nonumber\\
&+\Big[(\check{\divsl}\,\widetilde{\betabo})\mathfrak h - (\check{\curlsl}\,\widetilde{\betabo}){}^\star\mathfrak{h}\Big]
+\Big[(\mathfrak{k}\cdot \check{\nablasl}_4\, \widetilde{\betabo})\mathfrak{h}-(\mathfrak{k}\wedge \check{\nablasl}_4\, \widetilde{\betabo}){}^{\star}\mathfrak{h} \Big]\nonumber\\
&+\Big[(\mathfrak{h}\cdot \check{\nablasl}_3\, \widetilde{\betabo})\mathfrak{h}-(\mathfrak{h}\wedge \check{\nablasl}_3\, \widetilde{\betabo}){}^{\star}\mathfrak{h}\Big] +\widetilde{\fo}+\widetilde{\Gammao}+\widetilde{\psio}\, . \label{comput_4_betab}
\end{align}
In the last equality, we have grouped together terms so as to easily derive the inequality
\begin{align}
    \|\check{\nablasl}_4\widetilde{\betabo}\| 
    &\leq  |\mathfrak k|_\infty\,|\mathfrak h|_\infty\,\|\check{\nablasl}_4\widetilde{\betabo}\| 
    + |\mathfrak h|_\infty\,\|\check\nablasl\widetilde{\betabo}\| 
    +\big(\|\check\nablasl \rhoo\|^2+\|\check\nablasl \sigmao\|^2\big)^{\frac12}
    +|\mathfrak k|_\infty\, \big(\|\check{\nablasl}_4\rhoo\|^2+\|\check{\nablasl}_4\sigmao\|^2\big)^{\frac12} \nonumber
    \\
    &\qquad+ |\mathfrak h|_\infty^2\,\|\check{\nablasl}_3\widetilde{\betabo}\| +\|(\widetilde{\fo},\widetilde{\Gammao},\widetilde{\psio})\| \, . \label{eq:est_4_betab-int}
\end{align}
Second, from equations \eqref{4_rho}, \eqref{4_beta} and \eqref{3_rho}, we compute (we use $(\check{\nablasl}_3 \sigmao)(\mathfrak{h}\cdot{}^{\star}\mathfrak{h})=0$)
\begin{align}
\check{\nablasl}_4\rhoo =& \, \check{\slashed{\text{div}}}\, \widetilde{\betao} +\mathfrak{h}\cdot \check{\nablasl}\rhoo+ (\check{\nablasl}_4 \rhoo) (\mathfrak{h}\cdot\mathfrak{k})+(\check{\nablasl}_3 \rhoo)(\mathfrak{h}\cdot\mathfrak{h}) \nonumber\\
&+\mathfrak{h}\cdot{}^{\star}\check{\nablasl}\sigmao+ (\check{\nablasl}_4 \sigmao) (\mathfrak{h}\cdot{}^{\star}\mathfrak{k}) \nonumber\\
& +\mathfrak{k}\cdot \check{\nablasl}_4\widetilde{\betao}+\widetilde{\fo}+\widetilde{\Gammao}+\widetilde{\psio} \nonumber\\
=& \, \check{\slashed{\text{div}}}\, \widetilde{\betao} +\mathfrak{h}\cdot \check{\nablasl}\rhoo+\mathfrak{h}\cdot{}^{\star}\check{\nablasl}\sigmao+ (\check{\nablasl}_4 \rhoo) (\mathfrak{h}\cdot\mathfrak{k})+ (\check{\nablasl}_4 \sigmao) (\mathfrak{h}\cdot{}^{\star}\mathfrak{k}) \nonumber\\
&-(\check{\slashed{\text{div}}}\, \widetilde{\betabo})(\mathfrak{h}\cdot\mathfrak{h}) -(\mathfrak{k}\cdot \check{\nablasl}_4\, \widetilde{\betabo})(\mathfrak{h}\cdot\mathfrak{h})  \nonumber\\
& -(\mathfrak{h}\cdot \check{\nablasl}_3\, \widetilde{\betabo})(\mathfrak{h}\cdot\mathfrak{h}) +\mathfrak{k}\cdot \check{\nablasl}_4\widetilde{\betao}+\widetilde{\fo}+\widetilde{\Gammao}+\widetilde{\psio} \, . \label{comput_4_rho}
\end{align}
We also compute, from \eqref{4_sigma}, \eqref{4_beta} and \eqref{3_sigma} (we use $(\check{\nablasl}_3 \rhoo)(\mathfrak{h}\wedge\mathfrak{h})=0$),
\begin{align}
\check{\nablasl}_4\sigmao=&  -\check{\slashed{\text{curl}}}\, \widetilde{\betao}  -\mathfrak{h}\wedge{}^{\star}\check{\nablasl}\sigmao - (\check{\nablasl}_4 \sigmao) (\mathfrak{h}\wedge{}^{\star}\mathfrak{k})- (\check{\nablasl}_3 \sigmao) (\mathfrak{h}\wedge{}^{\star}\mathfrak{h}) \nonumber\\
&-\mathfrak{h}\wedge \check{\nablasl}\rhoo- (\check{\nablasl}_4 \rhoo) (\mathfrak{h}\wedge\mathfrak{k}) \nonumber\\
& -\mathfrak{k}\wedge\check{\nablasl}_4\widetilde{\betao}+\widetilde{\fo}+\widetilde{\Gammao}+\widetilde{\psio} \nonumber\\
=& \, -\check{\slashed{\text{curl}}}\, \widetilde{\betao} -\mathfrak{h}\wedge \check{\nablasl}\rhoo-\mathfrak{h}\wedge{}^{\star}\check{\nablasl}\sigmao- (\check{\nablasl}_4 \rhoo) (\mathfrak{h}\wedge\mathfrak{k})- (\check{\nablasl}_4 \sigmao) (\mathfrak{h}\wedge{}^{\star}\mathfrak{k}) \nonumber\\
&+(\check{\slashed{\text{curl}}}\, \widetilde{\betabo})(\mathfrak{h}\wedge{}^{\star}\mathfrak{h}) +(\mathfrak{k}\wedge \check{\nablasl}_4\, \widetilde{\betabo})(\mathfrak{h}\wedge{}^{\star}\mathfrak{h})  \nonumber\\
&+(\mathfrak{h}\wedge \check{\nablasl}_3\, \widetilde{\betabo})(\mathfrak{h}\wedge{}^{\star}\mathfrak{h}) -\mathfrak{k}\wedge \check{\nablasl}_4\widetilde{\betao}+\widetilde{\fo}+\widetilde{\Gammao}+\widetilde{\psio} \, .\label{comput_4_sigma}
\end{align}
From the identities \eqref{comput_4_rho} and \eqref{comput_4_sigma}, by noting $\mathfrak{h}\wedge\mathfrak k=\mathfrak{h}\cdot {}^{\star}\mathfrak{k}=0$, one obtains the estimate
\begin{align}
    \big(\|\check\nablasl_4 \rhoo\|^2+\|\check\nablasl_4 \sigmao\|^2\big)^{\frac12} 
    \leq &\, |\mathfrak h|_\infty|\mathfrak k|_\infty\, \big(\|\check\nablasl_4 \rhoo\|^2+\|\check\nablasl_4 \sigmao\|^2\big)^{\frac12}
    +|\mathfrak h|_\infty\, \big(\|\check\nablasl \rhoo\|^2+\|\check\nablasl \sigmao\|^2\big)^{\frac12}
    \nonumber\\
    &+\| \check\nablasl\widetilde{\betao}\|+|\mathfrak h|^2_\infty\,\|\check\nablasl\widetilde{\betabo}\|
    +|\mathfrak h|^2_\infty|\mathfrak k|_\infty\, \|\check\nablasl_4\widetilde{\betabo}\|  \nonumber\\
    &+ |\mathfrak h|^3_\infty\,\|\check\nablasl_3\widetilde{\betabo}\| +|\mathfrak{k}|_\infty\, \| \check\nablasl_4\widetilde{\betao}\|+\|(\widetilde{\fo},\widetilde{\Gammao},\widetilde{\psio})\| \, . \label{eq:est_4_rhosigma-int}
\end{align}
Finally, we combine the estimates \eqref{eq:est_4_betab-int} and \eqref{eq:est_4_rhosigma-int}. Using \eqref{eq:est_4_betab-int} to estimate the fifth term on the right hand side of \eqref{eq:est_4_rhosigma-int}, we obtain
\begin{align*}
    &\left(1-|\mathfrak{h}|_\infty|\mathfrak{k}|_\infty-\frac{|\mathfrak{h}|^2_\infty|\mathfrak{k}|^2_\infty}{1-|\mathfrak{h}|_\infty|\mathfrak{k}|_\infty}\right)\big(\|\check\nablasl_4 \rhoo\|^2+\|\check\nablasl_4 \sigmao\|^2\big)^{\frac12} \\&\quad\leq \|\check\nablasl \widetilde{\betao}\|+\frac{|\mathfrak{h}|_\infty}{1-|\mathfrak{h}|_\infty|\mathfrak{k}|_\infty}\big(\|\check\nablasl \rhoo\|^2+\|\check\nablasl \sigmao\|^2\big)^{\frac12}+ \frac{|\mathfrak h|^2_\infty}{1-|\mathfrak{h}|_\infty|\mathfrak{k}|_\infty}\|\check\nablasl \widetilde{\betabo}\|\\
    &\quad\qquad +\frac{|\mathfrak h|^3_\infty}{1-|\mathfrak{h}|_\infty|\mathfrak{k}|_\infty}\|\check\nablasl_3 \widetilde{\betabo}\|+|\mathfrak{k}|_\infty\,\|\check\nablasl_4 \widetilde{\betao}\|+\|(\widetilde{\fo},\widetilde{\Gammao},\widetilde{\psio})\| \, .
\end{align*}
After simplifying and plugging into the the right hand side of \eqref{eq:est_4_betab-int}, we deduce 
\begin{align}
 (1-2|\mathfrak{h}|_\infty|\mathfrak{k}|_\infty)\,\big(\|\check\nablasl_4 \rhoo\|^2+\|\check\nablasl_4 \sigmao\|^2\big)^{\frac12} &\leq (1-|\mathfrak{h}|_\infty|\mathfrak{k}|_\infty)\,\|\check\nablasl \widetilde{\betao}\|+|\mathfrak{h}|_\infty\,\big(\|\check\nablasl \rhoo\|^2+\|\check\nablasl \sigmao\|^2\big)^{\frac12}+ |\mathfrak h|^2_\infty\,\|\check\nablasl \widetilde{\betabo}\|\nonumber\\
    &\qquad +|\mathfrak h|^3_\infty\,\|\check\nablasl_3 \widetilde{\betabo}\|+|\mathfrak{k}|_\infty(1-|\mathfrak{h}|_\infty|\mathfrak{k}|_\infty)\,\|\check\nablasl_4 \widetilde{\betao}\|+\|(\widetilde{\fo},\widetilde{\Gammao},\widetilde{\psio})\| \, ,\label{eq:est_4_rhosigma-int2}\\
(1-2|\mathfrak{h}|_\infty|\mathfrak{k}|_\infty)\,\|\check\nablasl_4 \widetilde{\betabo}\| &\leq |\mathfrak{k}|_\infty\,\|\check\nablasl \widetilde{\betao}\|+\big(\|\check\nablasl \rhoo\|^2+\|\check\nablasl \sigmao\|^2\big)^{\frac12}+ |\mathfrak h|_\infty\,\|\check\nablasl \widetilde{\betabo}\|\nonumber\\
    &\qquad +|\mathfrak h|^2_\infty\,\|\check\nablasl_3 \widetilde{\betabo}\|+|\mathfrak{k}|^2_\infty\,\|\check\nablasl_4 \widetilde{\betao}\|+\|(\widetilde{\fo},\widetilde{\Gammao},\widetilde{\psio})\| \, .\label{eq:est_4_betab-int2}
\end{align}
Using the properties of $\mathfrak{k}$ and $\mathfrak h$, as well as the already proven estimates  \eqref{est_4_beta} and \eqref{est_3_betab}, from \eqref{eq:est_4_rhosigma-int2} and \eqref{eq:est_4_betab-int2} we derive the stated estimates \eqref{est_4_betab} and \eqref{est_4_rho_sigma}.

\medskip

Let us now turn to estimates \eqref{est_3_beta} and \eqref{est_3_rho_sigma}. Using the same approach as to derive \eqref{eq:est_4_betab-int} and \eqref{eq:est_4_rhosigma-int}, we obtain, from the $\mathbb{S}^2$-projected linearised Bianchi equations \eqref{3_rho} and \eqref{3_sigma}, the estimate
\begin{equation}
\big(\|\check\nablasl_3 \rhoo\|^2+\|\check\nablasl_3 \sigmao\|^2\big)^{\frac12} 
    \leq  \|\check\nablasl\widetilde{\betabo}\|+|\mathfrak k|_\infty\, \|\check\nablasl_4\widetilde{\betabo}\| + |\mathfrak h|_\infty\, \|\check\nablasl_3\widetilde{\betabo}\|+\|(\widetilde{\fo},\widetilde{\Gammao},\widetilde{\psio})\| \, , \label{eq:est_3_rhosigma-int}
\end{equation}
and, from the $\mathbb{S}^2$-projected linearised Bianchi equation \eqref{3_beta},
\begin{align}
    \|\check\nablasl_3\widetilde{\betao}\| &\leq \big(\|\check\nablasl \rhoo\|^2+\|\check\nablasl \sigmao\|^2\big)^{\frac12}+ |\mathfrak k|_\infty\,\big(\|\check\nablasl_4 \rhoo\|^2+\|\check\nablasl_4 \sigmao\|^2\big)^{\frac12} \nonumber \\
    &\qquad + |\mathfrak h|_\infty\,\big(\|\check\nablasl_3 \rhoo\|^2+\|\check\nablasl_3 \sigmao\|^2\big)^{\frac12} 
     + \|(\widetilde{\fo},\widetilde{\Gammao},\widetilde{\psio})\| \, .\label{eq:est_3_beta-int}
\end{align}
To conclude, we use \eqref{eq:est_4_betab-int2} to estimate the second term on the right hand side of \eqref{eq:est_3_rhosigma-int} and get
\begin{align}
    (1-2|\mathfrak h|_\infty|\mathfrak k|_\infty)\big(\|\check\nablasl_3 \rhoo\|^2+\|\check\nablasl_3 \sigmao\|^2\big)^{\frac12} &\leq |\mathfrak{k}|^2_\infty \,\|\check\nablasl\widetilde{\betao}\|+(1-|\mathfrak h|_\infty|\mathfrak k|_\infty)\,\|\check\nablasl\widetilde{\betabo}\|+|\mathfrak{k}|_\infty\,\big(\|\check\nablasl \rhoo\|^2+\|\check\nablasl \sigmao\|^2\big)^{\frac12} \nonumber\\
    &\qquad +|\mathfrak h|_\infty(1-|\mathfrak h|_\infty|\mathfrak k|_\infty)\,\|\check\nablasl_3\widetilde{\betabo}\|+|\mathfrak{k}|_\infty^3\,\|\check\nablasl_4\widetilde{\betao}\|+ \|(\widetilde{\fo},\widetilde{\Gammao},\widetilde{\psio})\| \, , \label{eq:est_3_rhosigma-int2}
\end{align}
and then combine the estimates \eqref{eq:est_3_rhosigma-int2} and \eqref{eq:est_4_rhosigma-int2} to control the right hand side of \eqref{eq:est_3_beta-int}. We deduce
\begin{align}
    (1-2|\mathfrak h|_\infty|\mathfrak k|_\infty)\|\check\nablasl_3\widetilde{\betao}\| 
    &\leq \big(\|\check\nablasl \rhoo\|^2+\|\check\nablasl \sigmao\|^2\big)^{\frac12}+ |\mathfrak k|_\infty\,\|\check\nablasl\widetilde{\betao}\|+|\mathfrak h|_\infty\,\|\check\nablasl\widetilde{\betabo}\|\nonumber\\
    &\qquad + |\mathfrak h|^2_\infty\,\|\check\nablasl_3\widetilde{\betabo}\|+|\mathfrak k|^2_\infty\,\|\check\nablasl_4\widetilde{\betao}\|+ \|(\widetilde{\fo},\widetilde{\Gammao},\widetilde{\psio})\| \, .\label{eq:est_3_beta-int2}
\end{align}
Finally, we appeal to the properties of $\mathfrak{k}$ and $\mathfrak h$ and the already proven estimates \eqref{est_4_beta} and \eqref{est_3_betab} to finish the proof.
\end{proof}

\medskip

We now state the commuted versions of Proposition \ref{prop_preliminary_null_derivatives} which we will need in the sequel. The propositions are stated under the assumptions and notation of Proposition \ref{prop_preliminary_null_derivatives}.

\medskip

\begin{prop} \label{prop_commutation_1}
The estimates
\begin{align*}
    \|\check\nablasl{}^2\check\nablasl_4\widetilde{\betao}\| 
    &\leq C\|\check\partial{}^{\leq 3}\widetilde{\alphao}\|_w +C_R\|\check\partial{}^{\leq 2}(\widetilde{\fo},\widetilde{\Gammao},\widetilde{\psio})\| \, , \\
    \left(1-2\frac{a^2}{M^2}\frac{\Delta }{r^2}l^2\right) \|\check\nablasl{}^2\check\nablasl_4\widetilde{\betabo}\| 
    &\leq \frac{|a|}{M}l\|\check\nablasl{}^3\widetilde{\betao}\|+\|\check\nablasl{}^3 (\rhoo,\sigmao)\| +\frac{|a|}{M}\frac{\Delta}{r^2}l\|\check\nablasl{}^3\widetilde{\betabo}\|\\
    &\qquad + C\frac{a^2}{M^2}l^2\left(\|\check\partial{}^{\leq 3}\widetilde{\alphao}\|_w+\frac{\Delta^2}{r^4}\|\check\partial{}^{\leq 3}\widetilde{\alphabo}\|_w\right)+C_R\|\check\partial{}^{\leq 2}(\widetilde{\fo},\widetilde{\Gammao},\widetilde{\psio})\| \, ,\nonumber\\
    \left(1-2\frac{a^2}{M^2}\frac{\Delta}{r^2}l^2\right) \|\check\nablasl{}^2\check\nablasl_4 (\rhoo,\sigmao)\| &\leq \left(1-\frac{a^2}{M^2}\frac{\Delta }{r^2}l^2\right)\|\check\nablasl{}^3\widetilde{\betao}\|+\frac{|a|}{M}\frac{\Delta}{r^2}l\|\check\nablasl{}^3 (\rhoo,\sigmao)\|+\frac{a^2}{M^2}\frac{\Delta^2}{r^4}l^2\|\check\nablasl{}^3\widetilde{\betabo}\| \\
    &\qquad+ C\frac{|a|}{M}l\left(\|\check\partial{}^{\leq 3}\widetilde{\alphao}\|_w+\frac{\Delta^3}{r^6}l^2\|\check\partial{}^{\leq 3}\widetilde{\alphabo}\|_w\right)+C_R\|\check\partial{}^{\leq 2}(\widetilde{\fo},\widetilde{\Gammao},\widetilde{\psio})\| \nonumber
\end{align*}
and
\begin{align*}
    \left(1-2\frac{a^2}{M^2}\frac{\Delta }{r^2}l^2\right) \|\check\nablasl{}^2\check\nablasl_3\widetilde{\betao}\| 
    &\leq \frac{|a|}{M}l\|\check\nablasl{}^3\widetilde{\betao}\|+\|\check\nablasl{}^3 (\rhoo,\sigmao)\| +\frac{|a|}{M}\frac{\Delta}{r^2}l\|\check\nablasl{}^3\widetilde{\betabo}\| \\
    &\qquad + C\frac{a^2}{M^2}l^2\left(\|\check\partial{}^{\leq 3}\widetilde{\alphao}\|_w+\frac{\Delta^2}{r^4}\|\check\partial{}^{\leq 3}\widetilde{\alphabo}\|_w\right)+C_R\|\check\partial{}^{\leq 2}(\widetilde{\fo},\widetilde{\Gammao},\widetilde{\psio})\| \, ,\nonumber\\
    \|\check\nablasl{}^2\check\nablasl_3\widetilde{\betabo}\| 
    &\leq C\|\check\partial{}^{\leq 3}\widetilde{\alphabo}\|_w+C_R\|\check\partial{}^{\leq 2}(\widetilde{\fo},\widetilde{\Gammao},\widetilde{\psio})\| \, ,\\
    \left(1-2\frac{a^2}{M^2}\frac{\Delta}{r^2}l^2\right) \|\check\nablasl{}^2\check\nablasl_3 (\rhoo,\sigmao)\| &\leq \frac{a^2}{M^2}l^2\|\check\nablasl{}^3\widetilde{\betao}\|+\frac{|a|}{M}l\|\check\nablasl{}^3 (\rhoo,\sigmao)\|+\left(1-\frac{a^2}{M^2}\frac{\Delta }{r^2}l^2\right)\|\check\nablasl{}^3\widetilde{\betabo}\| \\
    &\qquad+ C\frac{|a|}{M}l\left(l^2\|\check\partial{}^{\leq 3}\widetilde{\alphao}\|_w+\frac{\Delta}{r^2}\|\check\partial{}^{\leq 3}\widetilde{\alphabo}\|_w\right)+C_R\|\check\partial{}^{\leq 2}(\widetilde{\fo},\widetilde{\Gammao},\widetilde{\psio})\| \nonumber
\end{align*}
hold for all $\mathbb{S}^2_{t,r}$-spheres with $r_+\leq r \leq R$.
\end{prop}

\medskip

\begin{proof}
One applies $\check{\nablasl}{}^2$ to the $\mathbb{S}^2$-projected linearised Bianchi identities \eqref{4_beta}--\eqref{3_beta}, \eqref{3_rho}--\eqref{3_sigma}, \eqref{3_betab} and the identities \eqref{comput_4_betab}, \eqref{comput_4_rho}--\eqref{comput_4_sigma} derived in Proposition \ref{prop_preliminary_null_derivatives}.~Noting that
\begin{align*}
    \|\check\nablasl \check\divsl\,\varsigma\|^2+\|\check\nablasl \check\curlsl\,\varsigma\|^2 = \|\check{\nablasl}{}^2\varsigma\|^2 +\text{l.o.t.}
\end{align*}
for any $\mathbb S^2_{t,r}$ one-form $\varsigma$, where ``l.o.t.''\ denotes terms which contain at most one derivative of $\varsigma$, one then repeats the proof of Proposition \ref{prop_preliminary_null_derivatives}. 
\end{proof}

\medskip

\begin{prop} \label{prop_commutation_2}
The estimates
\begin{align}
        l\|\check\nablasl\check\nablasl{}^2_4 \widetilde{\betao}\|+\frac{\Delta}{r^2}l\|\check\nablasl\check\nablasl_3\check\nablasl_4 \widetilde{\betao}\|&\leq C\|\check\partial{}^{\leq 3}\widetilde{\alphao}\|_w + C_R\|\check\partial{}^{\leq 2}(\widetilde{\fo},\widetilde{\Gammao},\widetilde{\psio})\| \, ,\label{est_44_34_beta}\\
    \frac{\Delta}{r^2}\left(1-2\frac{a^2}{M^2}\frac{\Delta}{r^2}l^2\right)^2\|\check\nablasl\check\nablasl{}_3^2\widetilde{\betao}\|
    &\leq  2\frac{a^2}{M^2}\frac{\Delta}{r^2}l^2\|\check\nablasl{}^3\widetilde{\betao}\|+2\frac{|a|}{M}\frac{\Delta}{r^2}l\|\check\nablasl{}^3(\rhoo,\sigmao)\|+\frac{\Delta}{r^2}\|\check\nablasl{}^3\widetilde{\betabo}\| \label{est_33_beta}\\
    &\qquad + C\frac{|a|}{M}l\left(\|\check\partial{}^{\leq 3}\widetilde{\alphao}\|+\frac{\Delta^2}{r^4}\|\check\partial{}^{\leq 3}\widetilde{\alphabo}\|_w\right)+ C_R\|\check\partial{}^{\leq 2}(\widetilde{\fo},\widetilde{\Gammao},\widetilde{\psio})\| \nonumber
\end{align}
and
\begin{align}
    \left(1-2\frac{a^2}{M^2}\frac{\Delta}{r^2}l^2\right)^2\|\check\nablasl\check\nablasl{}_4^2\widetilde{\betabo}\|
    &\leq  \|\check\nablasl{}^3\widetilde{\betao}\|+2\frac{|a|}{M}\frac{\Delta}{r^2}l\|\check\nablasl{}^3(\rhoo,\sigmao)\|+2\frac{a^2}{M^2}\frac{\Delta^2}{r^4}l^2\|\check\nablasl{}^3\widetilde{\betabo}\| \label{est_44_betab}\\
    &\qquad + C\frac{|a|}{M}l\left(\|\check\partial{}^{\leq 3}\widetilde{\alphao}\|_w+\frac{\Delta^2}{r^4}\|\check\partial{}^{\leq 3}\widetilde{\alphabo}\|_w\right)+ C_R\|\check\partial{}^{\leq 2}(\widetilde{\fo},\widetilde{\Gammao},\widetilde{\psio})\| \, ,\nonumber\\
    \frac{\Delta}{r^2}l\|\check\nablasl\check\nablasl{}^2_3 \widetilde{\betabo}\|+ l\|\check\nablasl\check\nablasl_3\check\nablasl_4 \widetilde{\betabo}\|
    &\leq C\|\check\partial{}^{\leq 3}\widetilde{\alphabo}\|_w + C_R\|\check\partial{}^{\leq 2}(\widetilde{\fo},\widetilde{\Gammao},\widetilde{\psio})\| \label{est_34_33_betab}
\end{align}
and
\begin{align}
    \|\check\nablasl\check\nablasl{}^2_4(\rhoo,\sigmao)\|+\|\check\nablasl\check\nablasl{}^2_3(\rhoo,\sigmao)\|
    & \leq  C\left(\|\check\partial{}^{\leq 3}\widetilde{\alphao}\|_w+\|\check\partial{}^{\leq 3}\widetilde{\alphabo}\|_w\right) + C_R\|\check\partial{}^{\leq 2}(\widetilde{\fo},\widetilde{\Gammao},\widetilde{\psio})\| \, , \label{est_44_33_rho_sigma}\\
    \left(1-2\frac{a^2}{M^2}\frac{\Delta}{r^2}l^2\right)^2\|\check\nablasl\check\nablasl_4\check\nablasl_3(\rhoo,\sigmao)\|&\leq \frac{|a|}{M}l\left (1-\frac{a^2}{M^2}\frac{\Delta}{r^2}l^2\right)\|\check\nablasl{}^3\widetilde{\betao}\|+\|\check\nablasl{}^3(\rhoo,\sigmao)\|+\frac{|a|}{M}\frac{\Delta}{r^2}l\|\check\nablasl{}^3\widetilde{\betabo}\| \label{est_34_rho_sigma}\\
    &\qquad +C\frac{|a|}{M}\left(l^2\|\check\partial{}^{\leq 3}\widetilde{\alphao}\|_w+\frac{\Delta}{r^2}\|\check\partial{}^{\leq 3}\widetilde{\alphabo}\|_w\right)+C_R\|\check\partial{}^{\leq 2}(\widetilde{\fo},\widetilde{\Gammao},\widetilde{\psio})\| \nonumber
\end{align}
hold for all $\mathbb{S}^2_{t,r}$-spheres with $r_+\leq r \leq R$.
\end{prop}

\medskip

\begin{proof}
By applying $\check{\nablasl}\check\nablasl_3$  and $\check{\nablasl}\check\nablasl_4$ to \eqref{4_beta} and \eqref{3_betab}, we obtain the estimates
\begin{align*}
    \|\check{\nablasl}\check\nablasl{}_4^2\widetilde{\betao}\|&\leq \|\check\nablasl\check\nablasl_4\check\divsl\widetilde{\alphao}\|+|\mathfrak k|_\infty \,\|\check{\nablasl}\check\nablasl{}_4^2\widetilde{\alphao}\|+|\mathfrak h|_\infty\,\|\check{\nablasl}\check\nablasl_4\check\nablasl_3\widetilde{\alphao}\|+\sum_{0\leq i_1+i_2+i_3\leq 2}\|\check\nablasl{}^{i_1}_4\check\nablasl{}^{i_2}_3\check\nablasl{}^{i_3}(\widetilde{\fo},\widetilde{\Gammao},\widetilde{\psio})\| \, ,\\
    \|\check{\nablasl}\check\nablasl_4\check\nablasl_3\widetilde{\betao}\|&\leq \|\check\nablasl\check\nablasl_3\check\divsl\widetilde{\alphao}\|+|\mathfrak k|_\infty \,\|\check{\nablasl}\check\nablasl_4\check\nablasl_3\widetilde{\alphao}\|+|\mathfrak h|_\infty\,\|\check{\nablasl}\check\nablasl{}_3^2\widetilde{\alphao}\|+\sum_{0\leq i_1+i_2+i_3\leq 2}\|\check\nablasl{}^{i_1}_4\check\nablasl{}^{i_2}_3\check\nablasl{}^{i_3}(\widetilde{\fo},\widetilde{\Gammao},\widetilde{\psio})\| \, ,\\
    \|\check{\nablasl}\check\nablasl{}_3^2\widetilde{\betabo}\|&\leq \|\check\nablasl\check\nablasl_3\check\divsl\widetilde{\alphabo}\|+|\mathfrak k|_\infty \,\|\check{\nablasl}\check\nablasl_4\check\nablasl_3\widetilde{\alphabo}\|+|\mathfrak h|_\infty\,\|\check{\nablasl}\check\nablasl{}_3^2\widetilde{\alphabo}\|+\sum_{0\leq i_1+i_2+i_3\leq 2}\|\check\nablasl{}^{i_1}_4\check\nablasl{}^{i_2}_3\check\nablasl{}^{i_3}(\widetilde{\fo},\widetilde{\Gammao},\widetilde{\psio})\| \, ,\\
     \|\check{\nablasl}\check\nablasl_4\check\nablasl_3\widetilde{\betabo}\|&\leq \|\check\nablasl\check\nablasl_4\check\divsl\widetilde{\alphabo}\|+|\mathfrak k|_\infty \,\|\check{\nablasl}\check\nablasl{}_4^2\widetilde{\alphabo}\|+|\mathfrak h|_\infty\,\|\check{\nablasl}\check\nablasl_4\check\nablasl_3\widetilde{\alphabo}\|+\sum_{0\leq i_1+i_2+i_3\leq 2}\|\check\nablasl{}^{i_1}_4\check\nablasl{}^{i_2}_3\check\nablasl{}^{i_3}(\widetilde{\fo},\widetilde{\Gammao},\widetilde{\psio})\| \, ,  
\end{align*}
which yield \eqref{est_44_34_beta} and \eqref{est_34_33_betab}.
Next, we apply $\check\nablasl\check\nablasl_3$ and $\check\nablasl\check\nablasl_4$ to, respectively, the $\mathbb S^2$-projected linearised Bianchi equations \eqref{3_rho}--\eqref{3_sigma} and \eqref{4_rho}--\eqref{4_sigma}, and deduce \eqref{est_34_rho_sigma}. More precisely, we obtain
\begin{align*}
    \big(\|\check\nablasl\check\nablasl{}_4^2 \rhoo\|^2+\|\check\nablasl\check\nablasl{}_4^2 \sigmao\|^2\big)^{\frac12} &\leq \|\check\nablasl{}^2\check\nablasl_4\widetilde{\betao}\|+|\mathfrak{k}|_\infty\, \|\check\nablasl\check\nablasl{}^2_4\widetilde{\betao}\| +|\mathfrak{h}|_\infty \,\|\check\nablasl\check\nablasl_3\check\nablasl_4\widetilde{\betao}\| \\
    &\qquad +\sum_{0\leq i_1+i_2+i_3\leq 2}\|\check\nablasl{}^{i_1}_4\check\nablasl{}^{i_2}_3\check\nablasl{}^{i_3}(\widetilde{\fo},\widetilde{\Gammao},\widetilde{\psio})\| \, ,\\
    \big(\|\check\nablasl\check\nablasl{}_3^2 \rhoo\|^2+\|\check\nablasl\check\nablasl{}_3^2 \sigmao\|^2\big)^{\frac12} &\leq \|\check\nablasl{}^2\check\nablasl_3\widetilde{\betabo}\|+|\mathfrak{h}|_\infty\, \|\check\nablasl\check\nablasl{}^2_3\widetilde{\betabo}\| +|\mathfrak{k}|_\infty \,\|\check\nablasl\check\nablasl_4\check\nablasl_3\widetilde{\betabo}\| \\
    &\qquad+\sum_{0\leq i_1+i_2+i_3\leq 2}\|\check\nablasl{}^{i_1}_4\check\nablasl{}^{i_2}_3\check\nablasl{}^{i_3}(\widetilde{\fo},\widetilde{\Gammao},\widetilde{\psio})\| \, .   
\end{align*}
We now turn to the proof of \eqref{est_33_beta} and \eqref{est_44_betab}. We apply $\check{\nablasl}\check{\nablasl}_4$ to the identities \eqref{comput_4_betab}, \eqref{comput_4_rho}--\eqref{comput_4_sigma} derived in Proposition \ref{prop_preliminary_null_derivatives}. Repeating the proof there, we obtain the same estimates, now commuted with $\check{\nablasl}\check{\nablasl}_4$. For instance, we have
\begin{align*}
    (1-2|\mathfrak{h}|_\infty|\mathfrak{k}|_\infty)^2\,\|\check\nablasl\check\nablasl{}_4^2 \widetilde{\betabo}\| &\leq \big(\|\check\nablasl{}^2\check\nablasl_4 \rhoo\|^2+\|\check\nablasl{}^2\check\nablasl_4 \sigmao\|^2\big)^{\frac12}+ |\mathfrak h|_\infty\,\|\check\nablasl{}^2\check\nablasl_4 \widetilde{\betabo}\|\nonumber\\
    &\qquad +|\mathfrak h|^2_\infty\,\|\check\nablasl\check\nablasl_4\check\nablasl_3 \widetilde{\betabo}\|+|\mathfrak{k}|^2_\infty\,\|\check\nablasl\check\nablasl{}_4^2 \widetilde{\betao}\|+|\mathfrak{k}|_\infty\,\|\check\nablasl{}^2 \check\nablasl_4\widetilde{\betao}\|\\
    &\qquad+\sum_{0\leq i_1+i_2+i_3\leq 2}\|\check\nablasl{}^{i_1}_4\check\nablasl{}^{i_2}_3\check\nablasl{}^{i_3}(\widetilde{\fo},\widetilde{\Gammao},\widetilde{\psio})\| \, ,
\end{align*}
which implies, by Proposition~\ref{prop_commutation_1},
\begin{align*}
    (1-2|\mathfrak{h}|_\infty|\mathfrak{k}|_\infty)^2\,\|\check\nablasl\check\nablasl{}_4^2 \widetilde{\betabo}\| &\leq \|\check\nablasl{}^3\widetilde{\betao}\|+2|\mathfrak h|_\infty\,\big(\|\check\nablasl{}^3 \rhoo\|^2+\|\check\nablasl{}^3 \sigmao\|^2\big)^{\frac12} + 2|\mathfrak h|_\infty^2\, \|\check\nablasl{}^3\widetilde{\betabo}\| \nonumber\\
    &\qquad +2|\mathfrak h|^3_\infty \|\check\nablasl{}^2\check\nablasl_3 \widetilde{\betabo}\|+|\mathfrak h|^2_\infty(1-2|\mathfrak{h}|_\infty|\mathfrak{k}|_\infty)\,\|\check\nablasl\check\nablasl_4\check\nablasl_3 \widetilde{\betabo}\|\\
    &\qquad +2|\mathfrak k|_\infty(1-|\mathfrak h|_\infty|\mathfrak k|_\infty)\|\check\nablasl{}^2 \check\nablasl_4\widetilde{\betao}\|+|\mathfrak{k}|^2_\infty(1-2|\mathfrak{h}|_\infty|\mathfrak{k}|_\infty)\,\|\check\nablasl\check\nablasl{}_4^2 \widetilde{\betao}\|\\
    &\qquad +\sum_{0\leq i_1+i_2+i_3\leq 2}\|\check\nablasl{}^{i_1}_4\check\nablasl{}^{i_2}_3\check\nablasl{}^{i_3}(\widetilde{\fo},\widetilde{\Gammao},\widetilde{\psio})\| \, .
\end{align*}
Similarly, applying $\nablasl\nablasl_3$ to the $\mathbb{S}^2$-projected linearised Bianchi equations \eqref{3_beta} and \eqref{3_rho}--\eqref{3_sigma}, and then invoking Proposition~\ref{prop_commutation_1}, we obtain
\begin{align*}
(1-2|\mathfrak{h}|_\infty|\mathfrak{k}|_\infty)^2\,\|\check\nablasl\check\nablasl{}_3^2 \widetilde{\betao}\| &\leq \|\check\nablasl{}^3\widetilde{\betabo}\|+2|\mathfrak k|_\infty\,\big(\|\check\nablasl{}^3 \rhoo\|^2+\|\check\nablasl{}^3 \sigmao\|^2\big)^{\frac12} + 2|\mathfrak k|_\infty^2\, \|\check\nablasl{}^3\widetilde{\betao}\| \nonumber\\
    &\qquad +2|\mathfrak k|^3_\infty \|\check\nablasl{}^2\check\nablasl_4 \widetilde{\betao}\|+|\mathfrak k|^2_\infty(1-2|\mathfrak{h}|_\infty|\mathfrak{k}|_\infty)\,\|\check\nablasl\check\nablasl_3\check\nablasl_4 \widetilde{\betao}\|\\
    &\qquad +2|\mathfrak h|_\infty(1-|\mathfrak h|_\infty|\mathfrak k|_\infty)\|\check\nablasl{}^2 \check\nablasl_3\widetilde{\betabo}\|+|\mathfrak{h}|^2_\infty(1-2|\mathfrak{h}|_\infty|\mathfrak{k}|_\infty)\,\|\check\nablasl\check\nablasl{}_3^2 \widetilde{\betabo}\|\\
    &\qquad +\sum_{0\leq i_1+i_2+i_3\leq 2}\|\check\nablasl{}^{i_1}_4\check\nablasl{}^{i_2}_3\check\nablasl{}^{i_3}(\widetilde{\fo},\widetilde{\Gammao},\widetilde{\psio})\| \, .
\end{align*}
Finally, one applies $\nablasl\nablasl_4$ to the $\mathbb{S}^2$-projected linearised Bianchi equations \eqref{3_rho}--\eqref{3_sigma}, and then invokes Proposition~\ref{prop_commutation_1} again to deduce the estimate
\begin{align*}
    &(1-2|\mathfrak h|_\infty|\mathfrak k|_\infty)^2\big(\|\check\nablasl\check\nablasl_4\check\nablasl_3 \rhoo\|^2+\|\check\nablasl\check\nablasl_4\check\nablasl_3 \sigmao\|^2\big)^{\frac12} \\
    &\quad\leq |\mathfrak k|_\infty(1-|\mathfrak h|_\infty|\mathfrak k|_\infty)\, \|\check\nablasl{}^3\widetilde{\betao}\|+\big(\|\check\nablasl{}^3 \rhoo\|^2+\|\check\nablasl{}^3\sigmao\|^2\big)^{\frac12}+|\mathfrak h|_\infty \, \|\check\nablasl{}^3\widetilde{\betabo}\|\\
    &\quad\qquad +|\mathfrak h|^2_\infty\,\|\check\nablasl{}^2\check\nablasl_3\widetilde{\betabo}\|+|\mathfrak h|_\infty(1-|\mathfrak h|_\infty|\mathfrak k|_\infty)(1-2|\mathfrak h|_\infty|\mathfrak k|_\infty)\,\|\check\nablasl\check\nablasl_4\check\nablasl_3\widetilde{\betabo}\| \\
    &\quad\qquad+|\mathfrak k|_\infty^2(3-4|\mathfrak h|_\infty|\mathfrak k|_\infty)\,\|\check\nablasl{}^2\check\nablasl_4\widetilde{\betao}\|+|\mathfrak k|^3_\infty(1-2|\mathfrak h|_\infty|\mathfrak k|_\infty)\,\|\check\nablasl\check\nablasl{}_4^2\widetilde{\betao}\|\\
    &\quad\qquad +\sum_{0\leq i_1+i_2+i_3\leq 2}\|\check\nablasl{}^{i_1}_4\check\nablasl{}^{i_2}_3\check\nablasl{}^{i_3}(\widetilde{\fo},\widetilde{\Gammao},\widetilde{\psio})\| \, ,
\end{align*}
from which \eqref{est_34_rho_sigma} follows.
\end{proof}

\subsection{\texorpdfstring{$L^2(\mathbb{S}^2)$}{L2(S2)}-estimates for angular operators}

As a second step towards proving Proposition \ref{prop_top_order_curvature_horizon}, we employ Propositions \ref{prop_commutation_1} and \ref{prop_commutation_2} to estimate third order angular operators applied to the linearised curvature components in terms of third order angular derivatives of the linearised curvature components (plus third order mixed derivatives of $\alpha$ and $\alphab$ and additional lower order terms). All norms appearing in the proposition are $L^2(\mathbb{S}^2_{t,r},\check{\slashed{g}})$-norms. We also adopt the (projected version of the) notation \eqref{schematic_notation_th}--\eqref{schematic_notation_th_b}, with the $L^2_w(\mathbb{S}^2_{t,r},\check{\slashed{g}})$-norms denoted here by the subscript $w$, and the radial function \eqref{def_function_l}. We remark that the first term on the right hand side of the inequality \eqref{est_ang_op_rho_sigma} comes without overall $\Delta$-factor.

\medskip

\begin{prop} \label{prop_estimates_covariant_angular_derivatives}
Let $0\leq |a|<M$. Then, there exists a uniform constant $C>0$ such that the following holds. For any finite constant $R>r_+$, there exists a constant $C_{R}>0$ such that, for any solution to the $\mathbb{S}^2$-projected linearised system of equations, the estimates
\begin{align}
    &2\left(1-2\frac{a^2}{M^2}\frac{\Delta}{r^2}l^2\right)\|2\,\check{\slashed{\mathcal{D}}}{}_2^\star\check{\slashed{\mathrm{div}}}\check{\slashed{\mathcal D}}{}_2^\star \widetilde{\betao}\|\nonumber\\
    &\quad \leq \frac{a^2}{M^2}\frac{\Delta}{r^2}l^2\|\check\nablasl{}^3\widetilde{\betao}\|+\frac{|a|}{M}\frac{\Delta}{r^2}l\|\check\nablasl{}^3(\rhoo,\sigmao)\|+\frac{a^2}{M^2}\frac{\Delta^2}{r^4}l^2\|\check\nablasl{}^3\widetilde{\betabo}\|\nonumber\\
    &\quad \qquad + C\frac{|a|}{M}l\left(\|\check\partial{}^{\leq 3}\widetilde{\alphao}\|+\frac{\Delta^3}{r^6}l^2\|\check\partial{}^{\leq 3}\widetilde{\alphabo}\|_w\right)+C\|\check\nablasl{}^2\check\nablasl_3\widetilde{\alphao}\|_w+ C_R\|\check\partial{}^{\leq 2}(\widetilde{\fo},\widetilde{\Gammao},\widetilde{\psio})\| \, ,\label{est_ang_op_beta}\\
    &2\left(1-2\frac{a^2}{M^2}\frac{\Delta}{r^2}l^2\right)\|2\,\check{\slashed{\mathcal D}}{}_2^\star\check{\slashed{\mathrm{div}}}\check{\slashed{\mathcal D}}{}_2^\star \widetilde{\betabo}\|\nonumber\\
    &\quad \leq \frac{a^2}{M^2}l^2\|\check\nablasl{}^3\widetilde{\betao}\|+\frac{|a|}{M}l\|\check\nablasl{}^3(\rhoo,\sigmao)\|+\frac{a^2}{M^2}\frac{\Delta}{r^2}l^2\|\check\nablasl{}^3\widetilde{\betabo}\|\nonumber\\
    &\qquad\quad + C\frac{|a|}{M}l\left(l^2\|\check\partial{}^{\leq 3}\widetilde{\alphao}\|_w+\frac{\Delta}{r^2}\|\check\partial{}^{\leq 3}\widetilde{\alphabo}\|_w\right)+C\|\check\nablasl{}^2\check\nablasl_4\widetilde{\alphabo}\|+ C_R\|\check\partial{}^{\leq 2}(\widetilde{\fo},\widetilde{\Gammao},\widetilde{\psio})\| \, , \label{est_ang_op_betab}
\end{align}
and 
\begin{align}
    &\left(1-2\frac{a^2}{M^2}\frac{\Delta}{r^2}l^2\right)^2\|2\,\check{\slashed{\mathrm{div}}}\check{\slashed{\mathcal D}}{}_2^\star\check{\slashed{\mathcal D}}{}_1^\star(\rhoo,\sigmao)\| \nonumber\\
    &\quad \leq 2\frac{|a|}{M}l\left(1-\frac{a^2}{M^2}\frac{\Delta}{r^2}l^2\right)\|\check\nablasl{}^3\widetilde{\betao}\| +4\frac{a^2}{M^2}\frac{\Delta}{r^2}l^2 \left(1-2\frac{a^2}{M^2}\frac{\Delta}{r^2}l^2\right)\|\check\nablasl{}^3(\rhoo,\sigmao)\|+\frac{|a|}{M}\frac{\Delta}{r^2}l\|\check\nablasl{}^3\widetilde{\betabo}\|\nonumber\\
    &\qquad + C\frac{|a|}{M}\left(l\|\check\partial{}^{\leq 3}\widetilde{\alphao}\|_w+\frac{\Delta}{r^2}\|\check\partial{}^{\leq 3}\widetilde{\alphabo}\|_w\right)+C\|\check\nablasl\check\nablasl{}_4^2\widetilde{\alphabo}\|+ C_R\|\check\partial{}^{\leq 2}(\widetilde{\fo},\widetilde{\Gammao},\widetilde{\psio})\| \label{est_ang_op_rho_sigma}
\end{align}
hold for all $\mathbb{S}^2_{t,r}$-spheres with $r_+\leq r \leq R$, with notation \eqref{schematic_lin_q_1}--\eqref{schematic_lin_q_3} for the linearised quantities.
\end{prop}

\medskip

\begin{proof}
All products in the proof are taken relative to $\check{\slashed{g}}$. Recall that, for any scalar function $f$ and $\mathbb{S}^2_{t,r}$ one-forms $\varsigma,\xi$, one has
\begin{align*}
-2\,\check{\slashed{\mathcal{D}}}{}_2^{\star}(f\varsigma)&=\check{\nablasl}f\,\widehat{\otimes}\,\varsigma-2  f\check{\slashed{\mathcal{D}}}{}_2^{\star}\varsigma \, , \\
    \check{\slashed{\text{div}}}(\varsigma\,\widehat{\otimes}\,\xi)&=(\check{\slashed{\text{div}}}\,\varsigma)\xi-(\check{\slashed{\text{curl}}}\,\varsigma){}^{\star}\xi+(\check{\slashed{\text{div}}}\,\xi)\varsigma-(\check{\slashed{\text{curl}}}\,\xi){}^{\star}\varsigma \, , \\
    |\varsigma\,\widehat{\otimes}\,\xi|^2&=2|\varsigma|^2|\xi|^2 \, .
\end{align*}

\medskip

We apply $\check{\slashed{\mathcal{D}}}{}_2^{\star}\,\check{\slashed{\textup{div}}}$ to the $\mathbb{S}^2$-projected linearised Bianchi equation \eqref{3_alpha} to obtain (denoting the linearised quantities in the schematic notation of Section \ref{sec_proj_lin_system})
\begin{align*}
-2\, \check{\slashed{\mathcal{D}}}{}_2^{\star}\,\check{\slashed{\textup{div}}}\,\check{\slashed{\mathcal{D}}}{}_2^{\star}\,\widetilde{\betao}=& \, \frac{1}{2}\Big[\mathfrak{h}\,\widehat{\otimes}\check{\nablasl}\,\check{\slashed{\textup{div}}}\,\check{\nablasl}_3\widetilde{\betao}-{}^{\star}\mathfrak{h}\,\widehat{\otimes}\check{\nablasl}\,\check{\slashed{\textup{curl}}}\,\check{\nablasl}_3\widetilde{\betao}\Big]\\
&+\frac{1}{2}\Big[\mathfrak{k}\,\widehat{\otimes}\check{\nablasl}\,\check{\slashed{\textup{div}}}\,\check{\nablasl}_4\widetilde{\betao} -{}^{\star}\mathfrak{k}\,\widehat{\otimes}\check{\nablasl}\,\check{\slashed{\textup{curl}}}\,\check{\nablasl}_4\widetilde{\betao}\Big]\\
&+\check{\slashed{\mathcal{D}}}{}_2^{\star}\,\check{\slashed{\textup{div}}}\,\check{\nablasl}_3\widetilde{\alphao}+\sum_{0\leq i\leq 2}\check{\nablasl}{}^i\widetilde{\Gammao}+\sum_{0\leq i\leq 2}\check{\nablasl}{}^i\widetilde{\psio}  \, .
\end{align*}
We then estimate
\begin{align}
    \|2\, \check{\slashed{\mathcal{D}}}{}_2^{\star}\,\check{\slashed{\textup{div}}}\,\check{\slashed{\mathcal{D}}}{}_2^{\star}\,\widetilde{\betao}\| &\leq \frac12 |\mathfrak h|_\infty\|\check\nablasl{}^2\check\nablasl_3\widetilde{\betao}\| + \frac12 |\mathfrak k|_\infty\|\check\nablasl{}^2\check\nablasl_4\widetilde{\betao}\| \nonumber\\
    &\qquad+\|\check{\slashed{\mathcal{D}}}{}_2^{\star}\,\check{\slashed{\textup{div}}}\,\check{\nablasl}_3\widetilde{\alphao}\| +\sum_{0\leq i_1+i_2+i_3\leq 2}\|\check{\nablasl}{}^{i_1}_4\check{\nablasl}{}^{i_2}_3\check{\nablasl}{}^{i_3}(\widetilde{\fo},\widetilde{\Gammao},\widetilde{\psio})\| \label{est_ang_op_beta_inter}
\end{align}
and thus, by Proposition~\ref{prop_commutation_1},
\begin{align}
    &(1-2|\mathfrak h|_\infty |\mathfrak k|_\infty)\|2\, \check{\slashed{\mathcal{D}}}{}_2^{\star}\,\check{\slashed{\textup{div}}}\,\check{\slashed{\mathcal{D}}}{}_2^{\star}\,\widetilde{\betao}\| \nonumber\\
    &\quad \leq \frac12 |\mathfrak h|_\infty\,\big(\|\check\nablasl{}^3\rhoo\|^2+\|\check\nablasl{}^3\sigmao\|^2\big)^{\frac12}+ \frac12 |\mathfrak h|_\infty|\mathfrak k|_\infty\,\|\check\nablasl{}^3\widetilde{\betao}\|+\frac12 |\mathfrak h|_\infty^2\,\|\check\nablasl{}^3\widetilde{\betabo}\|\nonumber \\
    &\quad \qquad + \frac12|\mathfrak h|_\infty^3\,\|\check\nablasl{}^2\check\nablasl_3\widetilde{\betabo}\|+ \frac12 |\mathfrak k|_\infty(1-|\mathfrak h|_\infty |\mathfrak k|_\infty)\,\|\check\nablasl{}^2\check\nablasl_4\widetilde{\betao}\|\nonumber \\
    &\quad \qquad+(1-2|\mathfrak h|_\infty |\mathfrak k|_\infty)\,\|\check{\slashed{\mathcal{D}}}{}_2^{\star}\,\check{\slashed{\textup{div}}}\,\check{\nablasl}_3\widetilde{\alphao}\| +\sum_{0\leq i_1+i_2+i_3\leq 2}\|\check{\nablasl}{}^{i_1}_4\check{\nablasl}{}^{i_2}_3\check{\nablasl}{}^{i_3}(\widetilde{\fo},\widetilde{\Gammao},\widetilde{\psio})\| \, . \label{eq:est_ang_op_beta_int}
\end{align}
Similarly, by applying $\check{\slashed{\mathcal{D}}}{}_2^{\star}\,\check{\slashed{\textup{div}}}$ to the $\mathbb{S}^2$-projected linearised Bianchi equation \eqref{4_alphab} and combining the resulting estimate with Proposition~\ref{prop_commutation_1}, we have
\begin{align}
    &(1-2|\mathfrak h|_\infty |\mathfrak k|_\infty)\|2\, \check{\slashed{\mathcal{D}}}{}_2^{\star}\,\check{\slashed{\textup{div}}}\,\check{\slashed{\mathcal{D}}}{}_2^{\star}\,\widetilde{\betabo}\| \nonumber\\
    &\quad \leq \frac12 |\mathfrak k|_\infty\,\big(\|\check\nablasl{}^3\rhoo\|^2+\|\check\nablasl{}^3\sigmao\|^2\big)^{\frac12}+ \frac12 |\mathfrak h|_\infty|\mathfrak k|_\infty\,\|\check\nablasl{}^3\widetilde{\betabo}\|+\frac12 |\mathfrak k|_\infty^2\,\|\check\nablasl{}^3\widetilde{\betao}\|\nonumber\\
    &\quad \qquad + \frac12|\mathfrak k|_\infty^3\,\|\check\nablasl{}^2\check\nablasl_4\widetilde{\betao}\|+ \frac12 |\mathfrak h|_\infty(1-|\mathfrak h|_\infty |\mathfrak k|_\infty)\,\|\check\nablasl{}^2\check\nablasl_3\widetilde{\betabo}\|\nonumber\\
    &\quad \qquad+(1-2|\mathfrak h|_\infty |\mathfrak k|_\infty)\,\|\check{\slashed{\mathcal{D}}}{}_2^{\star}\,\check{\slashed{\textup{div}}}\,\check{\nablasl}_4\widetilde{\alphabo}\| +\sum_{0\leq i_1+i_2+i_3\leq 2}\|\check{\nablasl}{}^{i_1}_4\check{\nablasl}{}^{i_2}_3\check{\nablasl}{}^{i_3}(\widetilde{\fo},\widetilde{\Gammao},\widetilde{\psio})\| \, . \label{eq:est_ang_op_betab_int}
\end{align}
By using Proposition~\ref{prop_commutation_1}, the estimates \eqref{eq:est_ang_op_beta_int} and \eqref{eq:est_ang_op_betab_int} yield the inequalities \eqref{est_ang_op_beta} and \eqref{est_ang_op_betab}.

\medskip

We now apply $\check{\slashed{\textup{div}}}\check{\nablasl}_4$ to the $\mathbb{S}^2$-projected linearised Bianchi equation \eqref{4_alphab} to obtain (denoting the linearised quantities in the schematic notation of Section \ref{sec_proj_lin_system})
\begin{align*}
    \check\divsl\,\check\nablasl{}_4^2\,\widetilde{\alphabo} &= 2\,\check\divsl\,\check{\slashed{\mathcal D}}{}_2^\star\check\nablasl_4\widetilde{\betabo} 
    -\check\divsl\big(\mathfrak{h}\widehat{\otimes}\check\nablasl_4\check\nablasl_3\widetilde{\betabo}\big)
    -\check\divsl\big(\mathfrak{k}\widehat{\otimes}\check\nablasl{}_4^2\widetilde{\betabo}\big)+\sum_{0\leq i\leq 2}\check{\nablasl}{}^i\widetilde{\Gammao}+\sum_{0\leq i\leq 2}\check{\nablasl}{}^i\widetilde{\psio}
\end{align*}
and then, using the $\mathbb{S}^2$-projected linearised Bianchi equation \eqref{4_betab}, we deduce 
\begin{align*}
    - 2\,\check\divsl\,\check{\slashed{\mathcal D}}{}_2^\star\,\check{\slashed{\mathcal D}}{}_1^\star(\rhoo,\sigmao) &=\big[(\check{\slashed{\Delta}}\check{\nablasl}_4\rhoo) \mathfrak k-(\check{\slashed{\Delta}}\check{\nablasl}_4\sigmao) {}^\star\mathfrak k\big]+\big[(\check{\slashed{\Delta}}\check{\nablasl}_3\rhoo) \mathfrak h-(\check{\slashed{\Delta}}\check{\nablasl}_3\sigmao) {}^\star\mathfrak h\big]\\
    &\qquad -\big[\check\divsl(\check\nablasl{}_4^2\widetilde{\betabo})\mathfrak k-\check\curlsl(\check{\nablasl}{}_4^2\widetilde{\betabo}){}^\star \mathfrak k\big] -\big[\check\divsl(\check\nablasl_4\check\nablasl_3\widetilde{\betabo})\mathfrak h-\check\curlsl(\check\nablasl_4\check\nablasl_3\widetilde{\betabo}){}^\star \mathfrak h\big] \\
    &\qquad -\check\divsl\,\check\nablasl{}^2_4\widetilde{\alphabo}+\sum_{0\leq i\leq 2}\check{\nablasl}{}^i\widetilde{\Gammao}+\sum_{0\leq i\leq 2}\check{\nablasl}{}^i\widetilde{\psio} \, .
\end{align*}
We then estimate
\begin{align}
    \|2\,\check\divsl\,\check{\slashed{\mathcal D}}{}_2^\star\,\check{\slashed{\mathcal D}}{}_1^\star(\rhoo,\sigmao)\| &\leq |\mathfrak{k}|_\infty\,\big(\|\check{\slashed{\Delta}}\check\nablasl_4\rhoo\|^2+\|\check{\slashed{\Delta}}\check\nablasl_4\sigmao\|^2\Big)^{\frac12}+ |\mathfrak{h}|_\infty\,\big(\|\check{\slashed{\Delta}}\check\nablasl_3\rhoo\|^2+\|\check{\slashed{\Delta}}\check\nablasl_3\sigmao\|^2\big)^{\frac12} \nonumber\\
    &\qquad +|\mathfrak{k}|_\infty\|\check\nablasl\check\nablasl{}_4^2\widetilde{\betabo}\| +|\mathfrak{h}|_\infty\|\check\nablasl\check\nablasl_4\check\nablasl_3\widetilde{\betabo}\| \nonumber\\
    &\qquad +\|\check\divsl\,\check\nablasl{}_4^2\,\widetilde{\alphabo}\|+\sum_{0\leq i_1+i_2+i_3\leq 2}\|\check{\nablasl}{}^{i_1}_4\check{\nablasl}{}^{i_2}_3\check{\nablasl}{}^{i_3}(\widetilde{\fo},\widetilde{\Gammao},\widetilde{\psio})\|  \label{est_ang_op_rho_sigma_inter}
\end{align}
and thus, from Propositions~\ref{prop_commutation_1} and \ref{prop_commutation_2}, we obtain
\begin{align}
    &(1-2|\mathfrak h|_\infty |\mathfrak k|_\infty)^2\|2\,\check\divsl\,\check{\slashed{\mathcal D}}{}_2^\star\,\check{\slashed{\mathcal D}}{}_1^\star(\rhoo,\sigmao)\|\nonumber\\ 
    &\quad \leq 2|\mathfrak k|_\infty(1-|\mathfrak h|_\infty |\mathfrak k|_\infty)\,\|\check\nablasl{}^3\widetilde{\betao}\|+4|\mathfrak h|_\infty |\mathfrak k|_\infty(1-2|\mathfrak h|_\infty |\mathfrak k|_\infty)\, \big(\|\check\nablasl{}^3\rhoo\|^2+\|\check\nablasl{}^3\sigmao\|^2\Big)^{\frac12}+|\mathfrak h|_\infty\, \|\check\nablasl{}^3\widetilde{\betabo}\|\nonumber\\
    &\quad\qquad +|\mathfrak{h}|_\infty^2\,\|\check\nablasl{}^2\check\nablasl_3\widetilde{\betabo}\|+|\mathfrak{h}|_\infty(1-|\mathfrak{h}|_\infty|\mathfrak{k}|_\infty)(1-2|\mathfrak{h}|_\infty|\mathfrak{k}|_\infty)\,\|\check\nablasl\check\nablasl_4\check\nablasl_3\widetilde{\betabo}\|\nonumber \\
    &\quad \qquad + |\mathfrak k|^2_\infty\, (3-4|\mathfrak h|_\infty|\mathfrak k|_\infty)\,\|\check\nablasl{}^2\check\nablasl_4\widetilde{\betao}\|+ |\mathfrak k|^2_\infty (1-|\mathfrak h|_\infty|\mathfrak k|_\infty)\,\|\check\nablasl\check\nablasl{}^2_4\widetilde{\betao}\|\nonumber\\
    &\quad\qquad +(1-2|\mathfrak h|_\infty |\mathfrak k|_\infty)^2\|\check\divsl\,\check\nablasl{}_4^2\,\widetilde{\alphabo}\|+\sum_{0\leq i_1+i_2+i_3\leq 2}\|\check{\nablasl}{}^{i_1}_4\check{\nablasl}{}^{i_2}_3\check{\nablasl}{}^{i_3}(\widetilde{\fo},\widetilde{\Gammao},\widetilde{\psio})\| \, . \label{eq:est_ang_op_rho_sigma-int}
\end{align}
Estimate \eqref{est_ang_op_rho_sigma} then follows from  Propositions~\ref{prop_commutation_1} and~\ref{prop_commutation_2}.
\end{proof}

\subsection{Conclusion of the proof}

In this section, we prove Proposition \ref{prop_top_order_curvature_horizon}.~To this end, we first state the following lemma, which encodes the ellipticity of the angular operators considered in Proposition \ref{prop_estimates_covariant_angular_derivatives}. All norms appearing in the lemma and the following propositions are $L^2(\mathbb{S}^2_{t,r},\check{\slashed{g}})$-norms. We also adopt the (projected version of the) notation \eqref{schematic_notation_th}--\eqref{schematic_notation_th_b}, with the $L^2_w(\mathbb{S}^2_{t,r},\check{\slashed{g}})$-norms denoted here by the subscript $w$, and the radial function \eqref{def_function_l}.

\medskip

\begin{lemma} \label{lemma_angular_operators} 
For any finite constant $R>r_+$, there exists a constant $C_{R}>0$ such that the estimates 
\begin{align}
\sqrt{2}\|\check\nablasl{}^3\widetilde{\betao}\|&\leq 2\|2\,\check{\slashed{\mathcal D}}{}_2^\star\check{\slashed{\mathrm{div}}}\check{\slashed{\mathcal D}}{}_2^\star\widetilde{\betao}\|+C_R\|\check\partial{}^{\leq 2}\widetilde{\betao}\| \, , \label{est_elliptic_ang_op_beta}
\\
\sqrt{2}\|\check\nablasl{}^3\widetilde{\betabo}\|&\leq 2\|2\,\check{\slashed{\mathcal D}}{}_2^\star\check{\slashed{\mathrm{div}}}\check{\slashed{\mathcal D}}{}_2^\star\widetilde{\betabo}\|+C_R\|\check\partial{}^{\leq 2}\widetilde{\betabo}\| \, , \label{est_elliptic_ang_op_betab}
\\
    \|\check\nablasl{}^3(\rhoo,\sigmao)\|&\leq \|2\,\check{\slashed{\mathrm{div}}}\check{\slashed{\mathcal D}}{}_2^\star\check{\slashed{\mathcal D}}{}_1^\star(\rhoo,\sigmao)\|+C_R\|\check\partial{}^{\leq 2}(\rhoo,\sigmao)\| \label{est_elliptic_ang_op_rho_sigma}
\end{align}
hold for all $\mathbb{S}^2_{t,r}$-spheres with $r_+\leq r \leq R$.
\end{lemma}

\medskip

\begin{proof}
See Appendix \ref{sec_appendix_lemma}. By taking $\varsigma=\beta$ and $\varsigma=\betab$ in Lemma \ref{lemma_angular_operators_appendix}, one proves the inequalities \eqref{est_elliptic_ang_op_beta} and \eqref{est_elliptic_ang_op_betab}, respectively, in the lemma. By taking $f_1=\rho$ and $f_2=\sigma$ in Lemma \ref{lemma_angular_operators_appendix}, one proves the inequality \eqref{est_elliptic_ang_op_rho_sigma} in the lemma.   
\end{proof}

\medskip

By combining Proposition \ref{prop_estimates_covariant_angular_derivatives} with Lemma \ref{lemma_angular_operators}, we obtain the following proposition.

\medskip

\begin{prop} \label{prop_angular_before_assorbing} 
Let $0\leq |a|<M$. Then, there exists a uniform constant $C>0$ such that the following holds. For any finite constant $R>r_+$, there exists a constant $C_{R}>0$ such that, for any solution to the $\mathbb{S}^2$-projected linearised system of equations, the estimates
\begin{align}
&\left(\sqrt{2}-(2\sqrt{2}+1)\frac{a^2}{M^2}\frac{\Delta}{r^2}l^2\right)\,\|\check\nablasl{}^3\widetilde{\betao}\|\nonumber\\
    &\quad \leq  \frac{|a|}{M}\frac{\Delta}{r^2}l\,\|\check\nablasl{}^3(\rhoo,\sigmao)\|+  \frac{a^2}{M^2}\frac{\Delta^2}{r^4}l^2\,\|\check\nablasl{}^3\widetilde{\betabo}\|\nonumber \\
    &\quad \qquad + C\frac{|a|}{M}l\left(\|\check\partial{}^{\leq 3} \widetilde{\alphao}\|_w+\frac{\Delta^3}{r^6}l^2\|\check\partial{}^{\leq 3} \widetilde{\alphabo}\|_w\right) + C\|\check\nablasl{}^2\check\nablasl_3\widetilde{\alphao}\| + C_R\|\check\partial{}^{\leq 2}(\widetilde{\fo},\widetilde{\Gammao},\widetilde{\psio})\| \, ,\label{eq:est_before_assorbing_beta}
    \\
    &\left(\sqrt{2}-(2\sqrt{2}+1)\frac{a^2}{M^2}\frac{\Delta}{r^2}l^2\right)\,\|\check\nablasl{}^3\widetilde{\betabo}\|\nonumber\\
    &\quad \leq  \frac{|a|}{M}l\,\|\check\nablasl{}^3(\rhoo,\sigmao)\|+  \frac{a^2}{M^2}l^2\,\|\check\nablasl{}^3\widetilde{\betabo}\|\nonumber \\
    &\quad \qquad + C\frac{|a|}{M}l\left(l^2\|\check\partial{}^{\leq 3} \widetilde{\alphao}\|_w+\frac{\Delta}{r^2}\|\check\partial{}^{\leq 3} \widetilde{\alphabo}\|_w\right) + C\|\check\nablasl{}^2\check\nablasl_4\widetilde{\alphabo}\| + C_R\|\check\partial{}^{\leq 2}(\widetilde{\fo},\widetilde{\Gammao},\widetilde{\psio})\|\label{eq:est_before_assorbing_betab}
\end{align}
and
\begin{align}
    &\left(1-6\frac{a^2}{M^2}\frac{\Delta}{r^2}l^2\right)\left(1-2\frac{a^2}{M^2}\frac{\Delta}{r^2}l^2\right)\|\check\nablasl{}^3(\rhoo,\sigmao)\|\nonumber\\
    &\quad\leq 2\frac{|a|}{M}l\left(1-\frac{a^2}{M^2}\frac{\Delta}{r^2}l^2\right)\|\check\nablasl{}^3 \widetilde{\betao}\|+\frac{|a|}{M}\frac{\Delta}{r^2}l\|\check\nablasl{}^3\widetilde{\betabo}\|\nonumber\\
    &\quad\qquad + C\frac{|a|}{M}\left(l\|\check\partial{}^{\leq 3} \widetilde{\alphao}\|_w+\|\check\partial{}^{\leq 3} \widetilde{\alphabo}\|_w\right) + C\|\check\nablasl\check\nablasl{}^2_4\widetilde{\alphabo}\| + C_R\|\check\partial{}^{\leq 2}(\widetilde{\fo},\widetilde{\Gammao},\widetilde{\psio})\| \label{eq:est_before_assorbing_rho_sigma}
\end{align}
hold for all $\mathbb{S}^2_{t,r}$-spheres with $r_+\leq r \leq R$, with notation \eqref{schematic_lin_q_1}--\eqref{schematic_lin_q_3} for the linearised quantities.
\end{prop}

\medskip

\begin{proof} 
Using Lemma~\ref{lemma_angular_operators} to estimate the left hand side of estimates \eqref{eq:est_ang_op_beta_int}--\eqref{eq:est_ang_op_rho_sigma-int}  from the proof of Proposition~\ref{prop_estimates_covariant_angular_derivatives}, we obtain
\begin{align}
&[\sqrt{2}-(2\sqrt{2}+1)|\mathfrak h|_\infty|\mathfrak k|_\infty]\,\|\check\nablasl{}^3\widetilde{\betao}\|\nonumber\\
    &\quad \leq  |\mathfrak h|_\infty\,\big(\|\check\nablasl{}^3\rhoo\|^2+\|\check\nablasl{}^3\sigmao\|^2\big)^{\frac12}+ |\mathfrak h|_\infty^2\,\|\check\nablasl{}^3\widetilde{\betabo}\|\nonumber \\
    &\quad \qquad + |\mathfrak h|_\infty^3\,\|\check\nablasl{}^2\check\nablasl_3\widetilde{\betabo}\|+  |\mathfrak k|_\infty(1-|\mathfrak h|_\infty |\mathfrak k|_\infty)\,\|\check\nablasl{}^2\check\nablasl_4\widetilde{\betao}\|\nonumber \\
    &\quad \qquad+2(1-2|\mathfrak h|_\infty |\mathfrak k|_\infty)\,\|\check{\slashed{\mathcal{D}}}{}_2^{\star}\,\check{\slashed{\textup{div}}}\,\check{\nablasl}_3\widetilde{\alphao}\| +\sum_{0\leq i_1+i_2+i_3\leq 2}\|\check{\nablasl}{}^{i_1}_4\check{\nablasl}{}^{i_2}_3\check{\nablasl}{}^{i_3}(\widetilde{\fo},\widetilde{\Gammao},\widetilde{\psio})\| \, , \label{eq:est_before_assorbing_beta-int}
\\
&[\sqrt{2}-(2\sqrt{2}+1)|\mathfrak h|_\infty|\mathfrak k|_\infty]\,\|\check\nablasl{}^3\widetilde{\betabo}\|\nonumber\\
&\quad \leq |\mathfrak k|_\infty^2\,\|\check\nablasl{}^3\widetilde{\betao}\|+
 |\mathfrak k|_\infty\,\big(\|\check\nablasl{}^3\rhoo\|^2+\|\check\nablasl{}^3\sigmao\|^2\big)^{\frac12}\nonumber\\
    &\quad \qquad + |\mathfrak k|_\infty^3\,\|\check\nablasl{}^2\check\nablasl_4\widetilde{\betao}\|+  |\mathfrak h|_\infty(1-|\mathfrak h|_\infty |\mathfrak k|_\infty)\,\|\check\nablasl{}^2\check\nablasl_3\widetilde{\betabo}\|\nonumber\\
    &\quad \qquad+2(1-2|\mathfrak h|_\infty |\mathfrak k|_\infty)\,\|\check{\slashed{\mathcal{D}}}{}_2^{\star}\,\check{\slashed{\textup{div}}}\,\check{\nablasl}_4\widetilde{\alphabo}\| +\sum_{0\leq i_1+i_2+i_3\leq 2}\|\check{\nablasl}{}^{i_1}_4\check{\nablasl}{}^{i_2}_3\check{\nablasl}{}^{i_3}(\widetilde{\fo},\widetilde{\Gammao},\widetilde{\psio})\| \, ,\label{eq:est_before_assorbing_betab-int}
\\
    &(1-2|\mathfrak h|_\infty|\mathfrak k|_\infty)(1-6|\mathfrak h|_\infty|\mathfrak k|_\infty)\,\big(\|\check\nablasl{}^3\rhoo\|^2+\|\check\nablasl{}^3\sigmao\|^2\big)^{\frac12} \nonumber
    \\ 
    &\quad \leq 2|\mathfrak k|_\infty(1-|\mathfrak h|_\infty |\mathfrak k|_\infty)\,\|\check\nablasl{}^3\widetilde{\betao}\|+|\mathfrak h|_\infty\, \|\check\nablasl{}^3\widetilde{\betabo}\|\nonumber\\
    &\quad\qquad +|\mathfrak{h}|_\infty^2\,\|\check\nablasl{}^2\check\nablasl_3\widetilde{\betabo}\|+|\mathfrak{h}|_\infty(1-|\mathfrak{h}|_\infty|\mathfrak{k}|_\infty)(1-2|\mathfrak{h}|_\infty|\mathfrak{k}|_\infty)\,\|\check\nablasl\check\nablasl_4\check\nablasl_3\widetilde{\betabo}\|\nonumber \\
    &\quad \qquad + |\mathfrak k|^2_\infty\, (3-4|\mathfrak h|_\infty|\mathfrak k|_\infty)\,\|\check\nablasl{}^2\check\nablasl_4\widetilde{\betao}\|+ |\mathfrak k|^2_\infty (1-|\mathfrak h|_\infty|\mathfrak k|_\infty)\,\|\check\nablasl\check\nablasl{}^2_4\widetilde{\betao}\|\nonumber\\
    &\quad\qquad +(1-2|\mathfrak h|_\infty |\mathfrak k|_\infty)^2\|\check\divsl\,\check\nablasl{}_4^2\,\widetilde{\alphabo}\|+\sum_{0\leq i_1+i_2+i_3\leq 2}\|\check{\nablasl}{}^{i_1}_4\check{\nablasl}{}^{i_2}_3\check{\nablasl}{}^{i_3}(\widetilde{\fo},\widetilde{\Gammao},\widetilde{\psio})\| \label{eq:est_before_assorbing_rho_sigma-int}
\end{align}
and the conclusion follows immediately.
\end{proof}

\medskip

By summing the inequalities of Proposition \ref{prop_angular_before_assorbing}, we obtain the following proposition, which achieves the desired estimates for the $\check{\nablasl}{}^3$-derivatives of the linearised curvature components (i.e.~Proposition \ref{prop_top_order_curvature_horizon} for the third order angular-derivative terms on the left hand side).

\medskip

\begin{prop} \label{main_th_angular_version} 
Let $0\leq |a|<M$. Then, there exists a uniform constant $C>0$ such that the following holds. For any finite constant $R>r_+$, there exists a constant $C_{R}>0$ such that, for any solution to the $\mathbb{S}^2$-projected linearised system of equations, the estimates
\begin{align*}
   \|\check\nablasl{}^3\widetilde{\betao}\|&\leq C\left(\|\check\nablasl{}^2\check\nablasl_3\widetilde{\alphao}\|+\frac{|a|}{M}\frac{\Delta}{r^2}l\|\check\nablasl\check\nablasl{}_4^2\widetilde{\alphabo}\|\right)+C\frac{|a|}{M}l\left(\|\check\partial{}^{\leq 3}\widetilde{\alphao}\|_w+\frac{\Delta}{r^2}\|\check\partial{}^{\leq 3}\widetilde{\alphabo}\|_w\right)+ C_R\|\check\partial{}^{\leq 2}(\widetilde{\fo},\widetilde{\Gammao},\widetilde{\psio})\| \, ,
    \\
    \|\check\nablasl{}^3\widetilde{\betabo}\|&\leq C\left(\frac{a^2}{M^2}l^2\|\check\nablasl{}^2\check\nablasl_3\widetilde{\alphao}\|+\frac{|a|}{M}l|\check\nablasl\check\nablasl{}_4^2\widetilde{\alphabo}|+|\check\nablasl{}^2\check\nablasl_4\widetilde{\alphabo}|\right)\\
    &\qquad+C\frac{|a|}{M}l\left(l\|\check\partial{}^{\leq 3}\widetilde{\alphao}\|_w+\|\check\partial{}^{\leq 3}\widetilde{\alphabo}\|_w\right)+ C_R\|\check\partial{}^{\leq 2}(\widetilde{\fo},\widetilde{\Gammao},\widetilde{\psio})\| \, ,
\\
    \|\check\nablasl{}^3(\rhoo,\sigmao)\|&\leq C\left(\frac{|a|}{M}l\|\check\nablasl{}^2\check\nablasl_3\widetilde{\alphao}\|+\|\check\nablasl\check\nablasl{}_4^2\widetilde{\alphabo}\|\right)
    +C\frac{|a|}{M}\left(l^2\|\check\partial{}^{\leq 3}\widetilde{\alphao}\|_w+\|\check\partial{}^{\leq 3}\widetilde{\alphabo}\|_w\right)+ C_R\|\check\partial{}^{\leq 2}(\widetilde{\fo},\widetilde{\Gammao},\widetilde{\psio})\|
\end{align*}
hold for all $\mathbb{S}^2_{t,r}$-spheres with $r_+\leq r \leq R$, with notation \eqref{schematic_lin_q_1}--\eqref{schematic_lin_q_3} for the linearised quantities.
\end{prop}

\medskip

\begin{proof}
We combine  \eqref{eq:est_before_assorbing_beta-int} and      \eqref{eq:est_before_assorbing_betab-int} in the proof of Proposition~\ref{prop_angular_before_assorbing} to obtain
\begin{align*}
    q(r)\|\check\nablasl{}^3\widetilde{\beta}\|&\leq \sqrt{2}|\mathfrak h|_\infty\, \|\check\nablasl{}^3(\rhoo, \sigmao)\| \\
    &\qquad + (1+\sqrt 2)|\mathfrak h|^3_\infty\,\|\check\nablasl{}^2\check\nablasl_3\widetilde{\betabo}\|+|\mathfrak{k}|_\infty(\sqrt 2-(1+\sqrt 2)|\mathfrak{h}|_\infty|\mathfrak{k}|_\infty)\, \|\check\nablasl{}^2\check\nablasl_4\widetilde{\betao}\|\\
    &\qquad +2[\sqrt 2 -(1+2\sqrt{2})|\mathfrak{h}|_\infty|\mathfrak{k}|_\infty]\,\|\check{\slashed{\mathcal{D}}}{}_2^\star\, \check\divsl\, \check\nablasl_3\widetilde{\alphao}\|+ 2|\mathfrak h|_\infty^2\, \|\check{\slashed{\mathcal{D}}}{}_2^\star\, \check\divsl\, \check\nablasl_4\widetilde{\alphabo}\|\\
    &\qquad+\sum_{0\leq i_1+i_2+i_3\leq 2}\|\check{\nablasl}{}^{i_1}_4\check{\nablasl}{}^{i_2}_3\check{\nablasl}{}^{i_3}(\widetilde{\fo},\widetilde{\Gammao},\widetilde{\psio})\| \, ,
    \\
    q(r)\|\check\nablasl{}^3\widetilde{\betab}\|&\leq \sqrt{2}|\mathfrak k|_\infty\, \|\check\nablasl{}^3(\rhoo, \sigmao)\| \\
    &\qquad + (1+\sqrt 2)|\mathfrak k|^3_\infty\,\|\check\nablasl{}^2\check\nablasl_4\widetilde{\betao}\|+|\mathfrak{h}|_\infty(\sqrt 2-(1+\sqrt 2)|\mathfrak{h}|_\infty|\mathfrak{k}|_\infty)\, \|\check\nablasl{}^2\check\nablasl_3\widetilde{\betabo}\|\\
    &\qquad +2[\sqrt 2 -(1+2\sqrt{2})|\mathfrak{h}|_\infty|\mathfrak{k}|_\infty]\,\|\check{\slashed{\mathcal{D}}}{}_2^\star\, \check\divsl\, \check\nablasl_4\widetilde{\alphabo}\|+ 2|\mathfrak k|_\infty^2\, \|\check{\slashed{\mathcal{D}}}{}_2^\star\, \check\divsl\, \check\nablasl_3\widetilde{\alphao}\|\\
    &\qquad +\sum_{0\leq i_1+i_2+i_3\leq 2}\|\check{\nablasl}{}^{i_1}_4\check{\nablasl}{}^{i_2}_3\check{\nablasl}{}^{i_3}(\widetilde{\fo},\widetilde{\Gammao},\widetilde{\psio})\| \, ,
\end{align*}
where $q(r):=2[1-(2+\sqrt 2)|\mathfrak{h}|_\infty|\mathfrak{k}|_\infty]$. Similarly, from \eqref{eq:est_before_assorbing_rho_sigma-int}, we deduce
\begin{align*}
    &(1-2|\mathfrak{h}|_\infty|\mathfrak{k}|_\infty)(1-6|\mathfrak{h}|_\infty|\mathfrak{k}|_\infty)\, \|\check\nablasl{}^3(\rhoo,\sigmao)\|\\
    &\quad\leq 2|\mathfrak{k}|_\infty(1-|\mathfrak{h}|_\infty|\mathfrak{k}|_\infty)\,\|\check\nablasl{}^3\widetilde{\betao}\|+ |\mathfrak{h}|_\infty\,\|\check\nablasl{}^3\widetilde{\betabo}\|\\
  &\quad\qquad +|\mathfrak{h}|_\infty^2\,\|\check\nablasl{}^2\check\nablasl_3\widetilde{\betabo}\|+|\mathfrak{h}|_\infty(1-|\mathfrak{h}|_\infty|\mathfrak{k}|_\infty)(1-2|\mathfrak{h}|_\infty|\mathfrak{k}|_\infty)\,\|\check\nablasl\check\nablasl_4\check\nablasl_3\widetilde{\betabo}\|\nonumber \\
    &\quad \qquad + |\mathfrak k|^2_\infty\, (3-4|\mathfrak h|_\infty|\mathfrak k|_\infty)\,\|\check\nablasl{}^2\check\nablasl_4\widetilde{\betao}\|+ |\mathfrak k|^2_\infty (1-|\mathfrak h|_\infty|\mathfrak k|_\infty)\,\|\check\nablasl\check\nablasl{}^2_4\widetilde{\betao}\|\nonumber\\
    &\quad\qquad +(1-2|\mathfrak h|_\infty |\mathfrak k|_\infty)^2\|\check\divsl\,\check\nablasl{}_4^2\,\widetilde{\alphabo}\|+\sum_{0\leq i_1+i_2+i_3\leq 2}\|\check{\nablasl}{}^{i_1}_4\check{\nablasl}{}^{i_2}_3\check{\nablasl}{}^{i_3}(\widetilde{\fo},\widetilde{\Gammao},\widetilde{\psio})\| \, . 
\end{align*}
Notice that, in the three estimates above, all terms involving null derivatives of $\beta$ or $\betab$ can be estimated, by using Propositions~\ref{prop_commutation_1} and \ref{prop_commutation_2}, in terms of norms which at top order only involve $\alpha$ and $\alphab$. Then, to conclude the proof, we combine these three estimates and note that
$$q(r) (1-2|\mathfrak{h}|_\infty|\mathfrak{k}|_\infty)(1-6|\mathfrak{h}|_\infty|\mathfrak{k}|_\infty) -\sqrt{2}|\mathfrak{h}|_\infty|\mathfrak{k}|_\infty(3-2 |\mathfrak{h}|_\infty|\mathfrak{k}|_\infty) \geq \frac{610+191 \sqrt{2}}{512} >0$$ for all $r\geq r_+$, with equality attained for $|a|=M$ at $r=(1+\sqrt{2})r_+$.
\end{proof}

\medskip

The inequalities of Proposition \ref{main_th_angular_version} can be used to estimate the right hand side of the inequalities of Propositions \ref{prop_commutation_1} and \ref{prop_commutation_2} and thus achieve the deired estimates for all third order mixed derivatives of the linearised curvature components containing one or two angular derivatives.~One can then derive a commuted version of Proposition \ref{prop_preliminary_null_derivatives} for third order mixed derivatives of the linearised curvature components which do not contain any angular derivatives.~The top-order terms on right hand side of the estimates for these derivatives will only consist of third order mixed derivatives containing at least one angular derivative, for which one already derived the desired estimates.~This last step of the procedure concludes the proof of Proposition \ref{prop_top_order_curvature_horizon}.

\medskip

\begin{remark} \label{rmk_high_order_th}
Consider the angular operator $\mathcal{A}^{[k]}$ acting on symmetric traceless $\mathbb{S}^2_{t,r}$ two-tensors, with $k\in\mathbb{N}$, such that
\begin{align*}
\mathcal{A}^{[0]}&:=\textup{Id} \, , & \mathcal{A}^{[2k+1]}&:=\check{\slashed{\textup{div}}}\mathcal{A}^{[2k]} \, , & \mathcal{A}^{[2k+2]}&:=\check{\slashed{\mathcal{D}}}{}_2^{\star}\check{\slashed{\textup{div}}}\mathcal{A}^{[2k]} \, .
\end{align*}
To obtain Theorem \ref{main_th_projected} for $k>3$, one revisits the proof of Proposition \ref{prop_estimates_covariant_angular_derivatives} by now commuting both the $\mathbb{S}^2$-projected linearised Bianchi equations \eqref{3_alpha} and \eqref{4_alphab} with $\mathcal{A}^{[k-1]}$, and with $\check{\nablasl}_3\mathcal{A}^{[k-2]}$ and $\check{\nablasl}_4\mathcal{A}^{[k-2]}$ respectively. The angular operator $\mathcal{A}^{[k]}$ possesses ellipticity properties in the sense of Lemma \ref{lemma_angular_operators}.
\end{remark}

\appendix

\section{Ellipticity lemma for angular operators} \label{sec_appendix_lemma}

In this appendix, we prove a general lemma on the ellipticity of the angular operators considered in the proof of Theorem \ref{main_th}. The lemma is directly applied in the proof of Lemma \ref{lemma_angular_operators} for the linearised curvature components. We recall the notation \eqref{schematic_notation_th}, which is used in the statement of the lemma.

\medskip

\begin{lemma} \label{lemma_angular_operators_appendix}
For any finite constant $R>r_+$, there exists a constant $C_{R}>0$ such that, for any $\mathbb{S}^2_{t,r}$ one-form $\varsigma$ and any scalar functions $f_1,f_2$, the estimates
\begin{align}
\sqrt{2}\,\|\check{\nablasl}{}^3\varsigma\|_{L^2(\mathbb{S}^2_{t,r},\check{\slashed{g}})}\leq& \,  2\, \|2 \,\check{\slashed{\mathcal{D}}}{}_2^{\star}\check{\slashed{\textup{div}}}\check{\slashed{\mathcal{D}}}{}_2^{\star}\,\varsigma\|_{L^2(\mathbb{S}^2_{t,r},\check{\slashed{g}})}+C_R\sum_{i=0}^2\|\check{\nablasl}{}^{i}\varsigma\|_{L^2(\mathbb{S}^2_{t,r},\check{\slashed{g}})} \, , \label{est_elliptic_ang_op_appendix_1} 
\\
\|\check{\nablasl}{}^3(f_1,f_2)\|_{L^2(\mathbb{S}^2_{t,r},\check{\slashed{g}})} \leq& \,  \|2 \,\check{\slashed{\textup{div}}}\check{\slashed{\mathcal{D}}}{}_2^{\star}\check{\slashed{\mathcal{D}}}{}_1^{\star}(f_1,f_2)\|_{L^2(\mathbb{S}^2_{t,r},\check{\slashed{g}})} +C_R\sum_{i=1}^2\|\check{\nablasl}{}^{i}(f_1,f_2)\|_{L^2(\mathbb{S}^2_{t,r},\check{\slashed{g}})} \label{est_elliptic_ang_op_appendix_2}
\end{align}
hold for all $\mathbb{S}^2_{t,r}$-spheres with $r_+\leq r \leq R$.
\end{lemma}

\medskip

\begin{proof}
All integrations in the proof are over the $\mathbb{S}^2_{t,r}$-spheres and relative to the volume form $\slashed{\varepsilon}_{\check{\slashed{g}}}$.~All dot products, contractions and pointwise norms are taken relative to the metric $\check{\slashed{g}}$. 

\medskip

For any smooth scalar function $f$, one can compute (by repeated integration by parts)
\begin{equation}
\int |\check{\nablasl}{}^3 f|^2 = \int |\check{\nablasl}\check{\slashed{\Delta}}f|^2+\int|[\check{\slashed{\Delta}},\check{\nablasl}]f|^2-2\int \check{\slashed{\Delta}}f\cdot \check{\slashed{\textup{div}}}([\check{\slashed{\Delta}},\check{\nablasl}]f) -\int \check{\nablasl}{}^2 f\cdot [\check{\slashed{\Delta}},\check{\nablasl}]\check{\nablasl}f \label{aux_comp_1}
\end{equation}
and
\begin{align}
\int |\check{\nablasl}{}^4 f|^2 =& \, \int |\check{\slashed{\Delta}}\check{\slashed{\Delta}}f|^2 +\int |[\check{\slashed{\Delta}},\check{\nablasl}]\check{\nablasl}f|^2 +\int |\check{\nablasl}[\check{\slashed{\Delta}},\check{\nablasl}]f|^2 \nonumber\\
&-2\int \check{\nablasl}\check{\slashed{\Delta}}f \cdot \check{\slashed{\textup{div}}}([\check{\slashed{\Delta}},\check{\nablasl}]\check{\nablasl}f) +2\int [\check{\slashed{\Delta}},\check{\nablasl}]\check{\nablasl}f\cdot \check{\nablasl}[\check{\slashed{\Delta}},\check{\nablasl}]f \nonumber\\
&-2\int \check{\nablasl}\check{\slashed{\Delta}}f\cdot \check{\slashed{\textup{div}}}(\check{\nablasl}[\check{\slashed{\Delta}},\check{\nablasl}]f)-\int \check{\nablasl}{}^3 f\cdot [\check{\slashed{\Delta}},\check{\nablasl}]\check{\nablasl}{}^2 f-\int K |\check{\nablasl}\check{\slashed{\Delta}}f|^2 \, , \label{aux_comp_2}
\end{align}
where to compute the identity \eqref{aux_comp_2} one also uses the standard identity
\begin{equation*}
\int |\check{\nablasl}{}^2 \varphi|^2= \int |\check{\slashed{\Delta}}\varphi|^2-\int K |\check{\nablasl}\varphi|^2 
\end{equation*}
with $\varphi=\check{\slashed{\Delta}}f$. We denoted by $K$ the Gauss curvature of $\check{\slashed{g}}$. We remark that $K$ is a function of the angular coordinates on the $\mathbb{S}^2_{t,r}$-spheres (which, in particular, are not spheres of constant $K$).

\medskip

We now turn to treating the angular operators in the lemma, for which we will make use of the standard (and easy to check) identities
\begin{align*}
2\, \check{\slashed{\textup{div}}}\check{\slashed{\mathcal{D}}}{}_2^{\star}&=-\check{\slashed{\Delta}}-K \, , & \check{\slashed{\mathcal{D}}}{}_1^{\star}\check{\slashed{\mathcal{D}}}{}_1&=-\check{\slashed{\Delta}}+K \, , & \check{\slashed{\mathcal{D}}}{}_1\check{\slashed{\mathcal{D}}}{}_1^{\star}&=-\check{\slashed{\Delta}} \, .
\end{align*}
For any $\mathbb{S}^2_{t,r}$ one-form $\varsigma$, we compute 
\begin{align}
2\, \check{\slashed{\mathcal{D}}}{}_2^{\star}\check{\slashed{\textup{div}}}\check{\slashed{\mathcal{D}}}{}_2^{\star}\varsigma &=\check{\slashed{\mathcal{D}}}{}_2^{\star}(-\check{\slashed{\Delta}}+K-2K)\varsigma \nonumber\\
&=\check{\slashed{\mathcal{D}}}{}_2^{\star}\check{\slashed{\mathcal{D}}}{}_1^{\star}\check{\slashed{\mathcal{D}}}{}_1\varsigma-2\,\check{\slashed{\mathcal{D}}}{}_2^{\star}(K\varsigma) \, . \label{aux_1}
\end{align}
By the Hodge-decomposition $\varsigma=\check{\slashed{\mathcal{D}}}{}_1^{\star}(f_1,f_2)$ for some scalar functions $f_1,f_2$, one can compute (by repeated integration by parts)
\begin{equation}
\int |\check{\slashed{\mathcal{D}}}{}_2^{\star}\check{\slashed{\mathcal{D}}}{}_1^{\star}\check{\slashed{\mathcal{D}}}{}_1\varsigma|^2 = \frac{1}{2}\int |\check{\nablasl}{}^4f_1|^2+\frac{1}{2}\int|\check{\nablasl}{}^4f_2|^2 - \int K | \check{\slashed{\mathcal{D}}}{}_1^{\star}(-\check{\slashed{\Delta}}f_1,-\check{\slashed{\Delta}}f_2)|^2 +\mathbb{L}[f_1]+\mathbb{L}[f_2] \, ,
\end{equation}
where one also uses the identity \eqref{aux_comp_2} and defines the scalar functions ($i=1,2$)
\begin{align*}
\mathbb{L}[f_i](t,r) :=&\, -\int |[\check{\slashed{\Delta}},\check{\nablasl}]\check{\nablasl}f_i|^2 -\int |\check{\nablasl}[\check{\slashed{\Delta}},\check{\nablasl}]f_i|^2 \nonumber\\
&+2\int \check{\nablasl}\check{\slashed{\Delta}}f_i \cdot \check{\slashed{\textup{div}}}([\check{\slashed{\Delta}},\check{\nablasl}]\check{\nablasl}f_i) -2\int [\check{\slashed{\Delta}},\check{\nablasl}]\check{\nablasl}f_i\cdot \check{\nablasl}[\check{\slashed{\Delta}},\check{\nablasl}]f_i \nonumber\\
&+2\int \check{\nablasl}\check{\slashed{\Delta}}f_i\cdot \check{\slashed{\textup{div}}}(\check{\nablasl}[\check{\slashed{\Delta}},\check{\nablasl}]f_i)+\int \check{\nablasl}{}^3 f_i\cdot [\check{\slashed{\Delta}},\check{\nablasl}]\check{\nablasl}{}^2 f_i+\int K|\check{\nablasl}\check{\slashed{\Delta}}f_i|^2 \, .
\end{align*}
We note the identities
\begin{align*}
[\check{\slashed{\Delta}},\check{\nablasl}]f_i&=K\check{\nablasl}f_i \, , & [\check{\slashed{\Delta}},\check{\nablasl}]\check{\nablasl}f_i&= \check{\slashed{R}}\cdot\check{\nablasl}{}^2f_i+\check{\nablasl}\check{\slashed{R}}\cdot\check{\nablasl}f_i \, , & [\check{\slashed{\Delta}},\check{\nablasl}]\check{\nablasl}{}^2 f_i&= \check{\slashed{R}}\cdot\check{\nablasl}{}^3f_i+\check{\nablasl}\check{\slashed{R}}\cdot\check{\nablasl}{}^2f_i \, ,
\end{align*}
where the second and third identities are in schematic form, with $\check{\slashed{R}}$ denoting some curvature quantity of $\check{\slashed{g}}$. The terms $\mathbb{L}[f_i]$ are therefore lower order terms (in fact, from first to third order). In particular, there exists a constant $C_r>0$, depending only on $r$, such that
\begin{align*}
&|\mathbb{L}[f_1]|+|\mathbb{L}[f_2]|
+\left\lvert\int K | \check{\slashed{\mathcal{D}}}{}_1^{\star}(-\check{\slashed{\Delta}}f_1,-\check{\slashed{\Delta}}f_2)|^2\right\rvert  + \int|\check{\slashed{\mathcal{D}}}{}_2^{\star}(K \check{\slashed{\mathcal{D}}}{}_1^{\star}(f_1,f_2))|^2  \\
&\qquad \leq C_r \sum_{i=1}^3\left(\|\check{\nablasl}{}^{i}f_1\|_{L^2(\mathbb{S}^2_{t,r},\check{\slashed{g}})}^2+\|\check{\nablasl}{}^{i}f_2\|_{L^2(\mathbb{S}^2_{t,r},\check{\slashed{g}})}^2\right)  \\
&\qquad \leq C_r \sum_{i=0}^2\|\check{\nablasl}{}^{i}\varsigma\|_{L^2(\mathbb{S}^2_{t,r},\check{\slashed{g}})}^2 \, ,
\end{align*}
where the second inequality follows from the standard (and easy to check) identities
\begin{align}
\int |\check{\nablasl}f_1|^2+\int |\check{\nablasl}f_2|^2 &=\int |\varsigma|^2  \, ,  \label{aux_4} \\
\int |\check{\nablasl}{}^2 f_1|^2+\int |\check{\nablasl}{}^2 f_2|^2+\int K|\check{\nablasl}f_1|^2+\int K|\check{\nablasl}f_2|^2  &=\int |\check{\nablasl}\varsigma|^2+\int K|\varsigma|^2 \label{aux_5}
\end{align}
and the inequality
\begin{equation*}
\int |\check{\nablasl}{}^3 f_1|^2 +\int |\check{\nablasl}{}^3f_2|^2 \leq C_r \sum_{i=0}^2\|\check{\nablasl}{}^{i}\varsigma\|_{L^2(\mathbb{S}^2_{t,r},\check{\slashed{g}})}^2 \, ,
\end{equation*}
the latter obtained by combining the identities \eqref{aux_4} and \eqref{aux_5} with the identity \eqref{aux_comp_1}. By noting the inequality 
\begin{equation*}
\int |\check{\nablasl}{}^3\varsigma|^2 \leq \int |\check{\nablasl}{}^4f_1|^2+\int |\check{\nablasl}{}^4f_2|^2 +C_r \sum_{i=0}^2\|\check{\nablasl}{}^{i}\varsigma\|_{L^2(\mathbb{S}^2_{t,r},\check{\slashed{g}})}^2
\end{equation*}
and by recalling the identity \eqref{aux_1}, one proves the inequality \eqref{est_elliptic_ang_op_appendix_1} in the lemma.

\medskip

For any scalar functions $f_1$ and $f_2$, we compute 
\begin{align}
-2\, \check{\slashed{\textup{div}}}\check{\slashed{\mathcal{D}}}{}_2^{\star}\check{\slashed{\mathcal{D}}}{}_1^{\star}(f_1,f_2) &= (-\check{\slashed{\Delta}}+K-2K)\check{\slashed{\mathcal{D}}}{}_1^{\star}(f_1,f_2) \nonumber \\
&=\check{\slashed{\mathcal{D}}}{}_1^{\star}\check{\slashed{\mathcal{D}}}{}_1\check{\slashed{\mathcal{D}}}{}_1^{\star}(f_1,f_2)-2K\check{\slashed{\mathcal{D}}}{}_1^{\star}(f_1,f_2) \label{aux_3}
\end{align}
and (by repeated integration by parts) 
\begin{equation*}
\int |\check{\slashed{\mathcal{D}}}{}_1^{\star}\check{\slashed{\mathcal{D}}}{}_1\check{\slashed{\mathcal{D}}}{}_1^{\star}(f_1,f_2)|^2 = \int |\check{\nablasl}{}^3 f_1|^2+\int |\check{\nablasl}{}^3 f_2|^2+\mathbb{L}[f_1]+\mathbb{L}[f_2] \, ,
\end{equation*}
where in the latter identity we also used the identity \eqref{aux_comp_1} and defined the scalar functions ($i=1,2$)
\begin{equation*}
\mathbb{L}[f_i](t,r) := -\int|[\check{\slashed{\Delta}},\check{\nablasl}]f_i|^2+2\int \check{\slashed{\Delta}}f_i\cdot \check{\slashed{\textup{div}}}([\check{\slashed{\Delta}},\check{\nablasl}]f_i) +\int \check{\nablasl}{}^2f_i\cdot [\check{\slashed{\Delta}},\check{\nablasl}]\check{\nablasl}f_i \, .
\end{equation*}
The terms $\mathbb{L}[f_i]$ are lower order terms (in fact, first and second order). In particular, there exists a constant $C_r>0$, depending only on $r$, such that
\begin{equation*}
|\mathbb{L}[f_1]|+|\mathbb{L}[f_2]|+\int|K\check{\slashed{\mathcal{D}}}{}_1^{\star}(f_1,f_2)|^2 \leq C_r \sum_{i=1}^2\left(\|\check{\nablasl}{}^{i}f_1\|_{L^2(\mathbb{S}^2_{t,r},\check{\slashed{g}})}^2+\|\check{\nablasl}{}^{i}f_2\|_{L^2(\mathbb{S}^2_{t,r},\check{\slashed{g}})}^2\right) \, .
\end{equation*}
By recalling the identity \eqref{aux_3}, one proves the inequality \eqref{est_elliptic_ang_op_appendix_2} in the lemma. 
\end{proof}

\bibliography{elliptic_estimates_kerr} 
\bibliographystyle{hsiam}

\end{document}